\documentclass[a4paper,11pt]{article}
\usepackage{jheppub} 

\usepackage{multirow, multicol}
\usepackage{tabularx}
\usepackage[boxsize=1.1em,aligntableaux=top]{ytableau}
\usepackage{amsmath}
\usepackage{amssymb}
\usepackage{mathtools}
\title{Covariant Construction of Generalized Form Factors}

\author[b,c]{Hao Sun, }
\author[a,b]{Tuo Tan, }
\author[a,b,c,d]{Jiang-Hao Yu, }
\affiliation[a]{School of Fundamental Physics and Mathematical Sciences, Hangzhou Institute for Advanced Study, UCAS, Hangzhou 310024, China }
\affiliation[b]{Institute of Theoretical Physics, Chinese Academy of Sciences, Beijing 100190, China }
\affiliation[c]{School of Physical Sciences, University of Chinese Academy of Sciences,
Beijing 100190, China }
\affiliation[d]{International Centre for Theoretical Physics Asia-Pacific,
Beijing/Hangzhou, China }

\emailAdd{sunhao@itp.ac.cn, tantuo24@mails.ucas.ac.cn, jhyu@itp.ac.cn}

\abstract{We present a systematic technique for constructing the Lorentz-covariant structures of hadronic matrix elements of local operators. The spinor Young tableaux of the Lorentz group is employed to construct all possible structures for the matrix elements of arbitrary operators, using the relativistic wave functions and momenta of the initial and final state particles of arbitrary spin as building blocks. We obtain the form factor bases for the scalar, vector, and rank-2 tensor operators for particles with spin-$\frac{1}{2}$, $1$, $\frac{3}{2}$, and $2$, among which the general $P$ and $T$ form factors for spin-$\frac{3}{2}$ and spin-$2$ particles are presented for the first time. The independent form factor structures are also cross-checked by the non-relativistic counting and Hilbert Series method and we find there is redundant $P$ and $T$ conserved structure for spin-$2$ particles in literature. As an application, the matrix elements of general nonlocal operators can be expanded by towers of the matrix elements of local operators, and thus can be decomposed by the constructed form factor bases above. 
}

\begin{document}
\maketitle

\flushbottom

\section{Introduction}
\label{sec:intro}
Understanding the internal structure and fundamental interactions of composite particles is one of the core missions of modern physics. A primary approach to probing these properties involves studying the matrix elements of various local and nonlocal operators between composite particle states, which we will also refer to as local and nonlocal matrix elements for brevity. These matrix elements are standardly decomposed into linear combinations of independent tensor and spinor structures with scalar coefficients known as form factors (FFs).
Common examples are the electromagnetic and gravitational FFs, governed by the electromagnetic current~\cite{Sachs:1962zzc,Arrington:2006zm,Perdrisat:2006hj,Pacetti:2014jai,Sterman:1997sx,Jain:1999xc} and the energy-momentum tensor (EMT)~\cite{Pagels:1966zza,Burkert:2023wzr,Lorce:2024ipy}, which encode critical system properties. 
These studies are particularly relevant to quantum chromodynamics (QCD), where the elementary degrees of freedom, quarks and gluons, are never observed due to confinement, but only appear in experiments as hadronic states. Thus, the hadronic FFs, especially those involved with hard scattering probes, act as primary observables for probing interior and nonperturbative quark and gluon dynamics, tracking structural details from electromagnetic behavior to multidimensional spatial and momentum profiles of the constituents~\cite{Polyakov:2018zvc,Burkert:2018bqq,Perdrisat:2006hj,Lorce:2018egm,Leader:2013jra}.

Historically, a large amount of attention has been devoted to the matrix-element decomposition of spin-$\tfrac{1}{2}$ particles~\cite{Sachs:1962zzc,Arrington:2006zm,Perdrisat:2006hj,Burkert:2023wzr,Pacetti:2014jai}, especially the proton, because of its abundance as a stable particle and its central role in the structure of visible matter. Considerable effort has also been directed toward spin-1 particles~\cite{Efremov:1981vs,Brodsky:1992px,Gross:2002ge,Holstein:2006ud,Taneja:2011sy,Boer:2016xqr,Polyakov:2019lbq,Cosyn:2019aio,Pefkou:2021fni}.
Recently, beyond these low-spin cases, the theoretical exploration of higher-spin particles has drawn significant interest, such as spin-$\frac{3}{2}$~\cite{Nozawa:1990gt,Pascalutsa:2006up,Kim:2020lrs}, spin-$2$ and general spin~\cite{Jaffe:1989xy,Lorce2009Part1,Lorce:2009br,Lorce:2019sbq,Cotogno:2019xcl,Pefkou:2021fni,JiLiu2022}. 
On one hand, several counting rules for particles of arbitrary spin have been proposed to determine the number of the FFs within the matrix elements of scalar, vector, and rank-2 tensor currents~\cite{Scadron:1968zz, Williams:1970ms,  Cotogno:2019xcl,Cotogno:2019vjb,Lorce:2009bs}. On the other hand, explicit parameterizations for the matrix elements of condensate operator, electromagnetic current operator and EMT operator, have been provided as concrete examples for spin-$\frac{3}{2}$ and spin-$2$ systems.
Nevertheless, a systematic construction method has been lacking, and the derivations have typically been carried out on a case by case basis for specific particles or processes. In this context, a complete basis of the matrix elements of general local operators for such higher-spin systems (spin $>1$) has remained absent.
Furthermore, if $P$ and $T$ conservation are not assumed, as in matrix elements involving weak currents, there is currently no systematic method for determining the number of independent basis structures of them. As a result, as will be shown in this work, a rigorous re-examination of the existing spin-$2$ parameterization of matrix elements of rank-2 tensor operators under $P$ and $T$ conservation reveals a previously unnoticed redundancy in Ref.~\cite{Cotogno:2019vjb}.


On the other hand, the high-energy scattering reveals more precise parton structures, for example, generalized parton distributions (GPDs)~\cite{Cosyn:2018thq,Ji:1998pc,Muller:1994ses,Diehl:2003ny,Ji:1996ek,Radyushkin:1997ki,Goeke:2001tz,Boffi:2007yc,Burkardt:2002hr,Polyakov:2002yz} and transverse momentum-dependent distributions (TMDs)~\cite{Boussarie:2023izj,Collins:2011zzd}. These observables are described by nonlocal operators involving field insertions at distinct spacetime points. Essentially, any nonlocal operator and the associated matrix element can be Taylor expanded to an infinite series of the local ones. In particular, all the arising local operators are characterized by a unique geometric quantity known as twist.
The complete twist decomposition and tensor reduction for a variety of nonlocal operators, such as bilocal quark-aitiquark operator and bilocal gluon operator, has been systematically established, using tensor Young tableaux ~\cite{Geyer:1999uq, Geyer:2000ig}. Alongside these developments, the systematic enumeration of independent FFs has already been accomplished for the matrix elements of local twist-2 operators generated by the nonlocal quark-antiquark and gluon operators, using the tensor Young tableaux for the tensor reduction, and corresponding FF parameterization for spin-$\frac{1}{2}$ particles is achieved~\cite{Diehl:2003ny,Hagler:2004yt,Ji:2000id, Chen:2004cg}. However, the exact covariant matrix elements of these resulting local operators for higher-spin particles have not been systematically constructed, nor has a universal and systematic counting method been established. 

To address these challenges, we propose a comprehensive and systematic group-theoretical technique to construct the covariant bases for the local matrix elements, in terms of which the nonlocal matrix elements can be expanded systematically. 
Although the FFs can be enumerated via a non-relativistic method~\cite{Ji:2000id,Chen:2004cg} by constraining the relativistic states in the center-of-mass frame, it counts only the $P$- and $T$-conserving FFs. For the general case, we utilize the Hilbert series~\cite{Feng:2007ur,Lehman:2015via,Lehman:2015coa,Henning:2015daa,Henning:2017fpj,Marinissen:2020jmb,Henning:2015alf,Hanany:2010vu} to count all the FFs of different $P$ and $T$ properties. In particular, we present the counting results of the spin-1 particles explicitly.
In parallel, we construct the corresponding tensor structures by mapping the tensor indices to spinor indices. Such a Young tableaux method has been applied to the operator basis of effective field theories extensively~\cite{Li:2020gnx,Li:2020xlh,Li:2022tec}. In this paper, we generalize the Hilbert series method and the Young tableaux method to the construction of the Lorentz covariant basis associated with the FFs, since the tensor product in the spinor representations is easy to manipulate. Compared to the construction using tensor representations, the redundancies are much more transparent to identify and eliminate in the spinor representations, as is shown later. This simplification has two main origins. Firstly, extracting irreps in $SO(3,1)$ generally requires cumbersome trace subtractions, whereas the $SU(2)$ spinor framework treats outer products more directly and removes redundant components already at the level of the construction. Secondly, unlike in the tensor representation of $SO(3,1)$, the spinor Young tableaux associated with two conjugate representations are distinct, which makes the algebraic construction more transparent and substantially simplifies the classification of independent structures.

Equipped with these counting and construction methods, one can in principle construct the matrix elements of arbitrary operators between particle states of arbitrary spin. In this work, we reproduce the standard spin-$1/2$ and spin-$1$ results, consistent with the counting results obtained from both the nonrelativistic method and the Hilbert series. Furthermore, we present the complete and independent covariant tensor bases for spin-$3/2$ and spin-$2$ particles for the first time. In particular, all the structures of different eigenvalues under $T$ and $P$ transformations are considered. In addition, the bases of any higher spin can be obtained by following a similar procedure.
In terms of these covariant bases, any nonlocal matrix elements can also be expanded systematically in the bases after corresponding Taylor expansion, twist decomposition and tensor reduction using tensor Young tableaux, resulting in the complete decompositions of arbitrary nonlocal FFs for particles with any spin.

The remainder of this paper is organized as follows. In Section~\ref{sec:counting}, we introduce the physical objects of our study, establishing the definitions and kinematics of the FFs, and present the counting methods such as the non-relativistic method and Hilbert series. Section~\ref{sec:group} provides a necessary overview of group representation theory, focusing on the Lorentz group, tensor representations, and spinor formulations. The core of our methodology, detailing the specific construction procedure and the rigorous elimination of redundancies, is presented in Section~\ref{sec:method}, which also showcases the explicit construction results of low-spin targets (spin-$1/2$ and spin-$1$). Next in Section~\ref{sec:highspin}, we apply the construction method to the higher-spin cases (spin-$3/2$ and spin-$2$) directly, and present the covariant basis explicitly. In Section~\ref{sec:nonlocal}, we demonstrate the application of our constructed bases to the parameterization of nonlocal FFs. Finally, we summarize our findings and conclude in Section~\ref{sec:conclusion}.

\section{Generalized Form Factors and Their Countings}
\label{sec:counting}

Considering a local operator $\hat{O}^{\mu\nu\dots}(x)$, and two particle states of spin $s$ and momenta $p_1$ and $p_2$, respectively, the matrix element is
\begin{equation}
    \langle p_1,s|\hat{O}^{\mu\nu\dots}(x)|p_2,s\rangle\,. 
\end{equation}
The element vanishes unless the operator contains two field operators $\hat{\Phi}(x)$ interpolating the initial and final states,
\begin{equation}
    \hat{\Phi}(x)|p,s\rangle = u^s(p)\exp(-ip\cdot x)\,,
\end{equation}
where $u^s$ is the corresponding wave function. As a result, the matrix element takes the form
\begin{align}
    \langle p_1,s|\hat{O}^{\mu\nu\dots}(x)|p_2,s\rangle &= \int \tilde{dp}\exp(-iq\cdot x) T^{\mu\nu\dots}  \,,
\end{align}
where $q=p_2-p_1$ and $T^{\mu\nu\dots}$ is a tensor composed of the initial and final momenta and wave functions. In particular, the tensor $T^{\mu\nu\dots}$ can be decomposed as follows
\begin{equation}
    T^{\mu\nu\dots} = \sum_\mathbf{r} \sum_{i} F_{\mathbf{r},i} \times T^{\mu\nu\dots}_{\mathbf{r},i}\,,
\end{equation}
where $T^{\mu\nu\dots}_{\mathbf{r},i}$ represents the tensor basis element with Lorentz irreducible representations (irreps) labeled by $\mathbf{r}$ decomposed from the operator $\hat{O}^{\mu\nu\dots}(x)$, and $F_{\mathbf{r},i}$ are the corresponding coefficients. The coefficients $F_{\mathbf{r},i}$'s are Lorentz scalars and are referred to as the FFs~\cite{Pacetti:2014jai}. Thus, all these FFs are determined by their corresponding tensors.
Before the explicit constructions, it is helpful to count their number first. A method adopting a specific frame of reference counts the P- and T-conserved tensors, which is referred to as the non-relativistic counting method~\cite{Ji:2000id,Chen:2004cg}. While for the general ones, we use the Hilbert series~\cite{Grinstein:2023njq} to count them.

\subsection{Non-relativistic Counting Method}


Because of the cross symmetry, a 1-to-1 matrix element is equivalent to an annihilation matrix element 
\begin{equation}
    \langle p_1,s|\hat{O}^{\mu\nu\dots}(x)|p_2,s\rangle \sim \langle 0|\hat{O}^{\mu\nu\dots}(x)|\overline{p_1,s};p_2,s\rangle\,,
\end{equation}
where $|\overline{p_1,s};p_2,s\rangle = \overline{|p_1,s\rangle} \otimes |p_2,s\rangle$ is the two-particle state, and $\overline{|p_1,s\rangle}$ represents the anti-particle state.

This method is called non-relativistic because the two-particle state $|\overline{p_1,s};p_2,s\rangle $ is analysed in the center-of-mass (COM) frame, in which the two-particle state can be decomposed into LS-coupled states
\begin{equation}
    |\overline{p_1,s};p_2,s\rangle \rightarrow \sum_{L,S}\sum_{J} |pJM,LS\rangle\,,
\end{equation}
where $p_1=(p_0,\vec{p})$, $p_2=(p_0,-\vec{p})$, and 
\begin{equation}
    S=0,1,\dots,2s,\quad L = |J-S|,\dots,J+S-1,J+S.
\end{equation}
The LS coupled states are convenient since they are eigenstates of P and C,
\begin{align}
    P|pJM,LS\rangle &= \eta_P(-1)^L|pJM,LS\rangle\,,\notag \\
    C|pJM,LS\rangle &= (-1)^{L+S}|pJM,LS\rangle\,,
\end{align}
where $\eta_P=-1$ for two fermions and $\eta_P=1$ for two bosons. For the cases of two fermions $s=1/2$ and two bosons $s=1$, all the quantum numbers $J_L^{PC}$ up to $J=2$ are listed in Tab.~\ref{tab:JLPC}.

\begin{table}[]
    \centering
    \begin{tabular}{c|l}
\hline
\multicolumn{2}{c}{$s=1/2$} \\
\hline
$s=0$ & $0_0^{-+}\,,\quad 1_1^{+-}\,,\quad 2_2^{-+}$ \\
\hline
$s=1$ & $0_1^{++}\,,\quad 1_0^{--}\,,\quad 1_1^{++}\,,\quad 1_2^{--}\,,\quad 2_1^{++}\,,\quad 2_2^{--}\,,\quad 2_3^{++}$ \\
\hline
\multicolumn{2}{c}{$s=1$} \\
\hline
$s=0$ & $0_0^{++}\,,\quad 1_1^{--}\,,\quad 2_2^{++}$ \\
\hline
$s=1$ & $0_1^{-+}\,,\quad 1_0^{+-}\,,\quad 1_1^{-+}\,,\quad 1_2^{+-}\,,\quad 2_1^{-+}\,,\quad 2_2^{+-}\,,\quad 2_3^{-+}$ \\
\hline
$s=2$ & $0_2^{++}\,,\quad 1_1^{--}\,,\quad 1_2^{++}\,,\quad 1_3^{--}\,,\quad 2_0^{++}\,,\quad 2_1^{--}\,,\quad 2_2^{++}\,,\quad 2_3^{--}\,,\quad 2_4^{++}$ \\
\hline
    \end{tabular}
    \caption{The $J_L^{PC}$ for $s=1/2$ and $s=1$ cases up to $J=1$.}
    \label{tab:JLPC}
\end{table}

On the other hand, the irreducible tensors of the Lorentz group irreps $ T^{\mu\nu\dots}\in (j_L,j_R)$\footnote{The Lorentz irreps are labeled by the Casimir elements of its double-cover group $SU(2)_L\times SU(2)_R$. The representation theory of the Lorentz group is discussed in next section.} can be decomposed into tensors with angular momentum $J=|j_L-j_R|,\dots j_L+j_R$, because of the relation $\vec{J} = \vec{J}_L+\vec{J}_R$. At the same time, these tensors are also $C\,,P$ eigenstates. For irreps $(j,j)$, the contributing angular momenta are $J = 0,1,\dots,2j$, and the corresponding $P$ and $C$ charges are $(-1)^J$ and $(-1)^{2j}$. While for irreps $(j_1,j_2)$ with $j_1\neq j_2$, the $C$ charges are $(-1)^{2\times Max(j_1,j_2)}$, but for $P$, the eigenstates are combinations of the irreps
\begin{equation}
    (j_1,j_2) - (j_2,j_1)\,,\quad (j_1,j_2) + (j_2,j_1)\,,
\end{equation}
of which the former gives $P$ charge $(-1)^{J+1}$, and the latter gives $P$ charge $(-1)^{J}$. The contributing angular momenta and the corresponding quantum numbers $J^{PC}$ of the first several Lorentz group irreps are listed in Tab.~\ref{tab:tensors}.

\begin{table}[]
    \centering
    \begin{tabular}{c|c}
\hline
Irreps & $J^{PC}$ \\
\hline
$(0,0)$ & $0^{++}$ \\
\hline
$(\frac{1}{2},\frac{1}{2})$ & $0^{+-}\,,\quad 1^{--}$ \\
\hline
$(1,0)\oplus (0,1)$ & $1^{-+}\,,1^{++}$ \\
\hline
$(1,1)$ & $0^{++}\,,\quad 1^{-+}\,, \quad 2^{++}$ \\
\hline
    \end{tabular}
    \caption{The quantum numbers $J^{PC}$ of the Lorentz irreducible tensors up to irrep $(1,1)$.}
    \label{tab:tensors}
\end{table}

Given the two sets of quantum numbers of the two-particle states and the irreducible tensors, the matrix element vanishes unless the quantum numbers are the same, with the assumption that $P$ and $C$ are conserved. For example, we consider the spin-1/2 particles, the irreducible tensor of irrep $(1/2,1/2)$ contributes to two angular momenta $0^{+-}$ and $1^{--}$, where $0^{+-}$ have no correspondence in Tab.~\ref{tab:JLPC}, while $1^{--}$ corresponds to $1_0^{++}\,, 1_2^{++}$. Thus, there are two irreducible tensors conserving both $P$ and $C$, which are the well-known ones
\begin{equation}
    T^\mu_1 = \overline{u} p^\mu u\,,\quad T^\mu_2 = i\overline{u}\sigma^{\mu\nu}q_\nu u\,,
\end{equation}
where $p^\mu = p_1^\mu+p_2^\mu$ is the total momentum and $q^\mu=p_1^\mu-p_2^\mu$ is the momentum transfer. 
The matrix element is
\begin{equation}
    \langle p_1,\frac{1}{2}|\hat{O}^\mu (x)|p_2,\frac{1}{2}\rangle = \int \tilde{dp} \left[F_1 \left(\overline{u}_{\sigma_1} p^\mu u_{\sigma_2}\right) + i F_2 \left(\overline{u}_{\sigma_1}\sigma^{\mu\nu}q_\mu u_{\sigma_2}\right)\right]\exp(-iq\cdot x)\,,
\end{equation}
where $F_{1,2}$ are the two independent FFs. 

Similar arguments can be applied to any spin and any Lorentz group irreps. In Tab.~\ref{tab:numbers} the numbers of $P$- and $C$-conserving FFs of $s=1/2,1,3/2,2$ and irreps $(1/2,1/2), (1,0)\oplus(0,1)$ and $(1,1)$ are listed.
In particular, the numbers in Tab.~\ref{tab:numbers} count the independent FFs conserving $P$ and $C$ (or equivalently, conserving $P$ and $T$ due to the CPT theorem). If these conservation laws are not satisfied, more FFs arise.

\begin{table}[b]
    \centering
    \begin{tabular}{|c|c|c|c|c|}
\hline
& $(0,0)$ & $(\frac{1}{2},\frac{1}{2})$ & $(1,0)\oplus(0,1)$ & $(1,1)$ \\
\hline
$s=\frac{1}{2}$ & 1 & 2 & 1 & 3 \\
\hline
$s=1$ & 2 & 3 & 2 & 7 \\
\hline
$s=\frac{3}{2}$ & 2 & 4 & 4 & 8 \\
\hline
$s=2$ & 3 & 5 & 4 & 12 \\
\hline
    \end{tabular}
    \caption{The numbers of independent FFs conserving both $P$ and $C$ of different spins and different irreps.}
    \label{tab:numbers}
\end{table}

\subsection{Hilbert Series Counting Method}

Hilbert series is a systematic counting tool to count group invariants~\cite{Feng:2007ur,Lehman:2015via,Lehman:2015coa,Henning:2015daa,Henning:2017fpj,Marinissen:2020jmb,Henning:2015alf,Grinstein:2023njq,Graf:2020yxt,Sun:2022aag,Graf:2022rco}. Considering a set of covariants $\{q_i\}$ transforming under some irreps of a group $G$, the Hilbert series takes the form,
\begin{equation}
    \mathcal{HS}^{\text{inv}}_G(q) = 1 + n_1 q + n_2 q^2 + \dots\,,
\end{equation}
where on the right hand the superscripts indicate the degrees of the invariants, and the associated coefficients $n_i$ are the numbers of the independent invariants of degree $i$. Since all the invariants are closed under multiplication, they form a ring. For any ring, there exists a set of invariants, in terms of which any invariant can be expressed as a unique polynomial. These special ones are known as the generators of the ring. Furthermore, the generators can be classified into two kinds, one could be any power in the polynomials of the other ones, called primary invariants, while the other could be at most linear, called secondary invariants. With such a separation, the Hilbert series can be organized as a fractional form,
\begin{equation}
    \mathcal{HS}^{\text{inv}}_G(q) = \frac{N(q)}{D(q)}\,,
\end{equation}
where the numerator $N(q)$ counts the secondary invariants, and the denominator $D(q)$ counts the primary ones.

Essentially, the Hilbert series can be obtained not from the explicit invariants, but from the orthogonality of the characters of group irreps. For connected and compact groups, the representative matrices can be reduced to the maximal torus, which takes the diagonal form,
\begin{equation}
    g(z) = \text{diag}(f(z)\,, g(z)\,,\dots)\,,
\end{equation}
with $f\,,g$ symmetric functions of the complex variables $z=\exp(i\theta)$, then the Hilbert series is given by the integral that
\begin{equation}
    \mathcal{HS}^{\text{inv}}_G = \int_G d\mu_G \frac{1}{\prod_q\det[1-q g_q(z)] }\,,\label{eq:hs1}
\end{equation}
where $g_q(z)$ is the group elements corresponding to the irrep of $q$ expressed by the maximal-torus variables. This integral formula is known as the Molien-Weyl formula, one of whose advantages is that it is easy to generalize to covariants~\cite{Grinstein:2023njq}. For irrep $R$, the character can also be expressed by the maximal-torus variables, $\chi_R(z)$, then the Hilbert series counting the independent covariants in the irrep $R$ is obtained from the integral
\begin{equation}
    \mathcal{HS}^{R}_G = \int_G d\mu_G \frac{\chi_R(z)}{\prod_q\det[1-q g_q(z)]}\,,
\end{equation}
which contains the invariant case in Eq.~\eqref{eq:hs1} as a special case.
Mathematically, the covariants form a module. It also contains a set of generators, and any covariants can be expressed in terms of them, with the coefficients being the group invariants. 
Thus, the corresponding Hilbert series also takes the fractional form. 

Concentrating on the nonequivalent tensors, they are covariants of the Lorentz group. The maximal torus is $U(1) \times U(1)$, whose variables are noted as $x\,,y$, respectively, and the associated characters of the irreps are, for example,
\begin{align}
    \chi_{(0,0)} & =1,\quad \chi_{(\frac{1}{2},0)} = x+\frac{1}{x}\,,\quad \chi_{(0,\frac{1}{2})} = y+\frac{1}{y}\,,\\
    \chi_{(1,0)} &= x^2+1+\frac{1}{x^2}\,,\quad \chi_{(0,1)} = y^2+1+\frac{1}{y^2}\,,\\
    \chi_{(\frac{1}{2},\frac{1}{2})} &= xy + \frac{x}{y} + \frac{y}{x} + \frac{1}{xy}\,,\\
    \chi_{(1,1)} &= 1 + \frac{1}{x^2} + x^2 + \frac{1}{y^2} + \frac{1}{x^2 y^2} + \frac{x^2}{y^2} + y^2 + \frac{y^2}{x^2} + x^2 y^2\,.
\end{align}
Considering the spin-1 case, the building blocks of the tensors are the two wave functions $\varepsilon_1,\varepsilon_2^\dagger$, and the corresponding momenta, $p_1.p_2$. For convenience, we define, $P=p_2+p_1$ and $q=p_2-p_1$, there are equations
\begin{equation}
\label{eq:relation1}
    P \cdot q =0\,,\quad P \cdot \varepsilon_2^\dagger = - q \cdot \varepsilon_2^\dagger \,,\quad P \cdot \varepsilon_1 = - q \cdot \varepsilon_1\,.
\end{equation}
These relations are not complemented in the Molien-Weyl formula, but should be taken into account manually. The procedure is presented in App.~\ref{app:hs}. 
With these relations considered, the Hilbert series of the tensors of the different irreps takes the form
\begin{equation}
    \mathcal{HS}^{R}(P,q,\varepsilon_2^\dagger,\varepsilon_1) = \frac{N_R(P,q,\varepsilon_2^\dagger,\varepsilon_1)}{D(P,q,\varepsilon_2^\dagger,\varepsilon_1)}\,,
\end{equation}
where the denominator is universal,
\begin{equation}
    D(P,q,\varepsilon_2^\dagger,\varepsilon_1) = (1-q\varepsilon_1)(1-q \varepsilon_2^\dagger)(1-\varepsilon_1 \varepsilon_2^\dagger)\,,
\end{equation}
and the numerators of the various irreps are
\begin{align}
    N_{(0,0)} &= 1+Pq \varepsilon_1 \varepsilon_2^\dagger \,,\\
    N_{(\frac{1}{2},\frac{1}{2})} &= P+q + \varepsilon_1 + \varepsilon_2^\dagger + Pq \varepsilon_1 + Pq \varepsilon_2^\dagger + P \varepsilon_1\varepsilon_2^\dagger + q \varepsilon_1 \varepsilon_2^\dagger\,,\\
    N_{(1,0)\oplus (0,1)} &= 2Pq + 2 P \varepsilon_1 + 2 P\varepsilon_2^\dagger + 2 q \varepsilon_1 + 2 q\varepsilon_2^\dagger + 2 \varepsilon_1 \varepsilon_2^\dagger - P^2q^2 \varepsilon_1 \varepsilon_2^\dagger\,,\\
    N_{(1,1)} &= P^2 + q^2 + Pq + P\varepsilon_1 + P \varepsilon_2^\dagger + q\varepsilon_1 + q \varepsilon_2^\dagger + P^2q \varepsilon_1 + P^2q \varepsilon_2^\dagger \notag \\
    &+ q^2P \varepsilon_1 + q^2P \varepsilon_2^\dagger + \varepsilon_1 \varepsilon_2^\dagger + P^2 \varepsilon_1 \varepsilon_2^\dagger + 3Pq \varepsilon_1 \varepsilon_2^\dagger + q^2 \varepsilon_1 \varepsilon_2^\dagger\,.
\end{align}
Here are some remarks,
\begin{itemize}
    \item The $(0,0)$ representation is a special case of the covariant module, which contains the identity element $1$. The other term in the numerator $N_{(0,0)}$ corresponds to the $P$-odd tensor $T_p = \epsilon^{\mu\nu\rho\lambda} P_\mu q_\nu \varepsilon_1{}_\rho \varepsilon_2^\dagger{}_\lambda$.
    \item For the representation $(1,0)\oplus (0,1)$, the minus term means the tensor such as $\epsilon^{\mu\nu\rho\lambda}P_\rho q_\lambda T_p$ is redundant, since of $\epsilon^{\mu\nu\rho\lambda}\epsilon_{\mu'\nu'\rho'\lambda'} = -\delta^\mu_{\mu'} \delta^\nu_{\nu'} \delta^\rho_{\rho'}\delta^\lambda_{\lambda'} + \text{permutations}$.
    \item For the irreducible tensors corresponding to FFs, they are linear in the initial and final wave functions, respectively. Thus, the expansion of the Hilbert series is truncated. 
    For example, the expansion of the trivial representation is
    \begin{equation}
        \frac{N_{(0,0)}}{D} \sim 1+\varepsilon_1\varepsilon_2^\dagger + q^2\varepsilon_1 \varepsilon_2^\dagger+ Pq \varepsilon_1 \varepsilon_2^\dagger\,.
    \end{equation}
    Consequently, the tensors of each irrep can be obtained, which are listed in Tab.~\ref{tab:number1}.
\end{itemize}
\begin{table}[ht]
\renewcommand{\arraystretch}{1.8}
    \centering
    \begin{tabular}{|c|c|c|c|c|}
\hline
irrep & $(0,0)$ & $\left(\frac{1}{2},\frac{1}{2}\right)$ & $(1,0)\oplus (0,1)$ & $(1,1)$ \\
\hline
number& 3 & 12 & 15 & 23 \\
\hline
    \end{tabular}
    \caption{The numbers of the irreducible tensors of the irreps up to $(1,1)$ of the spin-1 FFs. All the tensors of different C, P eigenvalues are included.}
    \label{tab:number1}
\end{table}
The Hilbert series can be generalized to count invariants of different C, P eigenvalues~\cite{Graf:2020yxt,Sun:2022aag,Graf:2022rco}. But as discussed above, the relations such as Eq.~\eqref{eq:relation1} are difficult to implement. In this paper, we propose an alternative method using Young tableaux that systematically identifies and eliminates all such relations.

\section{Lorentz Group Representation Theory}
\label{sec:group}


To rigorously classify the kinematic structures and FFs of arbitrary-spin particles, we must return to the foundational principles of group representation theory.
We review some useful conclusions of the representation theory of the Lorentz group here, leaving more details to the textbook (see, e.g., Refs.~\cite {Ma:1989,Tung:1985iqd}).

\subsection{Tensor Representation}

The decomposition of tensor spaces is governed by the symmetric group $S_n$. For illustration, let $V$ be an $N$-dimensional vector space over the complex field $\mathbb{C}$. We define a primitive vector basis for $V$ as $\{ e_{\mu} \}_{\mu=1}^N$. The $n$-th rank tensor product space, denoted as $V^{\otimes n} = V \otimes V \otimes \dots \otimes V$, is a vector space of dimension $N^n$. The natural basis for this full tensor space is the primitive tensor basis, constructed by the direct products of the vector bases:
\begin{equation}
    | \mu_1 \mu_2 \dots \mu_n \rangle \equiv e_{\mu_1} \otimes e_{\mu_2} \otimes \dots \otimes e_{\mu_n}\,,
\end{equation}
where each tensor index $\mu_k \in \{1, 2, \dots, N\}$. 
A general physical tensor $T$ of rank $n$ is defined as a linear combination of all possible primitive tensor bases:
\begin{equation}
    T = \sum_{\mu_1, \dots, \mu_n} T^{\mu_1 \mu_2 \dots \mu_n} e_{\mu_1} \otimes e_{\mu_2} \otimes \dots \otimes e_{\mu_n}\,.
\end{equation}

There are two distinct but fundamentally related groups acting on this space. For both groups, we explicitly define their actions on the abstract basis vectors and their corresponding effects on the tensor components:
\begin{itemize}
    \item \textbf{The General Linear Group $GL(N, \mathbb{C})$} (or subgroups like $SU(N), SO(N)$): An element $R \in G$ acts linearly on the vector space. Its action on the tensor basis extends linearly to the direct product:
    \begin{equation}
        R \cdot T = \sum_{\mu_1, \dots, \mu_n} T^{\mu_1 \mu_2 \dots \mu_n} (R e_{\mu_1}) \otimes (R e_{\mu_2}) \otimes \dots \otimes (R e_{\mu_n})\,.
    \end{equation}
    Equivalently, when analyzing the transformation of the tensor itself, the components $T^{\mu_1 \dots \mu_n}$ transform via a multilinear direct product:
    \begin{equation}
        T^{\mu_1\ldots\mu_n} \quad \rightarrow \quad R^{\mu_1}\!_{\nu_1} \dots R^{\mu_n}\!_{\nu_n} T^{\nu_1 \dots \nu_n}\,,
    \end{equation}
    where $R^{\mu_i}{}_{\nu_i}$ are matrix representatives of the group element $R$ in the space $V$.   
    \item \textbf{The Symmetric Group $S_n$:} An element $\sigma \in S_n$ permutes the $n$ distinct \textit{positions} (or slots) of the tensor indices. Its action on the primitive tensor basis is defined as:
    \begin{equation}
        \sigma \cdot \left( e_{\mu_1} \otimes \dots \otimes e_{\mu_n} \right) = e_{\mu_{\sigma^{-1}(1)}} \otimes \dots \otimes e_{\mu_{\sigma^{-1}(n)}}\,.
    \end{equation}
    To maintain the invariance of the tensor $T$ under these permutations, the components must transform with the inverse permutation, meaning the action on the tensor components is:
    \begin{equation}
        \sigma \cdot T^{\mu_1 \mu_2 \dots \mu_n} = T^{\mu_{\sigma(1)} \mu_{\sigma(2)} \dots \mu_{\sigma(n)}}\,.
    \end{equation}
\end{itemize}
An essential property is that the actions of the continuous group $G$ and the symmetric group $S_n$ commute: $R\cdot (\sigma \cdot T) = \sigma \cdot (R\cdot T)$. This establishes the Schur-Weyl Duality.
This implies that the decompositions into irreps of both the continuous group and the symmetric group can be performed simultaneously. Actually, the two kinds of irreps correspond to each other one-by-one, which means we can label the irreps of $G$ by the irreps of $S_n$, while the latter is labeled by the SYDs.
We denote a Young diagram by $[a_1, a_2, \dots]$, where the $i$-th entry $a_i$ represents the number of boxes in the $i$-th row. Then the SYDs satisfy $a_i \geq a_{i+1}$.



In particular, the SYDs naturally generate a set of bases for the corresponding irreps by filling the boxes with the numbers ranging from $1$ to $N$, respecting the following rules: in each row, the filling numbers increase weakly, while in each column, the filling numbers increase strictly. The resultant tableaux are referred to as semi-standard Young tableaux (SSYTs). Because the maximum of the filling number is $N$, all the SYDs involving columns longer than $N$ vanishes, thus correspond to no irreps. 
 For example, the SYD $\ydiagram{2}$ corresponds to a $SU(2)$ irrep, whose SSYTs are
\begin{equation}
    \ytableaushort{11}\,,\quad \ytableaushort{12}\,,\quad \ytableaushort{22}\,,
\end{equation}
which means the irrep is of dimension 3, thus, is the $SU(2)$ adjoint representation. 
Moreover, any tableaux other than the SSYTs can be reduced to the latter following the Fock conditions. Besides, for any irrep, the dimension is determined by the hook length formula. More details can be found in Ref.~\cite{Ma:1989}.

Another advantage of the SYDs is that the decomposition of the tensor product can be performed via the outer product of the corresponding SYDs.
The outer product respects the Littlewood-Richardson (L-R) Rule. For $SU(2)$ group, the decomposition of two fundamental representations can be performed as
\begin{equation}
    \ydiagram{1} \otimes\ydiagram{1} = \ydiagram{1,1} \oplus \ydiagram{2}\,,
\end{equation}
where the rank-2 tensors are decomposed into symmetric and antisymmetric sectors.

Unlike the special unitary groups $SU(N)$, the orthogonal groups $SO(N)$ possess an invariant metric $\delta_{\mu\nu}$ (or $g_{\mu\nu}$ in Minkowski space). This allows for index contractions. Any tensor can be uniquely decomposed into a traceless part and a trace part. The trace part inherently contains the metric $g^{\mu_i \mu_j}$, which dynamically reduces the tensor to a lower-rank representation. Thus, the irreps of $SO(N)$ strictly consist of \textit{traceless} tensors with a specific permutation symmetry dictated by a valid Young tableau. A well-known theorem states that for $SO(N)$, a traceless tensor identically vanishes if the sum of the lengths of its first two columns exceeds $N$ ($c_1 + c_2 > N$).

Moreover, in an $SO(N)$ group, when the number of boxes in the first column of a SYD strictly exceeds $N/2$, the tensor basis can be contracted with the $N$-th rank completely antisymmetric tensor $\epsilon_{\alpha_1 \dots \alpha_N}$ to form a dual tensor basis. The resulting dual SYD defines an equivalent representation.
Two such SYDs are dual to each other, and their corresponding representations are strictly equivalent. 

Specializing to the proper orthochronous Lorentz group $SO^+(3,1)$, we operate in a spacetime where $N=4$. In addition to the invariant metric $g_{\mu\nu}$, which permits trace contractions, the Lorentz group is endowed with the completely antisymmetric Levi-Civita symbol $\epsilon_{\mu\nu\rho\sigma}$. The existence of this tensor naturally gives rise to the concept of the \textbf{Hodge duality}. Applying the duality formula to $N=4$, we discuss the duality relationship as follow:

\vspace{0.5em}
\noindent \textbf{1. Four-Row SYDs:} \\
The traceless constraint strictly requires $c_2 = 0$. The only valid SYD is a single column of 4 boxes. Contracting this with $\epsilon_{\mu\nu\rho\sigma}$ yields a scalar. Thus, the 4-row SYD duplicates the trivial scalar representation.

\vspace{0.5em}
\noindent \textbf{2. Three-Row SYDs:} \\
The traceless constraint $3 + c_2 \leqslant 4$ forces $c_2 \leqslant 1$, which means the diagram must take a hook-shape $[s, 1, 1]$. For $N=4$, Hodge duality maps a column of length 3 to a column of length $4-3=1$. Consequently, only the first row of the dual SYD survives:
\begin{equation}
    [s, 1, 1] \quad \xleftrightarrow{\text{Hodge Duality}} \quad [s]\,,
\end{equation}
where we have used an overline to indicate duality.
Visually, taking $s=3$ as an example:
\begin{equation}
    \ydiagram{3,1,1} \quad \xleftrightarrow{\epsilon_{\mu\nu\rho\sigma}} \quad \overline{\ydiagram{1,1,1}}\ydiagram{2} = \ydiagram{3}\,.
\end{equation}
This duality is formally realized by contracting the Levi-Civita tensor with the totally antisymmetric indices of the first column of the tensor basis. Specifically, the Young operator ensures that the basis $\Phi_{\mu\nu\rho, \sigma_1 \dots \sigma_{s-1}}$ is totally antisymmetric in its first three indices. Its dual basis $\tilde{\Phi}_{\kappa, \sigma_1 \dots \sigma_{s-1}}$, which corresponds to a single-row symmetric tensor, is constructed as:
\begin{equation}
    \tilde{\Phi}_{\kappa, \sigma_1 \dots \sigma_{s-1}} = \frac{1}{3!} \epsilon_{\kappa \mu\nu\rho} \Phi^{\mu\nu\rho}_{\quad \,\, \sigma_1 \dots \sigma_{s-1}}\,.
\end{equation}
In this framework, the physical tensor components $T$ follow the inverse permutation and remain totally symmetric in the indices of the first row. Thus, the three-row hook-shaped SYD $[s,1,1]$ is equivalent to a rank-$s$ totally symmetric tensor representation.

\vspace{0.5em}
\noindent \textbf{3. Two-Row SYDs:} \\
For a two-row SYD, the length of the first column is $c_1 = 2$. The traceless constraint $c_1 + c_2 \leqslant 4$ then strictly requires $c_2 \leqslant 2$, implying the second row can contain either 1 or 2 boxes. When Hodge duality is applied to the columns of length 2, the dual column length remains $4 - 2 = 2$, meaning the dual SYD possesses the exact same shape as the original one:
\begin{equation}
    [s, t] \quad \xleftrightarrow{\text{Hodge Duality}} \quad [s, t]\,, \quad (t =1,2)\,.
\end{equation}
The duality operation is fundamentally realized at the level of the tensor basis. Specifically, the Young operator ensures that the irreducible tensor basis is totally antisymmetric in its column indices. By contracting the Levi-Civita tensor $\epsilon_{\mu\nu\alpha\beta}$ with these antisymmetric basis indices, one induces the dual representation. To illustrate this, consider the basis $\Phi_{\mu\nu,\rho}$ corresponding to the SYD $[2,1]$, where the first two indices (the first column) are totally antisymmetric. Its dual basis $\tilde{\Phi}_{\mu\nu,\rho}$ is constructed as:
\begin{equation}
    \tilde{\Phi}_{\mu\nu,\rho} = \frac{1}{2} \epsilon_{\mu\nu \alpha\beta} \Phi^{\alpha\beta}_{\quad \rho}\,.
\end{equation}
While the dual basis $\tilde{\Phi}$ shares the same index symmetries, the negative square of the Hodge star operator in Minkowski spacetime ($\star^2 = -1$) splits the space into two distinct irreducible subspaces. We can define the self-dual and anti-self-dual bases $\Phi^\pm = \frac{1}{2}(\Phi \mp i\tilde{\Phi})$, which satisfy $\tilde{\Phi}^\pm = \pm i \Phi^\pm$. 

The dual representations $T^+$ and $T^-$, associated with the self-dual and anti-self-dual bases, are complex conjugates of each other. Under $P$ transformation, the Levi-Civita tensor behaves as a pseudo-tensor and acquires an overall minus sign. Consequently, the Hodge duality operation reverses sign ($\tilde{\Phi} \xrightarrow{P} -\tilde{\Phi}$), meaning a representation and its dual interchange up to a relative sign. For the self-dual and anti-self-dual bases specifically, 
they are of opposite $P$ charges, $\Phi^\pm \xrightarrow{P} \Phi^\mp$. In the spinor formalism, they perfectly map to $(j_L, j_R)$ and $(j_R, j_L)$, representing states with opposite chiralities that naturally interchange under $P$ transformation.

\subsection{Spinor Representation}

The complexified Lie algebra of the Lorentz group is isomorphic to the direct-sum algebra, $\mathfrak{so}(3,1)_\mathbb{C} \simeq \mathfrak{sl}(2,\mathbb{C}) \oplus \mathfrak{sl}(2,\mathbb{C})$. Thus, finite-dimensional irreps are uniquely labeled by a pair of half-integers $(j_L, j_R) = (\frac{n_1}{2}, \frac{n_2}{2})$, whose dimension is:
\begin{equation}
    \dim(j_L, j_R) = (2j_L + 1)(2j_R + 1)\,.
\end{equation}
We refer to the representation of the Lorentz group via its decomposed two $SU(2)$ groups as the spinor representation, since the quantities of the $SU(2)$ irrep $\frac{1}{2}$ are called spinors. 

Every spinor irrep of $SO(3,1)$ can be represented by a pair of SYDs, 
    \begin{equation}
    \ytableausetup{aligntableaux=center}
        (j_L\,,j_R) \longleftrightarrow \left(\underbrace{\ydiagram{1}\dots\ydiagram{1}}_{2j_L}\,,\underbrace{\ydiagram{1}\dots\ydiagram{1}}_{2j_R}\right)\,,
    \end{equation}
specifying the symmetry properties of the left-handed and right-handed spinor indices respectively. We will refer to these SYDs as spinor SYDs.
In this notation, the decomposition of the tensor product of two irreps is determined by the decompositions of the two $SU(2)$ tensor products,
    \begin{equation}
    \label{eq:lorentz_tensor_product}
        (j_L{}_1\,,j_R{}_1)\otimes (j_L{}_2\,,j_R{}_2) \rightarrow \bigoplus(j_L\,,j_R)\,,
    \end{equation}
    where $|j_L{}_1-j_L{}_2|\leq j_L\leq j_L{}_1+j_L{}_2$ and $|j_R{}_1-j_R{}_2|\leq j_R\leq j_R{}_1+j_R{}_2$. At the same time, the decomposition can also be expressed by the outer product of the SYDs via the L-R rule. For example, 
    \begin{align}
    \ytableausetup{boxsize=0.8em}
    (\frac{1}{2}\,,\frac{1}{2})\otimes (\frac{1}{2}\,,\frac{1}{2}) &\longrightarrow (0,0) \oplus (0,1) \oplus (1,0) \oplus (1,1): \notag \\
        \left(\ydiagram{1}\,,\ydiagram{1}\right)\otimes \left(\ydiagram{1}\,,\ydiagram{1}\right)& \longrightarrow \left(\ydiagram{1,1}\,,\ydiagram{1,1}\right) \oplus \left(\ydiagram{1,1}\,,\ydiagram{2}\right) \oplus \left(\ydiagram{2}\,,\ydiagram{1,1}\right) \oplus \left(\ydiagram{2}\,,\ydiagram{2}\right)\,.\label{eq:littlewood}
    \end{align}
In Tab.~\ref{tab:lorentz_irreps} we have listed the first several irreps of the Lorentz group and their related properties. In particular, the $(\frac{1}{2}\,,0)/(0\,,\frac{1}{2})$ is the left-/right-handed spinor representation, and the variables $\xi_a/\eta^{\dot{a}}$ is the left-/right-handed spinor, where we have used the van der Waerden symbols to distinguish them. The irrep $(\frac{1}{2}\,,\frac{1}{2})$ is of 4 dimensions, and is the fundamental representation of the Lorentz group, also called the vector representation, the variables $v^\mu$ of which are called 4-vector or vector. According to the tensor product rule in Eq.~\eqref{eq:lorentz_tensor_product}, the vector representation is equivalent to the tensor product of the two spinor representations,
\begin{equation}
    (\frac{1}{2}\,,0) \otimes (0\,,\frac{1}{2}) = (\frac{1}{2}\,,\frac{1}{2})\,,
\end{equation}
which is realized by the projection of $\sigma^\mu_{a\dot{a}}$, where $\sigma^\mu=(I,\vec{\sigma})$ and $\vec{\sigma}$ are the 3 Pauli matrices. In detail, the equivalence can be expressed by the variables as
\begin{equation}
\label{eq:transf}
    \xi^a \sigma^{\mu}_{a\dot{a}}\eta^{\dot{a}} = \xi \sigma^\mu \eta = v^\mu \in (\frac{1}{2}\,,\frac{1}{2})\,.
\end{equation}
Conversely, the vector $v^\mu$ can also be projected to be spinor form by $\sigma^\mu$ as
\begin{equation}
    v_\mu \sigma^{\mu}_{a\dot{a}} = v_{a\dot{a}}\,.
\end{equation}
It can be verified that $\sigma^\mu$ indeed behaves as a projector,
\begin{align}
    \sigma^{\mu}_{a\dot{a}}\sigma_\mu{}_{b\dot{b}} &= 2\epsilon_{ab}\epsilon_{\dot{a}\dot{b}}\,, \\
    \epsilon^{ab}\epsilon^{\dot{a}\dot{b}} \sigma^\mu_{a\dot{a}}\sigma^\nu_{b\dot{b}} &= 2 g^{\mu\nu}\,, \label{eq:projector_1}
\end{align}
where $\epsilon_{ab}/\epsilon_{\dot{a}\dot{b}}$ is the spinor metric, $g^{\mu\nu}$ is the Lorentz metric. It is convenient to define $\overline{\sigma}^\mu{}^{a\dot{a}} = \epsilon^{ab}\epsilon^{\dot{a}\dot{b}}\sigma^\mu_{b\dot{b}}$, and express the relation in Eq.~\eqref{eq:projector_1} as
\begin{equation}
    \text{tr}(\sigma^\mu \overline{\sigma}^\nu) = 2g^{\mu\nu}\,.
\end{equation}
Besides, we can define the tensors
\begin{align}
    \sigma^{\mu\nu}_{ab} & = \frac{i}{2}(\sigma^\mu_{a\dot{a}}\overline{\sigma}^\nu{}^{\dot{a}c}-\sigma^\nu_{a\dot{a}}\overline{\sigma}^\mu{}^{\dot{a}c})\epsilon_{cb}\,, \\
    \overline{\sigma}^{\mu\nu}{}^{\dot{a}\dot{b}} &= \frac{i}{2}(\overline{\sigma}^\mu{}^{\dot{a}a}\sigma^\nu_{a\dot{c}}-\overline{\sigma}^\nu{}^{\dot{a}a}\sigma^\mu_{a\dot{c}})\epsilon^{\dot{c}\dot{b}}\,,
\end{align}
which are antisymmetric for the vector indices $\mu\,,\nu$ and symmetric for the spinor indices $a\,,b$ or $\dot{a}\,,\dot{b}$. Thus, they project a pair of symmetric spinor indices onto a pair of antisymmetric vector indices, so the tensor of irreps $(1,0)$ or $(0,1)$ can be expressed in terms of a pair of antisymmetric vector indices. 

In our method for constructing irreducible tensors, we use the spinor indices of the building blocks and convert the result to vector indices. The projection is important, which means we need to manipulate various traces composed of two or more $\sigma^\mu$ matrices. Several important relations will be useful in the construction, and we list them here,

\begin{align}
    \text{tr}(\sigma^{\mu}\overline{\sigma}^\nu) &= 2g^{\mu\nu} \,,\\
    \text{tr}(\sigma^\mu\overline{\sigma}^\nu\sigma^\rho\overline{\sigma}^\kappa) &= 2g^{\mu\nu}g^{\rho\kappa} - 2g^{\mu\rho} g^{\nu\kappa} + 2g^{\nu\rho}g^{\mu\kappa} + 2i\epsilon^{\mu\nu\rho\kappa} \,,\\
    \text{tr}(\overline{\sigma}^\mu\sigma^\nu\overline{\sigma}^\rho\sigma^\kappa) &= 2g^{\mu\nu}g^{\rho\kappa} - 2g^{\mu\rho} g^{\nu\kappa} + 2g^{\nu\rho}g^{\mu\kappa} - 2i\epsilon^{\mu\nu\rho\kappa} \,.
\end{align}

We adopt the spinor representation because not only is it convenient to manipulate, and the redundancies are clearer to identify, but it is also necessary for the spin-half-integer cases. 
In principle, the two different representations are equivalent for the vector and tensor fields. But the tensor representation is difficult to eliminate redundancies, raising more subtleties. In App.~\ref{app:tensor} we will discuss the construction method utilizing tensor representation, and show its drawbacks by an example.

\begin{table}[]
    \centering
    \begin{tabular}{|c|c|c|c|}
\hline
irrep name & irrep dimension & spinor SYDs & variable \\
\hline
$(0,0)$ & 1 & $(1\,,1)$ & 1 \\
\hline
$(\frac{1}{2}\,,0)$ & 2 & $\left(\ydiagram{1}\,,1\right)$ & $\xi_a $ \\
\hline
$(0\,,\frac{1}{2})$ & 2 & $\left(1\,,\ydiagram{1}\right)$ & $\eta^{\dot{a}} $ \\
\hline
$(\frac{1}{2}\,,\frac{1}{2})$ & 4 & $\left(\ydiagram{1}\,,\ydiagram{1}\right)$ & $v^\mu$ \\
\hline
    \end{tabular}
    \caption{The first several irreps of the Lorentz group and their related properties.}
    \label{tab:lorentz_irreps}
\end{table}

\section{Spinor Young Tensor Basis}

\label{sec:method}

In this section, we present our method for constructing independent Lorentz tensors. First, we set up the wave functions for general spin, then illustrate the procedure explicitly. Throughout this section, we use spin-1/2 and spin-1 particles as two examples, which not only help to clarify every step in the construction but also justify our method by comparing them to the results of the previous counting result~\cite{Cotogno:2019vjb}. We find that even for these two simple cases, the construction method we propose reveals non-trivial results.
Since we adopt the spinor representations, we specify the Young tableaux and the corresponding tensors (in spinor space) as Young tensors. When referring to the tensor representations, we use tensor SYDs and tableaux for distinction.

\subsection{Relativistic Wave Functions of Any-spin Particles}


The spin generators are the sum of the generators of the two $SU(2)$ groups, $\vec{J} = \vec{J}_L + \vec{J}_R$, 
which means the wave functions of a spin-$j$ particles are of Lorentz group irrep $(j_L,j_R)$ if the $SU(2)$ irrep $j$ can be decomposed by the tensor product $j_L\otimes j_R$, or $|j_L-j_R|\leq j \leq j_L+j_R$.
Thus, the wave function of a particle with a specific spin is not unique, and some convention must be made. For spin-1/2 and (massive) spin-1 particles, the interpolating fields are the Dirac field and the Proca field, respectively. 

The Proca field is a vector field of the irrep $(\frac{1}{2},\frac{1}{2})$, thus the wave function of spin-1 particle is
\begin{equation}
    \varepsilon^\mu(p) \in (\frac{1}{2},\frac{1}{2})\,,
\end{equation}
and is called a polarization vector.
However, the vector irrep $(\frac{1}{2},\frac{1}{2})$ is of dimension 4, which is larger than the irrep 1 of the $SU(2)$, thus some relation is needed to eliminate one redundant component. This is realized by imposing a gauge symmetry and constructing a Lagrangian that
\begin{equation}
    \mathcal{L} = -\frac{1}{4}F_{\mu\nu}F^{\mu\nu} + \frac{1}{2}m^2A_\mu A^\mu\,,
\end{equation}
where $F_{\mu\nu} = \partial_\mu A_\nu - \partial_\nu A_\mu$ is gauge invariant. The equation of motion (EOM) of $A_\mu$ leads to
\begin{equation}
    \partial_\mu A^\mu=0\,,
\end{equation}
which corresponds to the relation of wave functions that
\begin{equation}
    \partial_\mu A^\mu = 0 \quad \Longrightarrow \quad p_\mu \varepsilon^\mu(p) = 0\,.
\end{equation}

For the bosons of general spin $n$, where $n$ is an integer, we generalize the spin-1 case that the wave function of a spin-$n$ particle is of irrep $(\frac{n}{2},\frac{n}{2})$, which means that the wave function takes the form
\begin{equation}
    \varepsilon^{\mu_1\dots\mu_n}(p) \equiv \varepsilon^{(\mu_1\dots\mu_n)}(p)\,,
\end{equation}
where $(\dots)$ means the indices between them are traceless and symmetrical, indicating it is of the irrep $\underbrace{\ydiagram{2}\dots\ydiagram{1}}_n$.

In most cases, we will not write the $(\dots)$ explicitly for convenience. There are $(n+1)^2$ components, and a similar relation
\begin{equation}
\label{eq:relation_11}
    p_{\mu_1} \varepsilon^{\mu_1\dots\mu_n} = 0\,,
\end{equation}
can eliminate the redundant ones exactly. This is because $p_{\mu_1} \varepsilon^{\mu_1\dots\mu_n} \in (\frac{n-1}{2},\frac{n-1}{2})$, there are $n^2$ relations in Eq.~\eqref{eq:relation_11}, and the remaining independent components are of number 
\begin{equation}
    (n+1)^2-n^2=2n+1\,,
\end{equation}
coincident with the independent components of a spin-$n$ particle.
The integer-spin wave functions can be explicitly constructed using the tensor representation. In this formalism, the wave function is described by a completely symmetric and traceless tensor, whose properties are uniquely determined by a single-row SYD. Thus, the wave function of a spin-$n$ particle corresponds to an $n$-rank symmetric tensor, which can be indicated by the following spinor SYD:
\begin{equation}
    \varepsilon^{\mu_1\mu_2\dots \mu_n}(p) \sim \left(\underbrace{\ydiagram{2}\dots\ydiagram{1}}_n\,,\underbrace{\ydiagram{2}\dots\ydiagram{1}}_n\right)\,.
\end{equation}

Turning to fermions, the wave functions of the spin-1/2 particle are of the representation
\begin{equation}
    u(p,\sigma) \in (\frac{1}{2},0)\oplus (0,\frac{1}{2})\,,\quad \sigma=\pm\frac{1}{2}\,,
\end{equation}
where both $(\frac{1}{2},0)$ and $(0,\frac{1}{2})$ are needed to amount the particle and antiparticle. The wave function $u(p,\sigma)$ is also called Dirac spinor, which can be expressed as the direct sum of two Weyl spinors,
\begin{equation}
    u = (\xi_a,\eta^{\dot{a}})^T\,.
\end{equation}
The corresponding field operators is the Dirac field $\psi(x)$, whose Lagrangian is
\begin{equation}
    \mathcal{L} = \overline{\psi}i\gamma^\mu\partial_\mu\psi + m\overline{\psi}\psi\,,
\end{equation}
and the EOM is
\begin{equation}
    (i\gamma^\mu\partial_\mu + m)\psi = 0 \rightarrow (\gamma^\mu p_\mu +m)u(p,\sigma) = 0\,.
\end{equation}

For the fermions with general half spin $n+\frac{1}{2}$, where $n$ is an integer, we follow the convention that the wave function is of representation $(\frac{n+1}{2},\frac{n}{2})\oplus (\frac{n}{2},\frac{n+1}{2})$, which means we can express the wave function as 
\begin{equation}
    \xi{}^{\mu_1\dots\mu_n}_a \oplus \eta{}^{\mu_1\dots\mu_n\dot{a}}\,, \quad \text{where }\xi{}^{\mu_1\dots\mu_n}_a\in (\frac{n+1}{2},\frac{n}{2})\,,\eta{}^{\mu_1\dots\mu_n\dot{a}} \in (\frac{n}{2},\frac{n+1}{2})\,.
\end{equation}
Focus on half of the representation, $\xi{}^{\mu_1\dots\mu_n}_a\in(\frac{n+1}{2},\frac{n}{2})$ for example, there are $(n+2)(n+1)$ components, and as shown in the bosonic case, the simultaneous decrease of $j_L\,,j_R$ by $1/2$ eliminates the redundant ones,
\begin{equation}
    (n+2)(n+1)-(n+1)n = 2n+2 = 2\left(n+\frac{1}{2}\right)+1\,.
\end{equation}
However, two different contraction ways exist for the fermionic case to decrease the $j_L\,,j_R$ simultaneously. The first one is independent of the free spinor index $a$, and the resultant relation is similar to that of the bosons,
\begin{equation}
    p_{\mu_1} \xi^{\mu_1\dots\mu_n}_a = 0\,.
\end{equation}
On the other hand, the second way involves the free spinor index, the resultant relation is
\begin{equation}
    \overline{\sigma}_{\mu_1}^{\dot{a}a}\xi^{\mu_1\dots\mu_n}_a = 0\,.
\end{equation}
Similar relations about $\eta$ is
\begin{equation}
    p_{\mu_1} \eta^{\mu_1\dots\mu_n}{}^{\dot{a}} = 0 \,,\quad \sigma_{\mu_1}{}_{a\dot{a}}\eta^{\mu_1\dots\mu_n}{}^{\dot{a}} = 0\,.
\end{equation}
Compared to the Dirac spinor, we define
\begin{equation}
    u^{\mu_1\dots\mu_n} = (\xi^{\mu_1\dots\mu_n}_a \,, \eta^{\mu_1\dots\mu_n}{}^{\dot{a}})^T\,,
\end{equation}
where the spinor indices on the left-hand side have been suppressed, the relations become
\begin{equation}
    p_{\mu_1}u^{\mu_1\dots\mu_n} = 0 \,,\quad \gamma_{\mu_1}u^{\mu_1\dots\mu_n} = 0\,,\label{eq:redundant_fermion_1}
\end{equation}
where 
\begin{equation}
    \gamma^\mu = \left(\begin{array}{cc}
0 & \sigma^\mu \\ \overline{\sigma}^\mu & 0
    \end{array}\right)\,.
\end{equation}
Including the EOM
\begin{equation}
\label{eq:eom_fermion}
    (p^\mu\gamma_\mu+m)u^{\mu_1\dots\mu_n} = 0\,,
\end{equation}
it means what we adopt for the fermions is the Rarita-Schwinger field~\cite{Rarita:1941mf}. The spinor SYDs corresponding to a spin-$n+\frac{1}{2}$ wave function is 
\begin{equation}
    u^{\mu_1\mu_2\dots\mu_n}(p,\sigma) \sim \left(\underbrace{\ydiagram{2}\dots\ydiagram{1}}_{n+1}\,,\underbrace{\ydiagram{2}\dots\ydiagram{1}}_{n}\right)\oplus\left(\underbrace{\ydiagram{2}\dots\ydiagram{1}}_{n}\,,\underbrace{\ydiagram{2}\dots\ydiagram{1}}_{n+1}\right)\,.
\end{equation}
\subsubsection*{Construction of Spinor Wave Functions}
The correspondence between the tensor and spinor forms of wave functions serves as an excellent example to illustrate the relationship between tensor and spinor representations. We now turn to the systematic construction of spinor wave functions.

Following the preceding analysis, to construct non-redundant wave functions, they must be traceless. Therefore, for a bosonic state of integer spin $s=n$, the wave function corresponds to the highest-weight representation $(n/2, n/2)$.
Similarly, for a fermionic state of half-integer spin $s=n+1/2$, the wave function resides in the $(\frac{n+1}{2}, \frac{n}{2}) \oplus (\frac{n}{2}, \frac{n+1}{2})$ representation.

At the beginning of the construction, we introduce the left- and right-handed two-component massive Weyl spinors, denoted by $\lambda_\alpha^{I}$ and $\tilde{\lambda}_{\dot{\alpha}}^{I}$ respectively, as the essential building blocks. Here, we incorporate the $SU(2)$ little group indices $I \in \{1, 2\}$ to represent the internal degrees of freedom. Higher-spin wave functions are then systematically constructed by taking the tensor products of these fundamental blocks.

For particle with integer spin, the $(n/2, n/2)$ maps to the following pair of spinor SYDs:
\begin{equation}
    \varepsilon_{\alpha_1\dots\alpha_n \dot{\beta}_1\dots\dot{\beta}_n}(p) \sim \left(\underbrace{\ydiagram{1}\dots\ydiagram{1}}_n\,,\underbrace{\ydiagram{1}\dots\ydiagram{1}}_n\right)\,,
\end{equation}
For the particle with half-integer spin, $(\frac{n+1}{2}, \frac{n}{2}) \oplus (\frac{n}{2}, \frac{n+1}{2})$ 
 maps to the spinor SYDs:
\begin{equation}
    u_{\alpha_1\dots\alpha_{n+1} \dot{\beta}_1\dots\dot{\beta}_n, \dots}(p,\sigma) \sim \left(\underbrace{\ydiagram{1}\dots\ydiagram{1}}_{n+1}\,,\underbrace{\ydiagram{1}\dots\ydiagram{1}}_{n}\right)\oplus\left(\underbrace{\ydiagram{1}\dots\ydiagram{1}}_{n}\,,\underbrace{\ydiagram{1}\dots\ydiagram{1}}_{n+1}\right)\,.
\end{equation}
Therefore, all the left-handed and right-handed spinor indices are totally symmetrized separately in the wave functions for both cases above.
To ensure that the wave functions correspond strictly to the physical states, they are required to satisfy specific constraints mentioned before in the spinor forms. 

For the integer spin $s=n$ case, these constraints can be expressed as:
\begin{equation}
    \label{eq:relation_1}
    p_{\mu_1} \varepsilon^{\mu_1\dots\mu_n}(p) = \frac{1}{2} p^{\dot{\beta}_1\alpha_1} \varepsilon_{\alpha_1\dots\alpha_n \dot{\beta}_1\dots\dot{\beta}_n}(p) = 0\,.
\end{equation}
For the fermionic state of half-integer spin $s=n+1/2$, the constraints are given by:
\begin{align}
    p_{\mu_1} \xi^{\mu_1\dots\mu_n}_{\alpha_{n+1}}(p) &= \frac{1}{2} p^{\dot{\beta}_1\alpha_1} \xi_{\alpha_1\dots\alpha_{n+1}\dot{\beta}_1\dots\dot{\beta}_n}(p) = 0\,, \\
    p_{\mu_1} \eta^{\mu_1\dots\mu_n}_{\dot{\beta}_{n+1}}(p) &= \frac{1}{2} p^{\dot{\beta}_1\alpha_1} \eta_{\alpha_1\dots\alpha_n\dot{\beta}_1\dots\dot{\beta}_{n+1}}(p) = 0\,, \\
    (\bar{\sigma}_{\mu_1})^{\dot{\beta}_1\alpha_1}\xi^{\mu_1\dots\mu_n}_{\alpha_1}(p) &= -\epsilon^{\alpha_1\alpha_2} \xi_{\alpha_1\alpha_2\dots\alpha_{n+1}\dot{\beta}_1\dots\dot{\beta}_n}(p) = 0\,, \\
    (\sigma_{\mu_1})_{\alpha_1\dot{\beta}_1}\eta^{\mu_1\dots\mu_n\dot{\beta}_1}(p) &= -\epsilon_{\dot{\beta}_1\dot{\beta}_2} \eta_{\alpha_1\dots\alpha_n\dot{\beta}_1\dot{\beta}_2\dots\dot{\beta}_{n+1}}(p) = 0\,.
\end{align}

To intrinsically satisfy both these constraints and the required index symmetries of the single-row SYDs, the construction is completed by completely symmetrizing all the little group indices of the fundamental spinors. 
Therefore, for the integer spin $s=n$ case, it is constructed by the product of $n$ left-handed and $n$ right-handed spinors, with the total symmetrization of the little group indices:
\begin{equation}
    \varepsilon_{\alpha_1\dots\alpha_n \dot{\beta}_1\dots\dot{\beta}_n}^{I_1\dots I_{2n}}(p) \sim \lambda_{\alpha_1}^{(I_1} \dots \lambda_{\alpha_n}^{I_n} \tilde{\lambda}_{\dot{\beta}_1}^{I_{n+1}} \dots \tilde{\lambda}_{\dot{\beta}_n}^{I_{2n})}\,.
\end{equation}
Similarly, for the half-integer spin $s=n+1/2$ case, the wave functions is
\begin{align}
    \xi_{\alpha_1\dots\alpha_{n+1} \dot{\beta}_1\dots\dot{\beta}_n}^{I_1\dots I_{2n+1}}(p) &\sim \lambda_{\alpha_1}^{(I_1} \dots \lambda_{\alpha_{n+1}}^{I_{n+1}} \tilde{\lambda}_{\dot{\beta}_1}^{I_{n+2}} \dots \tilde{\lambda}_{\dot{\beta}_n}^{I_{2n+1})}\,, \\
    \eta_{\alpha_1\dots\alpha_n \dot{\beta}_1\dots\dot{\beta}_{n+1}}^{I_1\dots I_{2n+1}}(p) &\sim \lambda_{\alpha_1}^{(I_1} \dots \lambda_{\alpha_n}^{I_n} \tilde{\lambda}_{\dot{\beta}_1}^{I_{n+1}} \dots \tilde{\lambda}_{\dot{\beta}_{n+1}}^{I_{2n+1})}\,.
\end{align}

Finally, the explicit correspondence between the conventional tensor wave functions and these purely spinorial wave functions is established through systematic contraction with the Pauli matrices for both integer and half-integer cases:
\begin{align}
\varepsilon_{\alpha_1\dot{\beta}_1\dots\alpha_n\dot{\beta}_n}(p) &= \varepsilon^{\mu_1\dots\mu_n}(p) (\sigma_{\mu_1})_{\alpha_1\dot{\beta}_1} (\sigma_{\mu_2})_{\alpha_2\dot{\beta}_2} \cdots (\sigma_{\mu_n})_{\alpha_n\dot{\beta}_n}\,, \\
    \xi_{\alpha_1\dots\alpha_{n+1}\dot{\beta}_1\dots\dot{\beta}_n}(p) &= \xi^{\mu_1\dots\mu_n}_{\alpha_{n+1}}(p) (\sigma_{\mu_1})_{\alpha_1\dot{\beta}_1} \cdots (\sigma_{\mu_n})_{\alpha_n\dot{\beta}_n}\,, \\
    \eta_{\alpha_1\dots\alpha_n\dot{\beta}_1\dots\dot{\beta}_{n+1}}(p) &= \eta^{\mu_1\dots\mu_n}_{\dot{\beta}_{n+1}}(p) (\sigma_{\mu_1})_{\alpha_1\dot{\beta}_1} \cdots (\sigma_{\mu_n})_{\alpha_n\dot{\beta}_n}\,.
\end{align}
To sum up, the mapping relationship between the tensor and spinor wave functions and the spinor Young's diagrams can be written as
\begin{align}\varepsilon^{\mu_1\mu_2\dots \mu_n}(p)
    &\sim \varepsilon_{\alpha_1\dots\alpha_n \dot{\beta}_1\dots\dot{\beta}_n}(p) \sim \left(\underbrace{\ydiagram{1}\dots\ydiagram{1}}_n\,,\underbrace{\ydiagram{1}\dots\ydiagram{1}}_n\right)\,,\\
    u^{\mu_1\mu_2\dots\mu_n}(p,\sigma)&\sim
    u_{\alpha_1\dots\alpha_{n+1} \dot{\beta}_1\dots\dot{\beta}_n, \dots}(p,\sigma) \sim \left(\underbrace{\ydiagram{1}\dots\ydiagram{1}}_{n+1}\,,\underbrace{\ydiagram{1}\dots\ydiagram{1}}_{n}\right)\oplus\left(\underbrace{\ydiagram{1}\dots\ydiagram{1}}_{n}\,,\underbrace{\ydiagram{1}\dots\ydiagram{1}}_{n+1}\right)\,,
\end{align}
where $u^{\mu_1\dots\mu_n} = (\xi^{\mu_1\dots\mu_n}_a \,, \eta^{\mu_1\dots\mu_n}{}^{\dot{a}})^T\,.$
\subsubsection*{Permutation Symmetries of Wave Functions}
Because every irreducible tensor contains two wave functions, one interpolates the initial state, called the initial wave function $\phi_1$, the other interpolates the final state, called the final wave function $\phi_2^\dagger$. Here we want to discuss the permutation symmetries of the two wave functions in various irreps composed by them, which are closely related to the $T$ symmetry of the irreducible tensors.

To distinguish the initial and final wave functions, we fill their corresponding SYDs with indices 1 and 2, respectively. For bosons, we obtain that
\begin{align}
\ytableausetup{mathmode, boxframe=normal, boxsize=1.1em}
    \text{initial wave function:} \quad & \varepsilon_1^{\mu_1\dots\mu_n} \sim \left(\underbrace{\ytableaushort{11}\dots \ytableaushort{1}}_n\,,\underbrace{\ytableaushort{11}\dots \ytableaushort{1}}_n\right)\,, \\
    \text{final wave function:} \quad & \varepsilon_2^\dagger {}^{\mu_1\dots\mu_n} \sim \left(\underbrace{\ytableaushort{22}\dots \ytableaushort{2}}_n\,,\underbrace{\ytableaushort{22}\dots \ytableaushort{2}}_n\right)\,, \label{eq:wave_functions_boson}
\end{align}
where the identical indices $1$ or $2$ in the boxes indicate that all the indices should be symmetric so that the wave functions are traceless. Such tableaux are actually SSYTs of the corresponding irreps. 
Similarly the fermionic wave functions are
\begin{align}
\ytableausetup{mathmode, boxframe=normal, boxsize=1.1em}
    \text{initial wave function:} \quad u_1^{\mu_1\dots\mu_n} &\sim\left(\underbrace{\ytableaushort{11}\dots \ytableaushort{1}}_{n+1}\,,\underbrace{\ytableaushort{11}\dots \ytableaushort{1}}_n\right) \notag \\
    &\oplus \left(\underbrace{\ytableaushort{11}\dots \ytableaushort{1}}_n\,,\underbrace{\ytableaushort{11}\dots \ytableaushort{1}}_{n+1}\right) \,, \notag\\
    \text{final wave function:} \quad \overline{u}_2{}^{\mu_1\dots\mu_n} &\sim \left(\underbrace{\ytableaushort{22}\dots \ytableaushort{2}}_{n+1}\,,\underbrace{\ytableaushort{22}\dots \ytableaushort{2}}_n\right)\notag \\
    & \oplus \left(\underbrace{\ytableaushort{22}\dots \ytableaushort{2}}_n\,,\underbrace{\ytableaushort{22}\dots \ytableaushort{2}}_{n+1}\right)\,, \label{eq:wave_function_fermion}
\end{align}
where $\overline{u}_2^{\mu_1\dots\mu_n} = u_2^\dagger{}^{\mu_1\dots\mu_n}\gamma^0$.
Next, we consider the tensor product of the two wave functions and argue that the outer product of the Young tableaux indicates their interchange symmetry naturally, which is important for the time reversal ($T$) or charge conjugation ($C$) of the corresponding tensors and will be discussed in detail subsequently.

Considering the bosons, because the vector indices of the initial or final wave function are symmetric separately, their interchange symmetry is determined by the interchange between them. 
Considering the transversion between the tensor and spinor representations, the interchange symmetry of the vector indices is determined by the symmetries of the left- and right-handed spinor indices at the same time. In detail, if the left-handed spinor indices and the right-handed spinor indices are symmetric or antisymmetric simultaneously, the two wave functions are symmetric, otherwise, they are antisymmetric. Expressing the wave functions as SSYTs, their tensor product corresponds to a set of pairs of SSYTs, of which one sub-SSYT represents the tensor product of left-handed spinors, and the other sub-SSYT represents the tensor product of right-handed spinors, both of which look like
\begin{equation}
\ytableausetup{mathmode, boxframe=normal, boxsize=1.1em}
    \begin{ytableau}
        1 & 1 & 1 & \none[\dots] & 1 & 2 & \none[\dots] & 2 \\
        2 & \none[\dots] & 2 
    \end{ytableau}\,,
\end{equation}
where the total numbers of indices 1 and 2 should be the same. The interchange of the two functions is reflected as the interchange of indices 1 and 2 in such a Young tableau, whose relations to the original one can be obtained by the Fock condition. When the indices 1 and 2 are interchanged, there are two different permutations of the Young tableaux, one is the interchange in a single row, and another is the interchange in a single column. According to the Fock condition, the interchange in the same column is equivalent to the original one with a minus sign,
\begin{equation}
\ytableausetup{mathmode, boxframe=normal, boxsize=1.1em}
    \cdots \ytableaushort{1,2} \cdots  \rightarrow \cdots \ytableaushort{2,1} \cdots = (-1)\times \cdots \ytableaushort{1,2} \cdots\,,
\end{equation}
while the interchange in the same row is equivalent to the original one,
\begin{equation}
\ytableausetup{mathmode, boxframe=normal, boxsize=1.1em}
\cdots \ytableaushort{1} \cdots\ytableaushort{2}\cdots \rightarrow \cdots \ytableaushort{2} \cdots\ytableaushort{1}\cdots = \cdots \ytableaushort{1} \cdots\ytableaushort{2} \cdots.
\end{equation}
Thus, the interchange symmetry of the Young tableau is determined by the boxes in the second row, $(-1)^p$, where $p$ is the box number in the second row. Furthermore, the interchange symmetry of the wave functions, which is determined by two Young tableaux, is $\eta_I = (-1)^{p_L+p_R}$, where $p_L\,,p_R$ are the numbers of the two tableaux respectively. Besides, the box number at the same row $p$ is related to the corresponding irrep $j$ as 
\begin{equation}
    j = n-p\,,
\end{equation}
thus the interchange symmetry of the wave functions is also related to the irreps
\begin{equation}
    \eta_I = (-1)^{p_L+p_R} = (-1)^{2n-p_L-p_R} = (-1)^{j_L+j_R}\,.
\end{equation}
For example, we can present the interchange symmetries of the spin-1 and spin-2 wave functions by their irreps in Tab.~\ref{tab:symmetry_boson}.
\begin{table}[]
    \centering
    \ytableausetup{smalltableaux}
    \renewcommand\arraystretch{1.5}
    \begin{tabular}{|c|c|c|}
\hline
irrep & Young tableau & symmetry \\
\hline
\multicolumn{3}{|c|}{spin-1} \\
\hline
$(1,1)$ & $\left(\ytableaushort{12}\,,\ytableaushort{12}\right)$ & 1 \\
\hline
$(0,1)$ & $\left(\ytableaushort{1,2}\,,\ytableaushort{12}\right)$ & -1 \\
\hline
$(1,0)$ & $\left(\ytableaushort{12}\,,\ytableaushort{1,2}\right)$ & -1 \\
\hline
$(0,0)$ & $\left(\ytableaushort{1,2}\,,\ytableaushort{1,2}\right)$ & 1 \\
\hline
\multicolumn{3}{|c|}{spin-2} \\
\hline
$(2,2)$ & $\left(\ytableaushort{1122}\,,\ytableaushort{1122}\right)$ & 1 \\
\hline
$(2,1)$ & $\left(\ytableaushort{1122}\,,\ytableaushort{112,2}\right)$ & -1 \\
\hline
$(2,0)$ & $\left(\ytableaushort{1122}\,,\ytableaushort{11,22}\right)$ & 1 \\
\hline
$(1,2)$ & $\left(\ytableaushort{112,2}\,,\ytableaushort{1122}\right)$ & -1 \\
\hline
$(1,1)$ & $\left(\ytableaushort{112,2}\,,\ytableaushort{112,2}\right)$ & 1 \\
\hline
$(1,0)$ & $\left(\ytableaushort{112,2}\,,\ytableaushort{11,22}\right)$ & -1 \\
\hline
$(0,2)$ & $\left(\ytableaushort{11,22}\,,\ytableaushort{1122}\right)$ & 1 \\
\hline
$(0,1)$ & $\left(\ytableaushort{11,22}\,,\ytableaushort{112,2}\right)$ & -1 \\
\hline
$(0,0)$ & $\left(\ytableaushort{11,22}\,,\ytableaushort{11,22}\right)$ & 1 \\
\hline
    \end{tabular}
    \caption{The permutation symmetries of the spin-1 and spin-2 wave functions in terms of their irreps.}
    \label{tab:symmetry_boson}
\end{table}

Applying to the fermions, there are two differences,
\begin{itemize}
    \item Because the irreps of the fermion wave functions are not symmetric between the two $SU(2)$ irreps, the interchange of the wave functions is not closed within a single irrep, which means that some combinations of the irreps are of specific interchange symmetries for fermions. 
    \item The fermions are antisymmetric under the interchange because of the Fermi-Dirac statistics. Thus, the interchange of the fermion wave functions possesses an extra minus sign.
\end{itemize}
For example, there are two $(\frac{1}{2}\,,\frac{1}{2})$ irreps in the tensor product of spin-1/2 fermions, 
\begin{equation}
\ytableausetup{mathmode, boxframe=normal, boxsize=1.2em}
    (\frac{1}{2}\,,\frac{1}{2})_1 = \left(\ytableaushort{1}\,,\ytableaushort{2}\right)\,,\quad (\frac{1}{2}\,,\frac{1}{2})_2 = \left(\ytableaushort{2}\,,\ytableaushort{1}\right)\,,
\end{equation}
and they are not closed under the interchange of indices, but related to each other. Considering the Fermi-Dirac statistics we get the combinations $(\frac{1}{2}\,,\frac{1}{2})_1\pm (\frac{1}{2}\,,\frac{1}{2})_2$ of specific interchange symmetries $\mp 1$ respectively. Accordingly, we list the interchange symmetries of the spin-1/2 and spin-3/2 cases in Tab.~\ref{tab:symmetry_fermion}.

\begin{table}[]
    \centering
    \ytableausetup{smalltableaux}
    \renewcommand\arraystretch{1.5}
    \begin{tabular}{|c|c|c|}
\hline
irrep & Young tableau & symmetry \\
\hline
\multicolumn{3}{|c|}{spin-1/2} \\
\hline
$(1,0)$ & $\left(\ytableaushort{12}\,,1\right)$ & -1 \\
\hline
$(0,0)_1$ & $\left(\ytableaushort{1,2}\,,1\right)$ & 1 \\
\hline
$(0,0)_2$ & $\left(1\,,\ytableaushort{1,2}\right)$ & 1 \\
\hline
$(0,1)$ & $\left(1\,,\ytableaushort{12}\right)$ & -1 \\
\hline
$(\frac{1}{2}\,,\frac{1}{2})_1\pm(\frac{1}{2}\,,\frac{1}{2})_2$ & $\left(\ytableaushort{1}\,,\ytableaushort{2}\right)\pm\left(\ytableaushort{2}\,,\ytableaushort{1}\right)$ & $\mp 1$\\
\hline
\multicolumn{3}{|c|}{spin-3/2} \\
\hline
$(0\,,0)_1$ & $\left(\ytableaushort{11,22}\,,\ytableaushort{1,2}\right)$ & -1 \\
\hline
$(0\,,0)_2$ & $\left(\ytableaushort{1,2}\,,\ytableaushort{11,22}\right)$ & -1 \\
\hline
$(0\,,1)_1$ & $\left(\ytableaushort{11,22}\,,\ytableaushort{12}\right)$ & 1 \\
\hline
$(0\,,1)_2$ & $\left(\ytableaushort{1,2}\,,\ytableaushort{112,2}\right)$ & 1 \\
\hline
$(1\,,0)_1$ & $\left(\ytableaushort{12}\,,\ytableaushort{11,22}\right)$ & 1 \\
\hline
$(1\,,0)_2$ & $\left(\ytableaushort{112,2}\,,\ytableaushort{1,2}\right)$ & 1 \\
\hline
$(1,1)_1$ & $\left(\ytableaushort{112,2}\,,\ytableaushort{12}\right)$ & -1 \\
\hline
$(1,1)_2$ & $\left(\ytableaushort{12}\,,\ytableaushort{112,2}\right)$ & -1 \\
\hline
$(2\,,0)$ & $\left(\ytableaushort{1122}\,,\ytableaushort{1,2}\right)$ & -1 \\
\hline
$(2\,,1)$ & $\left(\ytableaushort{1122}\,,\ytableaushort{12}\right)$ & 1 \\
\hline
$(0\,,2)$ & $\left(\ytableaushort{1,2}\,,\ytableaushort{1122}\right)$ & -1 \\
\hline
$(1\,,2)$ & $\left(\ytableaushort{12}\,,\ytableaushort{1122}\right)$ & 1 \\
\hline
$(\frac{1}{2}\,,\frac{1}{2})_1\pm (\frac{1}{2}\,,\frac{1}{2})_2$ & $\left(\ytableaushort{12,2}\,,\ytableaushort{11,2}\right) \pm \left(\ytableaushort{11,2}\,,\ytableaushort{12,2}\right)$ & $\mp 1$ \\
\hline
$(\frac{1}{2}\,,\frac{3}{2})_1\pm (\frac{1}{2}\,,\frac{3}{2})_2$ & $\left(\ytableaushort{12,2}\,,\ytableaushort{112}\right)\pm \left(\ytableaushort{11,2}\,,\ytableaushort{122}\right)$ & $\pm 1$ \\
\hline
$(\frac{3}{2}\,,\frac{1}{2})_1\pm (\frac{3}{2}\,,\frac{1}{2})_2$ & $\left(\ytableaushort{122}\,,\ytableaushort{11,2}\right) \pm \left(\ytableaushort{112}\,,\ytableaushort{12,2}\right)$ & $\pm1$ \\
\hline
$(\frac{3}{2}\,,\frac{3}{2})_1\pm (\frac{3}{2}\,,\frac{3}{2})_2$ & $\left(\ytableaushort{122}\,,\ytableaushort{112}\right)\pm \left(\ytableaushort{112}\,,\ytableaushort{122}\right)$ & $\mp 1$ \\
\hline
    \end{tabular}
    \caption{The interchange symmetries of the spin-1/2 and spin-3/2 wave functions in terms of their irreps.}
    \label{tab:symmetry_fermion}
\end{table}

\subsection{Construction (1): Building Blocks and Redundant Relations}

Denoting general wave functions as $\phi$, all the building blocks for the irreducible tensors are
\begin{equation}
\label{eq:building_blocks}
    \phi_1\,, \quad \phi_2^\dagger\,,\quad P\,,\quad q \,,
\end{equation}
where $\phi_1\,,\phi_2^\dagger$ are the wave functions of the initial and final states, respectively, and the momenta $P\,, q$ are defined as
\begin{equation}
\ytableausetup{mathmode, boxframe=normal, boxsize=1.1em}
\label{eq:p_q}
    P^\mu = p_1^\mu + p_2^\mu\,,\quad q^\mu = p_2^\mu - p_1^\mu\,. 
\end{equation}
In Eq.~\eqref{eq:wave_functions_boson} and Eq.~\eqref{eq:wave_function_fermion}, we have presented the SSYTs of the two wave functions for bosons and fermions, respectively. Similarly, the product of the momenta $P, q$ can also be expressed as
\begin{align}
\ytableausetup{mathmode, boxframe=normal, boxsize=1.1em}
    P^m_{(\alpha_1 \alpha_2 \dots \alpha_m), (\dot{\beta}_1 \dot{\beta}_2 \dots \dot{\beta}_m)} &\sim \left(\underbrace{\ytableaushort{33}\dots \ytableaushort{3}}_m\,,\underbrace{\ytableaushort{33}\dots \ytableaushort{3}}_m\right) \,, \label{eq:p_tableaux}\\
    q^l_{(\alpha_1 \alpha_2 \dots \alpha_l), (\dot{\beta}_1 \dot{\beta}_2 \dots \dot{\beta}_l)} &\sim \left(\underbrace{\ytableaushort{44}\dots \ytableaushort{4}}_l\,,\underbrace{\ytableaushort{44}\dots \ytableaushort{4}}_l\right) \,, \label{eq:q_tableaux}
\end{align}
where we use $3\,,4$ to represent the symmetrical indices and of the products of $P$ and $q$, respectively.
Before we construct irreducible tensors via the SSYTs above, some preliminaries should be considered. 

\subsubsection*{Redundant Relations}

Firstly, not all the SSYTs from the outer product of the building blocks are independent, because there are many redundant relations among the building blocks in Eq.~\eqref{eq:building_blocks}. Thus, we need to find all these relations and interpret them in the SSYT form.

According to the definition in Eq.~\eqref{eq:p_q}, there is an immediate relation that
\begin{equation}
    P^\mu q_\mu = 0 \,, \label{eq:pqrelation_1}\\
\end{equation}
which can be interpreted as 
\begin{equation}
\ytableausetup{mathmode, boxframe=normal, boxsize=1.1em}
P^\mu q_\mu =P_{\alpha \dot{\beta}} q^{\alpha \dot{\beta}} = 0 :\quad \left(\dots\ytableaushort{3,4}\dots\,,\dots\ytableaushort{3,4}\dots\right) = 0 \,.
\end{equation}
Besides, the relations in Eq.~\eqref{eq:relation_1} and Eq.~\eqref{eq:redundant_fermion_1} present two relations valid for both bosons and fermions,  
\begin{align}
    P_\mu \phi_1^{\mu\dots} &=  q_\mu \phi_1^{\mu\dots}\,, \\
    P_\mu \phi^\dagger_2{}^{\mu\dots} &= -q_\mu \phi^\dagger_2  {}^{\mu\dots}\,. \label{eq:pqrelation_2}
\end{align}
These two relations can not be interpreted respectively, since the second relation in Eq.~\eqref{eq:pqrelation_2} generates a minus sign, which means they can change the permutation symmetries of the wave functions. Thus these two relations should 
be interpreted together with the permutation symmetries, and we have
\begin{align}
\ytableausetup{mathmode, boxframe=normal, boxsize=1.1em}
    & P_\mu \phi_1^{\mu\dots} =  q_\mu \phi_1^{\mu\dots}\,,\quad P_\mu \phi^*_2{}^{\mu\dots} = -q_\mu \phi^*_2{}^{\mu\dots}:\notag \\
    & \left(\dots \ytableaushort{1,3}\dots\ytableaushort{2,3}\dots\,,\dots \ytableaushort{1,3}\dots\ytableaushort{2,3}\dots\right) = \frac{1+\eta_I}{2} \left(\dots \ytableaushort{1,4}\dots\ytableaushort{2,4}\dots\,,\dots \ytableaushort{1,4}\dots\ytableaushort{2,4}\dots\right)\,, \\
    & \left(\dots \ytableaushort{1,3}\dots\ytableaushort{2,4}\dots\,,\dots \ytableaushort{1,3}\dots\ytableaushort{2,4}\dots\right) = \frac{1-\eta_I}{2} \left(\dots \ytableaushort{1,4}\dots\ytableaushort{2,4}\dots\,,\dots \ytableaushort{1,4}\dots\ytableaushort{2,4}\dots\right)\,, \\
    & \left(\dots \ytableaushort{1,4}\dots\ytableaushort{2,3}\dots\,,\dots \ytableaushort{1,4}\dots\ytableaushort{2,3}\dots\right) = \frac{1-\eta_I}{2} \left(\dots \ytableaushort{1,4}\dots\ytableaushort{2,4}\dots\,,\dots \ytableaushort{1,4}\dots\ytableaushort{2,4}\dots\right)\,,
\end{align}
where $\eta_I$ is the permutation symmetry of the wave functions of the SSYT on the left hand.

While for the fermions, there are extra relations. Firstly, the EOM in Eq.~\eqref{eq:eom_fermion} implies
\begin{align}
    \overline{u}_2^{\nu\dots}P^\rho \gamma_\rho u_1^{\mu\dots} &= 2m \overline{u}_2^{\nu\dots}u_1^{\mu\dots} \,,\label{eq:relation_fermion_1}\\
    \overline{u}_2^{\nu\dots}q^\mu u_1^{\rho\dots} &= \overline{u}_2^{\nu\dots}\sigma^{\mu\lambda}P_\lambda u_1^{\rho\dots} \label{eq:fermion_relation_new}\\
    \overline{u}_2^{\nu\dots}q^\rho \gamma_\rho u_1^{\mu\dots} &= 0 \,,\label{eq:relation_fermion_2}
\end{align}
whose interpretations are
\begin{align}
\ytableausetup{mathmode, boxframe=normal, aligntableaux=top, boxsize=1.1em}
    & \overline{u}_2^{\nu\dots}P^\rho \gamma_\rho u_1^{\mu\dots} = 2m \overline{u}_2^{\nu\dots}u_1^{\mu\dots} :\notag \\
    & \left(\dots\ytableaushort{1,3}\dots\,,\dots\ytableaushort{2,3}\dots\right) = 2m\times\left(\dots\ytableaushort{1,2}\dots\,,\dots\right) \,,\\
    & \left(\dots\ytableaushort{2,3}\dots\,,\dots\ytableaushort{1,3}\dots\right) = 2m\times\left(\dots\,,\dots\ytableaushort{1,2}\dots\right) \,,\\
    &  \overline{u}_2^{\nu\dots}q^\mu u_1^{\rho\dots} = \overline{u}_2^{\nu\dots}\sigma^{\mu\lambda}P_\lambda u_1^{\rho\dots}:\notag \\
    & \left(\dots\ytableaushort{1,3}\dots\ytableaushort{2}\dots\,,\dots\ytableaushort{3}\dots\right)= \left(\dots\ytableaushort{1,2}\dots\ytableaushort{4}\dots\,,\dots\ytableaushort{4}\dots\right)\,, \\
    & \left(\dots\ytableaushort{3}\dots\,,\dots\ytableaushort{1,3}\dots\ytableaushort{2}\dots\right)= \left(\dots\ytableaushort{4}\dots\,,\dots\ytableaushort{1,2}\dots\ytableaushort{4}\dots\right) \,,\\
    & \overline{u}_2^{\nu\dots}q^\rho \gamma_\rho u_1^{\mu\dots} = 0:\notag \\
    & \left(\dots\ytableaushort{1,4}\dots\,,\dots\ytableaushort{2,4}\dots\right) = \left(\dots\ytableaushort{2,4}\dots\,,\dots\ytableaushort{1,4}\dots\right) = 0 \,.
\end{align}
In addition, there is another relation for the fermions called Gorden identity
\begin{equation}
\label{eq:gorden_identity}
    \overline{u}_2^{\nu\dots}\gamma^\mu u_1^{\rho\dots} = \frac{P^\mu}{2m}\overline{u}_2^{\nu\dots}u_1^{\rho\dots} + \frac{1}{2m}\overline{u}_2^{\nu\dots}i\sigma^{\mu\nu}q_\nu u_1^{\rho\dots}\,,
\end{equation}
which relates three different tensors to each other. To take the Gorden identity into account, we always identify the first two tensors as the same, $\overline{u}_2^{\nu\dots}\gamma^\mu u_1^{\rho\dots} = \frac{P^\mu}{2m}\overline{u}_2^{\nu\dots}u_1^{\rho\dots}$, which presents a relation that
\begin{align}
    & \overline{u}_2^{\nu\dots}\gamma^\mu u_1^{\rho\dots} = \frac{P^\mu}{2m}\overline{u}_2^{\nu\dots}u_1^{\rho\dots}:\notag \\
    & \left(\dots\ytableaushort{1}\dots\,,\dots\ytableaushort{2}\dots\right) =\frac{1}{2m}\times \left(\dots\ytableaushort{1,2}\dots\ytableaushort{3}\dots\,,\dots\ytableaushort{3}\dots\right)\,,\\
    & \left(\dots\ytableaushort{2}\dots\,,\dots\ytableaushort{1}\dots\right) =\frac{1}{2m}\times \left(\dots\ytableaushort{3}\dots\,,\dots\ytableaushort{1,2}\dots\ytableaushort{3}\dots\right)\,.
\end{align}

Apart from these relations, there is another relation of the bosons derived from the one in Eq.~\eqref{eq:fermion_relation_new} about fermions. To be specific we consider spin-$1/2$, then the relation and its SSYT correspondence are
\begin{align}
\ytableausetup{mathmode, boxframe=normal, boxsize=1.1em}
    &\overline{u}_2 q^\mu u_1 = \overline{u}_2\sigma^{\mu\lambda}P_\lambda u_1: \notag \\
    & \left(\ytableaushort{12,3}\,,\ytableaushort{3}\right)= \left(\ytableaushort{14,2}\,,\ytableaushort{4}\right) \,,\quad \left(\ytableaushort{3}\,,\ytableaushort{12,3}\right)= \left(\ytableaushort{4}\,,\ytableaushort{14,2}\right)\,. 
\end{align}
The tensor product of the above two relations and the SSYTs 
\begin{equation}
\ytableausetup{mathmode, boxframe=normal, boxsize=1.1em}
    \left(\ytableaushort{3}\,,\ytableaushort{12,3}\right)\,, \quad \left(\ytableaushort{12,3}\,,\ytableaushort{3}\right)\,,
\end{equation}
gives
\begin{align}
\ytableausetup{mathmode, boxframe=normal, boxsize=1.1em}
    & \left(\ytableaushort{123,3}\,,\ytableaushort{12,33}\right) = \left(\ytableaushort{134,2}\,,\ytableaushort{12,34}\right)\,, \label{eq:relation_boson_new_1}\\
    & \left(\ytableaushort{123,3}\,,\ytableaushort{12,33}\right) = \left(\ytableaushort{124,3}\,,\ytableaushort{13,24}\right)\,,\label{eq:relation_boson_new_2}
\end{align}
which are two relations about bosons, and will play an important role in the construction of irreducible SSYTs subsequently.

\subsubsection*{Maximal Numbers of the Momenta}

When constructing irreducible tensors, the two wave functions $\phi_1\,,\phi^\dagger_2$ are always needed, while the momenta $P\,, q$ are needed accordingly. However, the number of momenta can not increase infinitely, because if they are redundant, there would be factors such as $P^2\,, q^2$, which should be absorbed in the FFs in front of the tensors, and have been indicated by the expansion of the Hilbert series. Thus, there is a maximum of momenta. As discussed before, an irrep $(j_L\,,j_R)$ has $2\text{Max}(j_L\,,j_R)$ vector indices. Thus, for a boson of spin-$n$, when momenta contract all the vector indices of the wave functions, and the $2\text{Max}(j_L\,,j_R)$ indices are totally from the momenta, the maximum is reached, which is 
\begin{equation}
\label{eq:maximum}
    2n + 2\text{Max}(j_L\,,j_R)\,.
\end{equation}
The same argument applies to the fermions because although there would be vector indices coming from Dirac matrices such as $\gamma^\mu\,,\sigma^{\mu\nu}$, the relations such as Eq.~\eqref{eq:relation_fermion_1}, Eq.~\eqref{eq:redundant_fermion_1}, and the Gorden identity in Eq.~\eqref{eq:gorden_identity} implies the contractions between them and the momenta can all be converted. Thus, for a spin-$n+\frac{1}{2}$ particle, the momenta maximum is also determined by Eq.~\eqref{eq:maximum}.
For convenience, we list the momenta maximums of several irreps of the spin-1, spin-2, spin-$\frac{1}{2}$, and spin-$\frac{3}{2}$ cases in Tab.~\ref{tab:maximum}. 

\begin{table}
    \centering
    \renewcommand{\arraystretch}{1.5}
    \begin{tabular}{|c|c|c|c|c|}
\hline
\multirow{2}{*}{spin} & \multicolumn{4}{c|}{irreps} \\
\cline{2-5}
& $\left(0,0\right)$ & $\left(\frac{1}{2},\frac{1}{2}\right)$ & $\left(1,0\right)\oplus \left(0,1\right)$ & $\left(1,1\right)$ \\
\hline
1 & 2 & 3 & 4 & 4 \\
\hline
2 & 4 & 5 & 6 & 6 \\
\hline
$\frac{1}{2}$ & 0 & 1 & 2 & 2 \\
\hline
$\frac{3}{2}$ & 2 & 3 & 4 & 4 \\
\hline
    \end{tabular}
    \caption{The maximums of the momenta of different irreps for spin-1, spin-2, spin-1/2, and spin-3/2 cases.}
    \label{tab:maximum}
\end{table}

\subsection{Construction (2): Nonequivalent SSYT Selection}

The irreducible tensors are the tensor products of the wave functions and the momenta $P\,,q$. The building blocks and their corresponding SSYTs have been presented in Eq.~\eqref{eq:wave_functions_boson}, Eq.~\eqref{eq:wave_function_fermion}, Eq.~\eqref{eq:p_tableaux}, and Eq.~\eqref{eq:q_tableaux}, then the SSYTs corresponding to the wanted irreducible tensors are obtained by the outer products respecting the Littlewood-Richardson rule.

The Young tableaux of the tensor products of the two wave functions are listed in Tab.~\ref{tab:symmetry_boson} for spin-1 and spin-2 particles, in Tab.~\eqref{tab:symmetry_fermion} for spin-1/2 and spin-3/2 particles. The maximum of the momenta has also been worked out in Tab.~\ref{tab:maximum}. For spin-1 and spin-1/2 particles, at most 4 momenta are needed. For later convenience, we list all the SSYTs of their tensor product here,
During this procedure, all the relations derived before should be exploited. Although the constructions are the same in principle, we illustrate these procedures for bosonic and fermionic cases, respectively.

\subsubsection*{Bosonic Case}

\begin{description}
\ytableausetup{mathmode, boxframe=normal, aligntableaux=top, boxsize=1.1em}
    \item[Irrep $(0\,,0)$:] There is one SSYT with no momentum, 
    \begin{equation}
        \left(\ytableaushort{1,2}\,,\ytableaushort{1,2}\right)\,,
    \end{equation}
    while for two momenta, there is
    \begin{align}
        \left(\ytableaushort{12}\,,\ytableaushort{12}\right) \times \left\{\begin{array}{l}
\left(\ytableaushort{33}\,,\ytableaushort{33}\right) \\
\left(\ytableaushort{44}\,,\ytableaushort{44}\right) \\
\left(\ytableaushort{34}\,,\ytableaushort{34}\right)
        \end{array}\right. \supset \left\{\begin{array}{l}
\left(\ytableaushort{12,33}\,,\ytableaushort{12,33}\right) \\
\left(\ytableaushort{12,44}\,,\ytableaushort{12,44}\right) \\
\left(\ytableaushort{12,34}\,,\ytableaushort{12,34}\right) 
        \end{array}\right.\rightarrow \left(\ytableaushort{12,44}\,,\ytableaushort{12,44}\right)\,,
    \end{align}
    where there is only one nonequivalent. The arrow $\rightarrow$ means the redundant relations are used. The other cases are 
    \begin{align}
        & \left(\ytableaushort{12}\,,\ytableaushort{1,2}\right) \times \left(\ytableaushort{34}\,,\ytableaushort{3,4}\right) \supset \left(\ytableaushort{12,34}\,,\ytableaushort{13,24}\right)\,, \\
        & \left(\ytableaushort{1,2}\,,\ytableaushort{12}\right) \times \left(\ytableaushort{3,4}\,,\ytableaushort{34}\right) \supset \left(\ytableaushort{13,24}\,,\ytableaushort{12,34}\right)\,.
    \end{align}
    Thus, there are 4 nonequivalent SSYTs of this irrep.
    \item[Irrep $(\frac{1}{2}\,,\frac{1}{2})$:] The SSYTs of this irrep contain at least one momentum,
    \begin{align}
        \left(\ytableaushort{12}\,,\ytableaushort{12}\right)& \times \left\{\begin{array}{l}
            \left(\ytableaushort{3}\,,\ytableaushort{3}\right) \\
            \left(\ytableaushort{4}\,,\ytableaushort{4}\right)
        \end{array}\right. \supset \left\{\begin{array}{l}
            \left(\ytableaushort{12,3}\,,\ytableaushort{12,3}\right) \\
            \left(\ytableaushort{12,4}\,,\ytableaushort{12,4}\right)
        \end{array}\right. \,,\\
        \left(\ytableaushort{1,2}\,,\ytableaushort{1,2}\right)& \times \left\{\begin{array}{l}
            \left(\ytableaushort{3}\,,\ytableaushort{3}\right) \\
            \left(\ytableaushort{4}\,,\ytableaushort{4}\right)
        \end{array}\right. \supset \left\{\begin{array}{l}
            \left(\ytableaushort{13,2}\,,\ytableaushort{13,2}\right) \\
            \left(\ytableaushort{14,2}\,,\ytableaushort{14,2}\right)
        \end{array}\right. \,,\\
        \left(\ytableaushort{12}\,,\ytableaushort{1,2}\right)& \times \left\{\begin{array}{l}
            \left(\ytableaushort{3}\,,\ytableaushort{3}\right) \\
            \left(\ytableaushort{4}\,,\ytableaushort{4}\right)
        \end{array}\right. \supset \left\{\begin{array}{l}
            \left(\ytableaushort{12,3}\,,\ytableaushort{13,2}\right) \\
            \left(\ytableaushort{12,4}\,,\ytableaushort{14,2}\right)
        \end{array}\right.\,,
    \end{align}
    where the last two SSYTs contain two different sub-SSYTs of the same shape, so the SSYTs with the two sub-SSYTs interchanged are also in this irrep. The SSYTs related by interchanging the two sub-SSYTs are called the dual SSYTs. In particular, the SSYTs with two identical sub-SSYTs are called the self-dual SSYTs. Including the dual SSYTs, there are 8 nonequivalent SSYTs with one momentum. With three momenta, the self-dual SSYTs are
    \begin{align}
        &\left(\ytableaushort{12}\,,\ytableaushort{12}\right) \times \left\{\begin{array}{l}
\left(\ytableaushort{333}\,,\ytableaushort{333}\right)  \\
\left(\ytableaushort{334}\,,\ytableaushort{334}\right)  \\
\left(\ytableaushort{444}\,,\ytableaushort{444}\right)  \\
\left(\ytableaushort{344}\,,\ytableaushort{344}\right)  \\
    \end{array} \right. \supset \left\{\begin{array}{l}
\left(\ytableaushort{123,33}\,,\ytableaushort{123,33}\right) \\
\left(\ytableaushort{123,34}\,,\ytableaushort{123,34}\right) \\
\left(\ytableaushort{124,44}\,,\ytableaushort{124,44}\right) \\
\left(\ytableaushort{124,34}\,,\ytableaushort{124,34}\right) \\
    \end{array} \right.\notag \\
    & \rightarrow \left\{\begin{array}{l}
\left(\ytableaushort{123,44}\,,\ytableaushort{123,44}\right) \\
\left(\ytableaushort{124,44}\,,\ytableaushort{124,44}\right)
\end{array}\right. \,.
    \end{align}
    Other SSYTs are
    \begin{align}
    &\left(\ytableaushort{12}\,,\ytableaushort{1,2}\right) \times \left(\ytableaushort{334}\,,\ytableaushort{33,4}\right) \supset \left(\ytableaushort{123,34}\,,\ytableaushort{133,24}\right) \,,\\
    &\left(\ytableaushort{12}\,,\ytableaushort{1,2}\right) \times \left(\ytableaushort{344}\,,\ytableaushort{34,4}\right) \supset \left(\ytableaushort{124,34}\,,\ytableaushort{134,24}\right) \,,\\
    &\left(\ytableaushort{12}\,,\ytableaushort{1,2}\right) \times \left(\ytableaushort{334}\,,\ytableaushort{33,4}\right) \supset \left(\ytableaushort{123,34}\,,\ytableaushort{133,24}\right) \,,\\
    &\left(\ytableaushort{12}\,,\ytableaushort{1,2}\right) \times \left(\ytableaushort{344}\,,\ytableaushort{34,4}\right) \supset \left(\ytableaushort{124,34}\,,\ytableaushort{134,24}\right)\,,
\end{align}
and their dual SSYTs.
    \item[Irrep $(1\,,0)\oplus(0\,,1)$:] According to the previous discussion, the SSYTs of these two irreps are dual to each other. So we only present the SSYTs of the irrep $(1\,,0)$ here. Without momentum, there is one SSYT
    \begin{equation}
        \left(\ytableaushort{12}\,,\ytableaushort{1,2}\right)\,.
    \end{equation}
    while for the 2 momenta case, the nonequivalent SSYTs symmetric of the two wave functions are 
\begin{align}
   \left(\ytableaushort{12}\,,\ytableaushort{12}\right)& \times \left\{\begin{array}{l}
\left(\ytableaushort{33}\,,\ytableaushort{33}\right) \\
\left(\ytableaushort{44}\,,\ytableaushort{44}\right) \\
    \end{array}\right. \supset\left\{\begin{array}{l}
\left(\ytableaushort{123,3}\,,\ytableaushort{12,33}\right) \\
\left(\ytableaushort{124,4}\,,\ytableaushort{12,44}\right) \\
    \end{array}\right.\,,\\
    \left(\ytableaushort{12}\,,\ytableaushort{12}\right)& \times \left\{\begin{array}{l}
\left(\ytableaushort{34}\,,\ytableaushort{34}\right) \\
\left(\ytableaushort{3,4}\,,\ytableaushort{34}\right) \\
\left(\ytableaushort{34}\,,\ytableaushort{3,4}\right) 
    \end{array}\right. \supset \left\{\begin{array}{l}
\left(\ytableaushort{124,3}\,,\ytableaushort{12,34}\right) \\
\left(\ytableaushort{123,4}\,,\ytableaushort{12,34}\right)
\end{array}\right. \,,\\
    \left(\ytableaushort{1,2}\,,\ytableaushort{1,2}\right)& \times \left(\ytableaushort{34}\,,\ytableaushort{3,4}\right) \supset \left(\ytableaushort{134,2}\,,\ytableaushort{13,24}\right)\,, 
\end{align}
and the nonequivalent SSYTs antisymmetric of the two wave functions are 
\begin{align}
    & \left(\ytableaushort{12}\,,\ytableaushort{1,2}\right)\times \left(\ytableaushort{34}\,,\ytableaushort{3,4}\right) \supset \left(\ytableaushort{123,4}\,,\ytableaushort{13,24}\right) \,,\\
    & \left(\ytableaushort{1,2}\,,\ytableaushort{12}\right)\times \left(\ytableaushort{33}\,,\ytableaushort{33}\right) \supset \left(\ytableaushort{133,2}\,,\ytableaushort{12,33}\right) \,,\\
    & \left(\ytableaushort{1,2}\,,\ytableaushort{12}\right)\times \left(\ytableaushort{44}\,,\ytableaushort{44}\right) \supset \left(\ytableaushort{144,2}\,,\ytableaushort{12,44}\right) \,,\\
    & \left(\ytableaushort{1,2}\,,\ytableaushort{12}\right)\times \left(\ytableaushort{34}\,,\ytableaushort{34}\right) \supset \left(\ytableaushort{134,2}\,,\ytableaushort{12,34}\right)\,.
\end{align}
which are all redundant because of the relations in 
Eq.~\eqref{eq:relation_boson_new_1} and Eq.~\eqref{eq:relation_boson_new_2}. Thus, there are 5 nonequivalent SSYTs with two momenta. As for the 4-momentum case, there are
\begin{align}
    & \left(\ytableaushort{12}\,,\ytableaushort{12}\right)\times \left\{\begin{array}{ll}
\left(\ytableaushort{3334}\,,\ytableaushort{333,4}\right) \\
\left(\ytableaushort{3344}\,,\ytableaushort{334,4}\right) \\
\left(\ytableaushort{3444}\,,\ytableaushort{344,4}\right) 
    \end{array}\right. \supset \left\{\begin{array}{ll}
\left(\ytableaushort{1234,33}\,,\ytableaushort{123,334}\right)\\
\left(\ytableaushort{1233,34}\,,\ytableaushort{123,344}\right)\\
\left(\ytableaushort{1234,44}\,,\ytableaushort{123,444}\right)
    \end{array}\right. \notag \,,\\
    &\rightarrow \left(\ytableaushort{1234,44}\,,\ytableaushort{123,444}\right) \\
    & \left(\ytableaushort{1,2}\,,\ytableaushort{1,2}\right)\times \left(\ytableaushort{334,4}\,,\ytableaushort{33,44}\right) = \left(\ytableaushort{1334,24}\,,\ytableaushort{133,244}\right) \rightarrow 0 \,, \\
    & \left(\ytableaushort{12}\,,\ytableaushort{1,2}\right)\times \left(\ytableaushort{3344}\,,\ytableaushort{33,44}\right) \rightarrow \left(\ytableaushort{1234,34}\,,\ytableaushort{133,244}\right)\,,
\end{align}
so there are only 2 independent ones. There are 8 nonequivalent SSYTs in total of the irrep $(1\,,0)$, which is the same as the irrep $(0\,,1)$. 
    \item[Irrep $(1\,,1)$:] There is only 1 self-dual SSYT without momentum,
    \begin{equation}
        \left(\ytableaushort{12}\,,\ytableaushort{12}\right)\,.
    \end{equation}
    For the 2-momenta case, the self-dual SSYTs can be obtained similarly,
    \begin{align}
\left(\ytableaushort{12}\,,\ytableaushort{12}\right) \times \left\{\begin{array}{l}
\left(\ytableaushort{33}\,,\ytableaushort{33}\right)\\
\left(\ytableaushort{34}\,,\ytableaushort{34}\right)\\
\left(\ytableaushort{44}\,,\ytableaushort{44}\right)
\end{array}\right. \rightarrow \left\{\begin{array}{l}
\left(\ytableaushort{123,4}\,,\ytableaushort{123,4}\right) \\
\left(\ytableaushort{124,4}\,,\ytableaushort{124,4}\right) \\
\left(\ytableaushort{123,3}\,,\ytableaushort{123,3}\right) \\
\left(\ytableaushort{124,3}\,,\ytableaushort{124,3}\right) 
\end{array}\right.\,,
    \end{align} 
    \begin{align}
        &\left(\ytableaushort{1,2}\,,\ytableaushort{1,2}\right) \times\left\{\begin{array}{l}
\left(\ytableaushort{33}\,,\ytableaushort{33}\right)\\
\left(\ytableaushort{34}\,,\ytableaushort{34}\right)\\
\left(\ytableaushort{44}\,,\ytableaushort{44}\right)
\end{array}\right. \rightarrow \left\{\begin{array}{l}
\left(\ytableaushort{133,2}\,,\ytableaushort{133,2}\right) \\
\left(\ytableaushort{144,2}\,,\ytableaushort{144,2}\right) \\
\left(\ytableaushort{134,2}\,,\ytableaushort{134,2}\right)
\end{array}
\right.\,. 
    \end{align}
    Other SSYTs are 
    \begin{align}
        \left(\ytableaushort{12}\,,\ytableaushort{12}\right) &\times 
\left(\ytableaushort{34}\,,\ytableaushort{3,4}\right)
 \rightarrow 
\left(\ytableaushort{123,4}\,,\ytableaushort{124,3}\right) \,,\\
\left(\ytableaushort{12}\,,\ytableaushort{1,2}\right) &\times \left\{\begin{array}{l}
\left(\ytableaushort{33}\,,\ytableaushort{33}\right)\\
\left(\ytableaushort{34}\,,\ytableaushort{34}\right)\\
\left(\ytableaushort{44}\,,\ytableaushort{44}\right)
\end{array}\right. \rightarrow 
\left\{\begin{array}{cc}
\left(\ytableaushort{123,3}\,,\ytableaushort{133,2}\right) \\
\left(\ytableaushort{124,4}\,,\ytableaushort{144,2}\right) \\
\left(\ytableaushort{123,4}\,,\ytableaushort{134,2}\right)
\end{array}\right. \,,\\
\left(\ytableaushort{12}\,,\ytableaushort{1,2}\right) &\times \left(\ytableaushort{3,4}\,,\ytableaushort{34}\right) \rightarrow \left(\ytableaushort{124,3}\,,\ytableaushort{134,2}\right)\,,
    \end{align}
    and their dual SSYTs. While for the 4 momenta case, the nonequivalent SSYTs are not too massive because of the relations in Eq.~\eqref{eq:pqrelation_1} to Eq.~\eqref{eq:pqrelation_2}, The self-dual SSYTs are
    \begin{align}
        \left(\ytableaushort{12}\,,\ytableaushort{12}\right)\times \left\{\begin{array}{l}
\left(\ytableaushort{3333}\,,\ytableaushort{3333}\right)\\
\left(\ytableaushort{3334}\,,\ytableaushort{3334}\right)\\
\left(\ytableaushort{3344}\,,\ytableaushort{3344}\right)\\
\left(\ytableaushort{3444}\,,\ytableaushort{3444}\right)\\
\left(\ytableaushort{4444}\,,\ytableaushort{4444}\right)\\
        \end{array}\right.\rightarrow \left\{\begin{array}{l}
\left(\ytableaushort{1233,44}\,,\ytableaushort{1233,44}\right)\\
\left(\ytableaushort{1234,44}\,,\ytableaushort{1234,44}\right)\\
\left(\ytableaushort{1244,44}\,,\ytableaushort{1244,44}\right)
        \end{array}\right.\,,
    \end{align}
    and others are 
    \begin{align}
    \left(\ytableaushort{12}\,,\ytableaushort{12}\right)&\times \left\{\begin{array}{l}
\left(\ytableaushort{3334}\,,\ytableaushort{333,4}\right)\\
\left(\ytableaushort{3344}\,,\ytableaushort{334,4}\right)\\
\left(\ytableaushort{3444}\,,\ytableaushort{344,4}\right)\\
        \end{array}\right.\rightarrow \left\{
        \begin{array}{l}
\left(\ytableaushort{1234,33}\,,\ytableaushort{1233,34}\right) \\
\left(\ytableaushort{1234,34}\,,\ytableaushort{1233,44}\right) \\
\left(\ytableaushort{1244,34}\,,\ytableaushort{1234,44}\right) 
        \end{array}
        \right. \,,\\
    \left(\ytableaushort{12}\,,\ytableaushort{12}\right)&\times \left(\ytableaushort{3344}\,,\ytableaushort{33,44}\right) \supset \left(\ytableaushort{1244,33}\,,\ytableaushort{1233,44}\right)\,, \\
    \left(\ytableaushort{12}\,,\ytableaushort{1,2}\right)&\times \left\{\begin{array}{l}
\left(\ytableaushort{3334}\,,\ytableaushort{333,4}\right)\\
\left(\ytableaushort{3344}\,,\ytableaushort{334,4}\right)\\
\left(\ytableaushort{3444}\,,\ytableaushort{344,4}\right)\\
        \end{array}\right.\rightarrow \left\{
        \begin{array}{l}
\left(\ytableaushort{1233,34},\ytableaushort{1333,24}\right) \\
\left(\ytableaushort{1234,34},\ytableaushort{1334,24}\right) \\
\left(\ytableaushort{1244,34},\ytableaushort{1344,24}\right) 
        \end{array}
        \right.\,,
\end{align}
and their dual SSYTs. Thus, there are 11 self-dual SSYTs and 24 SSYTs dual to each other.
\end{description}
In particular, all these SSYTs above are not the minimal ones, because the T and P symmetries will offer new equivalent relations. Before discussing them, we work out the SSYTs of the fermions first. Different from the bosonic case, there are no self-dual SSYTs in the fermionic case.

\subsubsection*{Fermionic Case}

\begin{description}
    \item [Irrep $(0,0)$:] According to Tab.~\ref{tab:maximum}, there is no momentum in this irrep, and there are only two nonequivalent SSYTs,
    \begin{equation}
        \left(\ytableaushort{1,2},1\right)\,,\quad \left(1,\ytableaushort{1,2}\right)\,,
    \end{equation}
    According to Tab.~\ref{tab:symmetry_fermion}, both of them are symmetric under the permutation.
    \item [Irrep $(\frac{1}{2},\frac{1}{2})$:] Without momentum, there are two SSYTs
    \begin{equation}
        \left(\ytableaushort{1}\,,\ytableaushort{2}\right)\sim \left(\ytableaushort{13,2}\,,\ytableaushort{3}\right)\,,\quad \left(\ytableaushort{2}\,,\ytableaushort{1}\right) \sim \left(\ytableaushort{3}\,,\ytableaushort{13,2}\right)\,, 
    \end{equation}
    where the equivalence $\sim$ is from the Gorden identity in Eq.~\eqref{eq:gorden_identity}. With one momentum, there are
    \begin{align}
        \left(\ytableaushort{12}\,,1\right) \times \left(\ytableaushort{4}\,,\ytableaushort{4}\right) &\supset \left(\ytableaushort{12,4}\,,\ytableaushort{4}\right) \,,\\
        \left(\ytableaushort{12}\,,1\right) \times \left(\ytableaushort{3}\,,\ytableaushort{3}\right) &\supset \left(\ytableaushort{12,3}\,,\ytableaushort{3}\right) \rightarrow  \left(\ytableaushort{14,2}\,,\ytableaushort{4}\right) \,,\\
        \left(\ytableaushort{1,2}\,,1\right) \times \left(\ytableaushort{3}\,,\ytableaushort{3}\right) &\supset \left(\ytableaushort{13,2}\,,\ytableaushort{3}\right) \,,\\
        \left(\ytableaushort{1,2}\,,1\right) \times \left(\ytableaushort{4}\,,\ytableaushort{4}\right) &\supset \left(\ytableaushort{14,2}\,,\ytableaushort{4}\right)\,,
        \end{align}
    and their dual SSYTs. Thus, there are 6 nonequivalent SSYTs, including 
    \begin{equation}
        \left(\ytableaushort{12,4}\,,\ytableaushort{4}\right)\,, \quad \left(\ytableaushort{13,2}\,,\ytableaushort{3}\right)\,,\quad \left(\ytableaushort{14,2}\,,\ytableaushort{4}\right)\,,
    \end{equation}
    and their dual SSYTs.
    
    \item [Irrep $(1,0)\oplus(0,1)$:] Similar to the boson case, we only present the $(1,0)$ representation here. With no momentum, there is only one SSYT
    \begin{equation}
        \left(\ytableaushort{12}\,,1\right)\,,
    \end{equation}
    which is antisymmetric, while for one momentum, there are 4 SSYTs but none of them is independent because of the Gordon identity in Eq.~\eqref{eq:gorden_identity},
    \begin{align}
        \left(\ytableaushort{1}\,,\ytableaushort{2}\right) \times \left(\ytableaushort{3}\,,\ytableaushort{3}\right) & \supset \left(\ytableaushort{13}\,,\ytableaushort{2,3}\right) \rightarrow \left(\ytableaushort{133,2}\,,\ytableaushort{3,3}\right) \rightarrow 0  \,,\\
        \left(\ytableaushort{2}\,,\ytableaushort{1}\right) \times \left(\ytableaushort{3}\,,\ytableaushort{3}\right) & \supset \left(\ytableaushort{23}\,,\ytableaushort{1,3}\right) \rightarrow \left(\ytableaushort{33}\,,\ytableaushort{13,23}\right) \rightarrow 0 \,,\\
        \left(\ytableaushort{1}\,,\ytableaushort{2}\right) \times \left(\ytableaushort{4}\,,\ytableaushort{4}\right) & \supset \left(\ytableaushort{14}\,,\ytableaushort{2,4}\right) \rightarrow \left(\ytableaushort{134,2}\,,\ytableaushort{3,4}\right) \,,\\
        \left(\ytableaushort{2}\,,\ytableaushort{1}\right) \times \left(\ytableaushort{4}\,,\ytableaushort{4}\right) & \supset \left(\ytableaushort{24}\,,\ytableaushort{1,4}\right) \rightarrow \left(\ytableaushort{34}\,,\ytableaushort{13,24}\right)\,.
    \end{align}
    For two momenta, there are
    \begin{align}
        \left(\ytableaushort{1,2}\,,1\right) \times \left(\ytableaushort{34}\,,\ytableaushort{3,4}\right) &\supset \left(\ytableaushort{134,2}\,,\ytableaushort{3,4}\right) \,,\\
        \left(1\,,\ytableaushort{1,2}\right) \times \left(\ytableaushort{34}\,,\ytableaushort{3,4}\right) &\supset \left(\ytableaushort{34}\,,\ytableaushort{13,24}\right) \,,\\
        \left(\ytableaushort{12}\,,1\right) \times \left(\ytableaushort{34}\,,\ytableaushort{3,4}\right) &\supset \left\{\begin{array}{l}
        \left(\ytableaushort{123,4}\,,\ytableaushort{3,4}\right) \\
        \left(\ytableaushort{124,3}\,,\ytableaushort{3,4}\right) \rightarrow \left(\ytableaushort{144,2}\,,\ytableaushort{4,4}\right) \rightarrow 0
        \end{array}\right. \,,\\
        \left(1\,,\ytableaushort{12}\right) \times \left(\ytableaushort{33}\,,\ytableaushort{33}\right) &\supset \left(\ytableaushort{33}\,,\ytableaushort{12,33}\right) \rightarrow \left(\ytableaushort{34}\,,\ytableaushort{13,24}\right) \,,\\
        \left(1\,,\ytableaushort{12}\right) \times \left(\ytableaushort{44}\,,\ytableaushort{44}\right) &\supset \left(\ytableaushort{44}\,,\ytableaushort{12,44}\right) \,,\\
        \left(1\,,\ytableaushort{12}\right) \times \left(\ytableaushort{34}\,,\ytableaushort{34}\right) &\supset \left(\ytableaushort{34}\,,\ytableaushort{12,34}\right) \rightarrow \left(\ytableaushort{34}\,,\ytableaushort{13,24}\right)\,,
    \end{align}
    so there are 4 independent SSYTs, of which two are symmetric, 
    \begin{equation}
        \left(\ytableaushort{134,2}\,,\ytableaushort{3,4}\right) \,,\quad \left(\ytableaushort{34}\,,\ytableaushort{13,24}\right)\,,
    \end{equation}
    and the others are antisymmetric,
    \begin{equation}
        \left(\ytableaushort{12}\,,1\right) \,,\quad \left(\ytableaushort{123,4}\,,\ytableaushort{3,4}\right)\,.
    \end{equation}
    The SSYTs of irrep $(0\,,1)$ are the dual ones.
    \item [Irrep $(1,1)$:] There is at least 1 momentum in this irrep, but just as shown before, these SSYTs are not independent because of Gordon identity in Eq.~\eqref{eq:gorden_identity},
    \begin{align}
        \left(\ytableaushort{1}\,,\ytableaushort{2}\right) \times \left(\ytableaushort{3}\,,\ytableaushort{3}\right) &\supset \left(\ytableaushort{13}\,,\ytableaushort{23}\right) \rightarrow \left(\ytableaushort{133,2}\,,\ytableaushort{33}\right) \,,\\
        \left(\ytableaushort{1}\,,\ytableaushort{2}\right) \times \left(\ytableaushort{4}\,,\ytableaushort{4}\right) &\supset \left(\ytableaushort{14}\,,\ytableaushort{24}\right) \rightarrow \left(\ytableaushort{134,2}\,,\ytableaushort{34}\right) \,,\\
        \left(\ytableaushort{2}\,,\ytableaushort{1}\right) \times \left(\ytableaushort{3}\,,\ytableaushort{3}\right) &\supset \left(\ytableaushort{23}\,,\ytableaushort{13}\right) \rightarrow \left(\ytableaushort{33}\,,\ytableaushort{133,2}\right) \,,\\
        \left(\ytableaushort{2}\,,\ytableaushort{1}\right) \times \left(\ytableaushort{4}\,,\ytableaushort{4}\right) &\supset \left(\ytableaushort{24}\,,\ytableaushort{14}\right) \rightarrow \left(\ytableaushort{34}\,,\ytableaushort{134,2}\right) \,.
    \end{align}
    With two momenta, there are 
    \begin{align}
        \left(\ytableaushort{12}\,,1\right) \times \left(\ytableaushort{33}\,,\ytableaushort{33}\right) &\supset \left(\ytableaushort{123,3}\,,\ytableaushort{33}\right) \rightarrow \left(\ytableaushort{134,2}\,,\ytableaushort{34}\right)  \,,\\
        \left(\ytableaushort{12}\,,1\right) \times \left(\ytableaushort{44}\,,\ytableaushort{44}\right) &\supset \left(\ytableaushort{124,4}\,,\ytableaushort{44}\right) \,,\\
        \left(\ytableaushort{12}\,,1\right) \times \left\{\begin{array}{l}
        \left(\ytableaushort{34}\,,\ytableaushort{34}\right) \\
        \left(\ytableaushort{3,4}\,,\ytableaushort{34}\right)
        \end{array}\right. & \supset \left\{\begin{array}{l}
\left(\ytableaushort{123,4}\,,\ytableaushort{34}\right)  \\
\left(\ytableaushort{124,3}\,,\ytableaushort{34}\right)\rightarrow \left(\ytableaushort{144,2}\,,\ytableaushort{44}\right)
\end{array}\right. \,,
    \end{align}
    \begin{align}
        \left(\ytableaushort{1,2}\,,1\right) \times \left(\ytableaushort{33}\,,\ytableaushort{33}\right) &\supset \left(\ytableaushort{133,2}\,,\ytableaushort{33}\right)\,, \\
        \left(\ytableaushort{1,2}\,,1\right) \times \left(\ytableaushort{44}\,,\ytableaushort{44}\right) &\supset \left(\ytableaushort{144,2}\,,\ytableaushort{44}\right)\,, \\
        \left(\ytableaushort{1,2}\,,1\right) \times \left(\ytableaushort{34}\,,\ytableaushort{34}\right) &\supset \left(\ytableaushort{134,2}\,,\ytableaushort{34}\right)\,. 
    \end{align}
    Besides, the dual SSYTs must be included, thus we get 10 independent SSYTs, of which 6 are symmetric and 4 are antisymmetric.
\end{description}

In summary, we have illustrated the construction and reduction of the nonequivalent SSYTs by the spin-1 boson and spin-1/2 fermion. As shown before, although these SSYTs are independent to each other algebraically, the corresponding tensors are generally not. To reduce these SSYTs completely, we need to consult other constraints, which will be discussed next, namely the $P$ and $T$ symmetries.

\subsubsection*{Parity and Time-Reversal}

The $P$ and $T$ (or $C$ equivalent by the CPT principle) have been used to count independent P- and T-conserving tensors before, while here we will apply them to the SSYTs. It turns out that both $P$ and $T$ are naturally complemented by the SSYTs, thus they help to classify all the SSYTs obtained before in terms of their P and T eigenvalues. Besides, the P- and T-symmetries also offer a further constraint on the SSYTs, which reduces them so that their corresponding tensors are complete and independent.

The $P$ interchanges the $SU(2)_L$ indices and the $SU(2)_R$ indices of the Lorentz group, which means if a Lorentz irrep $(j_1\,,j_2)$ is symmetric, $(j\,,j)$, it is closed under $P$ transformation, otherwise, it is not, but the combinations $(j_1\,,j_2)\oplus (j_2\,,j_1)$ is. In detail, every tensor in a irrep $(j\,,j)$ possesses an specific $P$ charge, $\pm1$, otherwise only the combinations of two tensors of $(j_1\,,j_2)$ and $(j_2\,,j_1)$ are of specific $P$ charge. Expressed by SSYTs, the argument above is interpreted as 
\begin{itemize}
    \item The $P$ transform on a tensor corresponding to an SSYT leads to another tensor corresponding to the SSYT with the two sub-SSYTs, belonging to $SU(2)_L$ and $SU(2)_R$ respectively, interchanged, which is the dual SSYTs. 
    \item If the two sub-SSYTs are identical, it is invariant under the $P$ transformation, or its $P$ eigenvalue is $+1$, we say it or its corresponding tensors are $P$-even. Otherwise, the two SSYTs dual to each other add to generate a $P$-even one, and subtract to generate a $P$-odd one. For example, the SSYTs such as 
    \begin{equation}
        \left(\ytableaushort{12,44}\,,\ytableaushort{12,44}\right)\,, \quad \left(\ytableaushort{12,3}\,,\ytableaushort{12,3}\right) \,,\quad \left(\ytableaushort{123,4}\,,\ytableaushort{123,4}\right)\,,
    \end{equation}
    are $P$-even, while 
    \begin{equation}
        \left(\ytableaushort{12,4}\,,\ytableaushort{4}\right) \,,\quad \left(\ytableaushort{4}\,,\ytableaushort{12,4}\right)\,,
    \end{equation}
    are neither $P$-even nor $P$-odd, but their combinations are
    \begin{align}
        \text{$P$-even:}&\quad \left(\ytableaushort{12,4}\,,\ytableaushort{4}\right) +\left(\ytableaushort{4}\,,\ytableaushort{12,4}\right) \,,\\
        \text{$P$-odd:}&\quad \left(\ytableaushort{12,4}\,,\ytableaushort{4}\right) -\left(\ytableaushort{4}\,,\ytableaushort{12,4}\right)\,.
    \end{align}
    \item In particular, both of the situations above emerge in the bosonic case, which means in the symmetric irrep $(j\,,j)$, even though the two sub SSYTs shapes are the same, their fillings can be different, for example, 
    \begin{equation}
        \left(\ytableaushort{124,3}\,,\ytableaushort{124,2}\right) \quad \text{and} \quad \left(\ytableaushort{124,2}\,,\ytableaushort{124,3}\right)\,.
    \end{equation}
    In the fermionic case, only the latter situation is possible, since the Lorentz irreps of the fermions are always antisymmetric.
\end{itemize}

The time reversal is different, which interchanges the initial and final states and other quantities of them, thus the building blocks in Eq.~\eqref{eq:building_blocks} are transformed into 
\begin{equation}
    \left\{\begin{array}{l}
\phi_1 \\\phi_2^\dagger\\P\\q
    \end{array}\right. \quad \longrightarrow \quad \left\{\begin{array}{l}
\phi_2 \\\phi_1^\dagger\\P\\-q
    \end{array}\right. \,,
\end{equation}
by the time reversal. Regarding SSYTs, it means time reversal interchanges the indices 1 and 2 and times a $-1$ for every index 4. As shown in Tab.~\ref{tab:symmetry_boson} and Tab.~\ref{tab:symmetry_fermion}, all the SSYTs or their combinations are of specific permutation symmetries, thus they are of specific time reversal eigenvalues. Including the number of index $4$, the time reversal eigenvalue of an SSYT with specific wave functions permutation symmetry $\eta_I$ is
\begin{equation}
    \eta_T = (-1)^{\# 4} \times \eta_I\,.
\end{equation}
For example, the following SSYTs are $T$-even,
\begin{align}
    \left(\ytableaushort{12,44}\,,\ytableaushort{12,44}\right) \quad & \eta_T = (-1)^2\times 1 = 1 \,,\\
    \left(\ytableaushort{12,34}\,,\ytableaushort{13,24}\right) \quad & \eta_T = (-1)\times (-1) = 1 \,,\\
    \left(\ytableaushort{133,2}\,,\ytableaushort{33}\right) \quad & \eta_T = (-1)^0\times 1 = 1 \,.
\end{align}

The $P$ symmetry helps to reduce the SSYTs further. In detail, it can reduce the SSYTs of symmetric irrep $(j\,,j)$ in the bosonic case. Because the symmetric irrep is closed under the $P$ transformation, we can divide all the SSYTs into two parts, the first part contains all the SSYTs with two identical sub-SSYTs, which are self-dual and $P$-even, and the second part contains all the SSYTs with two different sub-SSYTs, which should be combined to form $P$-even or $P$-odd ones. The additions in the second part give $P$-even SSYTs, which can not be independent because all the $P$-even ones have been included in the first part, since we take the most general SSYTs obtained by the Littlewood-Richardson rule into account. Thus, we conclude the $P$-even ones obtained from the second part are all redundant. For example, we have presented 4 SSYTs for the irrep $(0,0)$ of spin-1 particles, and now it is understood that the sum of the last two is redundant, and only their minus is independent. Thus, there are 3 independent SSYTs rather than 4, of which 2 are $P$-even, and 1 is $P$-odd.
\begin{align}
    \text{$P$-even:}\quad & \left(\ytableaushort{1,2}\,,\ytableaushort{1,2}\right)\,,\quad \left(\ytableaushort{12,44}\,,\ytableaushort{12,44}\right)\,,\\
    \text{$P$-odd:}\quad & \left(\ytableaushort{12,34}\,,\ytableaushort{13,24}\right) - \left(\ytableaushort{13,24}\,,\ytableaushort{12,34}\right)\,.
\end{align}
Similar reductions exist for the other irrep of the bosonic and fermionic cases. Because these irreps are not symmetric, the combinations are needed to form SSYTs with specific $P$ eigenvalues. If an SSYT can be factorized as a product of a singlet SSYT and another SSYT of the same irrep as the original one, for example,
\begin{equation}
    \left(\ytableaushort{1234,34}\,,\ytableaushort{133,244}\right) = \left(\ytableaushort{12,34}\,,\ytableaushort{13,24}\right) \times \left(\ytableaushort{34}\,,\ytableaushort{3,4}\right)\,,
\end{equation}
where the first singlet SSYT is not symmetric. The combinations of this SSYT and its dual one can also be factorized,
\begin{align}
    \left(\ytableaushort{1234,34}\,,\ytableaushort{133,244}\right) + \text{dual} &= \left[\left(\ytableaushort{12,34}\,,\ytableaushort{13,24}\right) + \text{dual}\right] \times \left[\left(\ytableaushort{34}\,,\ytableaushort{3,4}\right) + \text{dual}\right] \notag \\
    &+ \left[\left(\ytableaushort{12,34}\,,\ytableaushort{13,24}\right) - \text{dual}\right] \times \left[\left(\ytableaushort{34}\,,\ytableaushort{3,4}\right) - \text{dual}\right]  \,,\\
    \left(\ytableaushort{1234,34}\,,\ytableaushort{133,244}\right) - \text{dual} &= \left[\left(\ytableaushort{12,34}\,,\ytableaushort{13,24}\right) + \text{dual}\right] \times \left[\left(\ytableaushort{34}\,,\ytableaushort{3,4}\right) - \text{dual}\right] \notag \\
    &+ \left[\left(\ytableaushort{12,34}\,,\ytableaushort{13,24}\right) - \text{dual}\right] \times \left[\left(\ytableaushort{34}\,,\ytableaushort{3,4}\right) + \text{dual}\right] \,.
\end{align}
As discussed before, the $P$-even combinations of the singlet SSYTs are not independent, thus, the first two on the right-hand side are not independent. While the remaining $P$-even one is a product of two $P$-odd ones, which is not independent because of the relation
\begin{equation}
    \epsilon^{\mu\nu\rho\lambda}\epsilon_{\alpha\beta\kappa\delta} =  - \delta^{\mu\nu\rho\lambda}_{\alpha\beta\kappa\delta}\,,
\end{equation}
so only the $P$-odd combination survives,
\begin{equation}
    \left(\ytableaushort{1234,34}\,,\ytableaushort{133,244}\right) - \text{dual} = \left[\left(\ytableaushort{12,34}\,,\ytableaushort{13,24}\right) - \text{dual}\right] \times \left[\left(\ytableaushort{34}\,,\ytableaushort{3,4}\right) + \text{dual}\right]\,.
\end{equation}
This relation also applies to cases with high spin.

At this stage, we have obtained all the independent SSYTs, and we would like to organize them by their $P$ and $T$ eigenvalues. The nonequivalent SSYTs of spin-1 boson are presented in Tab.~\ref{tab:SSYTs_boson_1} and Tab.~\ref{tab:SSYTs_boson_2}, while the nonequivalent SSYTs of spin-1/2 fermion are presented in Tab.~\ref{tab:SSYTs_fermion}. The $P$-even and $T$-even sectors of each irreps are consistent with the counting result presented in Tab.~\ref{tab:numbers}. In particular, the tensors for spin-1 cases are consistent with the counting result in Tab.~\ref{tab:number1} from the Hilbert series.

\begin{table}[]
\renewcommand\arraystretch{1.5}
\ytableausetup{smalltableaux}
    \centering
\begin{tabularx}{\linewidth}{c|>{\centering\arraybackslash}X|>{\centering\arraybackslash}X|}
    \cline{2-3}
     & \multicolumn{2}{c|}{\textbf{spin-$1$, irrep $(0\,,0)$}} \\
     \cline{2-3}
     \textbf{$T$-even} & $\left(\ytableaushort{1,2}\,,\ytableaushort{1,2}\right)$,$\left(\ytableaushort{12,44}\,,\ytableaushort{12,44}\right)$ & $\left(\ytableaushort{12,34}\,,\ytableaushort{13,24}\right)-\text{dual}$\\
     \cline{2-3}
     \textbf{$T$-odd} & & \\
     \cline{2-3}
     \multicolumn{2}{c}{\textbf{$P$-even}} & \multicolumn{1}{c}{\textbf{$P$-odd}} 
\end{tabularx}
\begin{tabularx}{\linewidth}{c| >{\centering\arraybackslash}X|>{\centering\arraybackslash}X|}
    \cline{2-3}
     & \multicolumn{2}{c|}{\textbf{spin-$1$, irrep $(\frac{1}{2}\,,\frac{1}{2})$}} \\
     \cline{2-3}
      \multirow{3}{*}{\textbf{$T$-even}} & $\left(\ytableaushort{12,3}\,,\ytableaushort{12,3}\right)$ & $\left(\ytableaushort{12,4}\,,\ytableaushort{14,2}\right) - \text{dual}$ \\
      & $\left(\ytableaushort{13,2}\,,\ytableaushort{13,2}\right)$ & $\left(\ytableaushort{124,34}\,,\ytableaushort{123,44}\right)-\text{dual}$ \\
      & $\left(\ytableaushort{123,44}\,,\ytableaushort{123,44}\right)$ & $\left(\ytableaushort{123,34}\,,\ytableaushort{133,24}\right)-\text{dual}$ \\
     \cline{2-3}
      \multirow{3}{*}{\textbf{$T$-odd}}& $\left(\ytableaushort{12,4}\,,\ytableaushort{12,4}\right)$ & 
$\left(\ytableaushort{12,3}\,,\ytableaushort{13,2}\right) - \text{dual}$ \\
     & $\left(\ytableaushort{14,2}\,,\ytableaushort{14,2}\right)$ & $\left(\ytableaushort{124,33}\,,\ytableaushort{123,34}\right)-\text{dual}$ \\
     & $\left(\ytableaushort{124,44}\,,\ytableaushort{124,44}\right)$ & $\left(\ytableaushort{124,34}\,,\ytableaushort{134,24}\right)-\text{dual}$ \\
     \cline{2-3}
     \multicolumn{2}{c}{\textbf{$P$-even}} & \multicolumn{1}{c}{\textbf{$P$-odd}} 
\end{tabularx}
\begin{tabularx}{\linewidth}{c| >{\centering\arraybackslash}X|>{\centering\arraybackslash}X|}
    \cline{2-3}
     & \multicolumn{2}{c|}{\textbf{spin-$1$, irrep $(1\,,0)\oplus(0\,,1)$}} \\
     \cline{2-3}
      \multirow{2}{*}{\textbf{$T$-even}} & $\left(\ytableaushort{123,3}\,,\ytableaushort{12,33}\right) + \text{dual}$ & $\left(\ytableaushort{123,3}\,,\ytableaushort{12,33}\right) - \text{dual}$ \\
      & $\left(\ytableaushort{124,4}\,,\ytableaushort{12,44}\right) + \text{dual}$ & $\left(\ytableaushort{124,4}\,,\ytableaushort{12,44}\right) - \text{dual}$ \\
     \cline{2-3}
     \multirow{5}{*}{\textbf{$T$-odd}}& $\left(\ytableaushort{12}\,,\ytableaushort{1,2}\right)+\text{dual}$ & $\left(\ytableaushort{12}\,,\ytableaushort{1,2}\right)-\text{dual}$ \\
     &  $\left(\ytableaushort{124,3}\,,\ytableaushort{12,34}\right)+\text{dual}$ & $\left(\ytableaushort{124,3}\,,\ytableaushort{12,34}\right)-\text{dual}$ \\
     &  $\left(\ytableaushort{123,4}\,,\ytableaushort{12,34}\right)+\text{dual}$ & $\left(\ytableaushort{123,4}\,,\ytableaushort{12,34}\right)-\text{dual}$ \\
     & $\left(\ytableaushort{134,2}\,,\ytableaushort{13,24}\right)+\text{dual}$ & $\left(\ytableaushort{134,2}\,,\ytableaushort{13,24}\right)-\text{dual}$ \\
     & $\left(\ytableaushort{1234,44}\,,\ytableaushort{123,444}\right)+\text{dual}$ & $\left(\ytableaushort{1234,44}\,,\ytableaushort{123,444}\right)-\text{dual}$ \\
     & 
     & $\left(\ytableaushort{1234,34}\,,\ytableaushort{133,244}\right)-\text{dual}$ \\
     \cline{2-3}
     \multicolumn{2}{c}{\textbf{$P$-even}} & \multicolumn{1}{c}{\textbf{$P$-odd}} 
\end{tabularx}
    \caption{All the nonequivalent SSYTs of spin-1 boson organized by their $P$ and $T$ eigenvalues. This tabular contains the irreps $(0\,,0)\,,(\frac{1}{2}\,,\frac{1}{2})$, and $(1\,,0)\oplus(0\,,1)$, while the irrep $(1\,,1)$ is presented in Tab.~\ref{tab:SSYTs_boson_2}.}
    \label{tab:SSYTs_boson_1}
\end{table}

\begin{table}
\begin{center}
\renewcommand\arraystretch{1.5}
\ytableausetup{smalltableaux}
    \begin{tabularx}{\linewidth}{c| >{\centering\arraybackslash}X|>{\centering\arraybackslash}X|}
    \cline{2-3}
     & \multicolumn{2}{c|}{\textbf{spin-$1$, irrep $(1\,,1)$}} \\
     \cline{2-3}
     \multirow{7}{*}{\textbf{$T$-even}} & $\left(\ytableaushort{12}\,,\ytableaushort{12}\right)$ & $\left(\ytableaushort{123,4}\,,\ytableaushort{134,2}\right)- \text{dual}$ \\
     & $\left(\ytableaushort{124,4}\,,\ytableaushort{124,4}\right)$ & $\left(\ytableaushort{124,3}\,,\ytableaushort{134,2}\right)- \text{dual}$ \\
     & $\left(\ytableaushort{123,3}\,,\ytableaushort{123,3}\right)$ & $\left(\ytableaushort{1234,34}\,,\ytableaushort{1233,44}\right)- \text{dual}$ \\
     & $\left(\ytableaushort{133,2}\,,\ytableaushort{133,2}\right)$ & $\left(\ytableaushort{1244,33}\,,\ytableaushort{1233,44}\right)- \text{dual}$ \\
     & $\left(\ytableaushort{144,2}\,,\ytableaushort{144,2}\right)$ & $\left(\ytableaushort{1233,34}\,,\ytableaushort{1333,24}\right)- \text{dual}$\\
     & $\left(\ytableaushort{1233,44}\,,\ytableaushort{1233,44}\right)$ & $\left(\ytableaushort{1244,34}\,,\ytableaushort{1344,24}\right)- \text{dual}$\\
     & $\left(\ytableaushort{1244,44}\,,\ytableaushort{1244,44}\right)$ & \\
     \cline{2-3}
     \multirow{6}{*}{\textbf{$T$-odd}}& $\left(\ytableaushort{123,4}\,,\ytableaushort{123,4}\right)$ & $\left(\ytableaushort{123,4}\,,\ytableaushort{124,3}\right) - \text{dual}$\\
     & $\left(\ytableaushort{124,3}\,,\ytableaushort{124,3}\right)$ & $\left(\ytableaushort{123,3}\,,\ytableaushort{133,2}\right) - \text{dual}$ \\
     & $\left(\ytableaushort{134,2}\,,\ytableaushort{134,2}\right)$ & $\left(\ytableaushort{124,4}\,,\ytableaushort{144,2}\right) - \text{dual}$ \\
     & $\left(\ytableaushort{1234,44}\,,\ytableaushort{1234,44}\right)$ & $\left(\ytableaushort{1234,33}\,,\ytableaushort{1233,34}\right) - \text{dual}$\\
     & & $\left(\ytableaushort{1244,34}\,,\ytableaushort{1234,44}\right) - \text{dual}$ \\
     & & $\left(\ytableaushort{1234,34}\,,\ytableaushort{1334,24}\right) - \text{dual}$ \\
     \cline{2-3}
     \multicolumn{2}{c}{\textbf{$P$-even}} & \multicolumn{1}{c}{\textbf{$P$-odd}} 
\end{tabularx}
\end{center}
\caption{All the nonequivalent SSYTs of spin-1 boson organized by their $P$ and $T$ eigenvalues. This tabular contains the irrep $(1\,,1)$, while the irreps $(0\,,0)\,,(\frac{1}{2}\,,\frac{1}{2})$, and $(1\,,0)\oplus(0\,,1)$ are presented in Tab.~\ref{tab:SSYTs_boson_1}}
    \label{tab:SSYTs_boson_2}
\end{table}

\begin{table}
\begin{center}
\renewcommand\arraystretch{1.5}
\ytableausetup{smalltableaux}
    \begin{tabularx}{\linewidth}{c|>{\centering\arraybackslash}X|>{\centering\arraybackslash}X|}
     \multicolumn{3}{c}{\textbf{spin-$1/2$, irrep $(0\,,0)$}} \\
     \cline{2-3}
     \textbf{$T$-even} & $\left(\ytableaushort{1,2}\,,1\right)+\text{dual}$ & $\left(\ytableaushort{1,2}\,,1\right)-\text{dual}$\\
     \cline{2-3}
     \textbf{$T$-odd} & & \\
     \cline{2-3}
     \multicolumn{2}{c}{\textbf{$P$-even}} & \multicolumn{1}{c}{\textbf{$P$-odd}} 
\end{tabularx}
\begin{tabularx}{\linewidth}{c|>{\centering\arraybackslash}X|>{\centering\arraybackslash}X|}
    \cline{2-3}
     & \multicolumn{2}{c|}{\textbf{spin-$1/2$, irrep $(\frac{1}{2}\,,\frac{1}{2})$}} \\
     \cline{2-3}
     \multirow{2}{*}{\textbf{$T$-even}} & $\left(\ytableaushort{12,4}\,,\ytableaushort{4}\right)+\text{dual}$ & $\left(\ytableaushort{12,4}\,,\ytableaushort{4}\right)-\text{dual}$\\
     & $\left(\ytableaushort{13,2}\,,\ytableaushort{3}\right)+\text{dual}$ & $\left(\ytableaushort{13,2}\,,\ytableaushort{3}\right)-\text{dual}$\\
     \cline{2-3}
     \textbf{$T$-odd} & $\left(\ytableaushort{14,2}\,,\ytableaushort{4}\right)+\text{dual}$ & $\left(\ytableaushort{14,2}\,,\ytableaushort{4}\right)-\text{dual}$ \\
     \cline{2-3}
     \multicolumn{2}{c}{\textbf{$P$-even}} & \multicolumn{1}{c}{\textbf{$P$-odd}} 
\end{tabularx}
\begin{tabularx}{\linewidth}{c|>{\centering\arraybackslash}X|>{\centering\arraybackslash}X|}
    \cline{2-3}
     & \multicolumn{2}{c|}{\textbf{spin-$1/2$, irrep $(1\,,0)\oplus (0\,,1)$}} \\
     \cline{2-3}
     \textbf{$T$-even} & $\left(\ytableaushort{123,4}\,,\ytableaushort{3,4}\right)+\text{dual}$ & $\left(\ytableaushort{123,4}\,,\ytableaushort{3,4}\right)-\text{dual}$\\
     \cline{2-3}
     \multirow{4}{*}{\textbf{$T$-odd}} & $\left(\ytableaushort{12}\,,1\right) + \text{dual}$ & $\left(\ytableaushort{12}\,,1\right) - \text{dual}$\\
     & $\left(\ytableaushort{134,2}\,,\ytableaushort{3,4}\right) + \text{dual}$ & $\left(\ytableaushort{134,2}\,,\ytableaushort{3,4}\right) - \text{dual}$\\
     & $\left(\ytableaushort{34}\,,\ytableaushort{13,24}\right) + \text{dual}$ & $\left(\ytableaushort{34}\,,\ytableaushort{13,24}\right) - \text{dual}$\\
     & $\left(\ytableaushort{44}\,,\ytableaushort{12,44}\right) + \text{dual}$ & $\left(\ytableaushort{44}\,,\ytableaushort{12,44}\right) - \text{dual}$\\
     \cline{2-3}
     \multicolumn{2}{c}{\textbf{$P$-even}} & \multicolumn{1}{c}{\textbf{$P$-odd}} 
\end{tabularx}
\begin{tabularx}{\linewidth}{c|>{\centering\arraybackslash}X|>{\centering\arraybackslash}X|}
    \cline{2-3}
     & \multicolumn{2}{c|}{\textbf{spin-$1/2$, irrep $(1\,,1)$}} \\
     \cline{2-3}
     \multirow{3}{*}{\textbf{$T$-even}} & $\left(\ytableaushort{123,4}\,,\ytableaushort{34}\right)+\text{dual}$ & $\left(\ytableaushort{123,4}\,,\ytableaushort{34}\right)-\text{dual}$\\
     & $\left(\ytableaushort{133,2}\,,\ytableaushort{33}\right)+\text{dual}$ & $\left(\ytableaushort{133,2}\,,\ytableaushort{33}\right)-\text{dual}$\\
     & $\left(\ytableaushort{144,2}\,,\ytableaushort{44}\right)+\text{dual}$ & $\left(\ytableaushort{144,2}\,,\ytableaushort{44}\right)-\text{dual}$\\
     \cline{2-3}
     \multirow{2}{*}{\textbf{$T$-odd}} & $\left(\ytableaushort{134,2}\,,\ytableaushort{34}\right)+\text{dual}$ & $\left(\ytableaushort{134,2}\,,\ytableaushort{34}\right)-\text{dual}$\\
     & $\left(\ytableaushort{124,4}\,,\ytableaushort{44}\right)+\text{dual}$ & $\left(\ytableaushort{124,4}\,,\ytableaushort{44}\right)-\text{dual}$\\
     \cline{2-3}
     \multicolumn{2}{c}{\textbf{$P$-even}} & \multicolumn{1}{c}{\textbf{$P$-odd}} 
\end{tabularx}
\end{center}
\caption{All the nonequivalent SSYTs of spin-1/2 fermion organized by their $P$ and $T$ eigenvalues. }
    \label{tab:SSYTs_fermion}
\end{table}

\subsection{Construction (3): Conversion to Tensor Representation}

What remaining to do is to interpret the obtained SSYTs to tensors. 
The interpretation is realized by the tranversion from the spinor representation to the tensor representation such as Eq.~\eqref{eq:transf}. 
Thus, every momentum can be projected as
\begin{align}
    P^\mu \rightarrow P_{a\dot{a}} = P^\mu\sigma_\mu{}_{a\dot{a}}\,, \\
    q^\mu \rightarrow q_{a\dot{a}} = q^\mu\sigma_\mu{}_{a\dot{a}}\,,
\end{align}
besides, the boson and fermion wave functions can also be projected,
\begin{align}
    \text{boson:} \quad& \varepsilon^{\mu_1\dots \mu_n} \rightarrow \varepsilon_{a_1\dots a_n}^{\dot{a}_1\dots \dot{a}_n} = \varepsilon^{\mu_1\dots\mu_n}\left(\sigma_{\mu_1}{}_{a_1}^{\dot{a}_1} \dots \sigma_{\mu_n}{}_{a_n}^{\dot{a}_n}\right) \label{eq:boson_projection}\,,\\
    \text{fermion:} \quad& u^{\mu_1\dots\mu_n} = \xi^{\mu_1\dots\mu_n}_a \oplus \eta^{\mu_1\dots\mu_n}{}^{\dot{a}} \notag \\
    &\rightarrow \xi_{a_1\dots a_n a}^{\dot{a}_1\dots \dot{a}_n} \oplus \eta_{a_1\dots a_n}^{\dot{a}_1\dots \dot{a}_n\dot{a}} = \xi^{\mu_1\dots\mu_n}_a \oplus \eta^{\mu_1\dots\mu_n}{}^{\dot{a}}\left(\sigma_{\mu_1}{}_{a_1}^{\dot{a}_1} \dots \sigma_{\mu_n}{}_{a_n}^{\dot{a}_n}\right)\,,\label{eq:fermion_projection}
\end{align}
where $\sigma_\mu{}_a^{\dot{a}} = \sigma_\mu{}_{a\dot{b}}\epsilon^{\dot{b}\dot{a}}$ and the symmetrization of the vector indices are understood.
With these projections, the interpretations are technically equivalent to the evaluations of some sigma matrice chains look like $\text{tr}(\sigma^\mu\overline{\sigma}^\nu\dots)$.
Next, we evaluate these traces of the bosons and fermions, respectively, and work out all the irreducible tensors of the SSYTs obtained previously. Although these calculations are direct, these examples will offer us some experience that will be useful when considering the FFs of higher-spin particles subsequently.

\subsubsection*{Boson Case}
\begin{description}
    \item[Irrep $(0\,,0)$:] The SSYT without momentum can be interpreted as 
    \begin{equation}
        \left(\ytableaushort{1,2}\,,\ytableaushort{1,2}\right) = \varepsilon_1^\mu \varepsilon_2^\dagger {}^\nu \sigma_\mu{}_{a\dot{a}}\sigma_\nu{}_{b\dot{b}} \times \epsilon^{ab}\epsilon^{\dot{a}\dot{b}} = \varepsilon_1^\mu \varepsilon_2^\dagger {}^\nu \text{tr}(\sigma^\mu\overline{\sigma}^\mu) = 2\varepsilon_1^\mu \varepsilon_2^\dagger {}^\nu g_{\mu\nu} = 2(\varepsilon_1\cdot \varepsilon_2^\dagger )\,,
    \end{equation}
    thus the corresponding tensor is $(\varepsilon_1\cdot \varepsilon_2^\dagger )$, whose permutation symmetry, $P$ and time reversal eigenvalues are the same as our respect. The second SSYT's interpretation is similar,
    \begin{align}
        \left(\ytableaushort{12,44}\,,\ytableaushort{12,44}\right) &= \varepsilon_1^\mu \varepsilon_2^\dagger {}^\nu q^\rho q^\kappa \sigma_\mu{}_{a\dot{a}}\sigma_\nu{}_{b\dot{b}}\sigma_\rho{}_{c\dot{c}}\sigma_\kappa{}_{d\dot{d}} \times \epsilon^{ac}\epsilon^{\dot{a}\dot{c}}\epsilon^{bd}\epsilon^{\dot{b}\dot{d}} \notag \\
        &= \varepsilon_1^\mu \varepsilon_2^\dagger {}^\nu q^\rho q^\kappa \text{tr}(\sigma^\mu\overline{\sigma}^\rho)\text{tr}(\sigma^\nu\overline{\sigma}^\kappa) = 4(\varepsilon_2^\dagger \cdot q)(\varepsilon_1\cdot q)\,,
    \end{align}
    thus the corresponding tensor is $(\varepsilon_2^\dagger \cdot q)(\varepsilon_1\cdot q)$, where although there are 4 sigma matrice, the SSYT determines their trace is actually a product of two traces of two sigma matrices. For the last two SSYTs, the situation is more complicated,
    \begin{align}
        \left(\ytableaushort{12,34}\,,\ytableaushort{13,24}\right) &= \varepsilon_1^\mu \varepsilon_2^\dagger {}^\nu P^\rho q^\kappa \sigma_\mu{}_{a\dot{a}}\sigma_\nu{}_{b\dot{b}}\sigma_\rho{}_{c\dot{c}}\sigma_\kappa{}_{d\dot{d}} \times \epsilon^{ac}\epsilon^{\dot{c}\dot{d}}\epsilon^{db}\epsilon^{\dot{b}\dot{a}} \notag \\
        &= \varepsilon_1^\mu \varepsilon_2^\dagger {}^\nu P^\rho q^\kappa \text{tr}(\sigma_\mu\overline{\sigma}_\nu\sigma_\kappa \overline{\sigma}_\rho) \notag \\
        &= -2(\varepsilon_1\cdot P)(\varepsilon_2^\dagger \cdot q) + 2(\varepsilon_1\cdot q)(\varepsilon_2^\dagger \cdot P) + 2i \epsilon_{\mu\nu\rho\kappa} \varepsilon_1^\mu\varepsilon_2^\dagger {}^\nu P^\rho q^\kappa \notag \\
        &= -4(\varepsilon_2^\dagger \cdot q)(\varepsilon_1\cdot q) + 2i \epsilon_{\mu\nu\rho\kappa} \varepsilon_1^\mu\varepsilon_2^\dagger {}^\nu P^\rho q^\kappa\,,\\
        \left(\ytableaushort{13,24}\,,\ytableaushort{12,34}\right) &= \varepsilon_1^\mu \varepsilon_2^\dagger {}^\nu P^\rho q^\kappa \sigma_\mu{}_{a\dot{a}}\sigma_\nu{}_{b\dot{b}}\sigma_\rho{}_{c\dot{c}}\sigma_\kappa{}_{d\dot{d}} \times \epsilon^{\dot{a}\dot{c}}\epsilon^{cd}\epsilon^{\dot{d}\dot{b}}\epsilon^{ba} \notag \\
        &= \varepsilon_1^\mu \varepsilon_2^\dagger {}^\nu P^\rho q^\kappa \text{tr}(\overline{\sigma}_\mu\sigma_\nu\overline{\sigma}_\kappa \sigma_\rho) \notag \\
        &= -2(\varepsilon_1\cdot P)(\varepsilon_2^\dagger \cdot q) + 2(\varepsilon_1\cdot q)(\varepsilon_2^\dagger \cdot P) - 2i \epsilon_{\mu\nu\rho\kappa} \varepsilon_1^\mu\varepsilon_2^\dagger {}^\nu P^\rho q^\kappa \notag \\
        &= -4(\varepsilon_2^\dagger \cdot q)(\varepsilon_1\cdot q) - 2i \epsilon_{\mu\nu\rho\kappa} \varepsilon_1^\mu\varepsilon_2^\dagger {}^\nu P^\rho q^\kappa\,.
    \end{align}
    As expected, the sum of two SSYTs generates no independent tensor, but their minus, $\epsilon_{\mu\nu\rho\kappa} \varepsilon_1^\mu\varepsilon_2^\dagger {}^\nu P^\rho q^\kappa$, is independent, whose $P$ and time reversal eigenvalues together with the permutation symmetry are the same its corresponding SSYT. 
    \item[Irrep $(\frac{1}{2}\,,\frac{1}{2})$:] When going beyond the trivial irrep, there are free spinor indices, which leads to two consequences. Firstly, the free spinor indices should be projected to vector indices, secondly, the permutation symmetry of the wave functions should be imposed manually. 
    the SSYTs with one momentum, the traces contain 4 sigma matrices, for example
    \begin{align}
        \left(\ytableaushort{12,3}\,,\ytableaushort{12,3}\right) &= (\varepsilon_1^\nu \varepsilon_2^\dagger {}^\rho P^\lambda + \varepsilon_1^\rho \varepsilon_2^\dagger {}^\nu P^\lambda) \text{tr}(\sigma_\nu\overline{\sigma}_\lambda)\text{tr}(\sigma_\rho\overline{\sigma}_\mu) \notag \\
        &\rightarrow (\varepsilon_1\cdot P)\varepsilon_2^\dagger {}^\mu + (\varepsilon_2^\dagger \cdot P)\varepsilon_1^\mu \notag \\
        &= (\varepsilon_1\cdot q)\varepsilon_2^\dagger {}^\mu - (\varepsilon_2^\dagger \cdot q)\varepsilon_1^\mu\,,
    \end{align}
    where we have imposed the symmetry of the wave function during the interpretation, and the $\overline{\sigma}_\mu$ in the trace is used to project the free spinor indices to vector indices. Other self-dual SSYTs' interpretations are the same. As for the non-self-dual ones, they also involve 4 sigma matrices, but the trace can not be factorized, for example, 
    \begin{align}
        & \left(\ytableaushort{12,4}\,,\ytableaushort{14,2}\right) - \text{dual} = (\varepsilon_1^\nu \varepsilon_2^\dagger {}^\rho q^\lambda - \varepsilon_1^\rho \varepsilon_2^\dagger {}^\nu q^\lambda)\left[\text{tr}(\sigma_\nu\overline{\sigma}_\lambda\sigma^\mu \overline{\sigma}_\rho) - \text{tr}(\overline{\sigma}_\nu\sigma_\lambda\overline{\sigma}^\mu \sigma_\rho)\right] \notag \\
        &\rightarrow \epsilon^{\mu\nu\rho\lambda} \varepsilon_1{}_\nu \varepsilon_2^\dagger {}_\rho q_\lambda\,.
    \end{align}
    Generally speaking, the interpretations of the other SSYTs with 3 momenta involve 6 sigma matrices, but fortunately, their traces can be factorized, which is transparent via SSYTs, for example,
    \begin{align}
        \left(\ytableaushort{123,44}\,,\ytableaushort{123,44}\right) &\supset \left(\ytableaushort{12,44}\,,\ytableaushort{12,44}\right) \times \left(\ytableaushort{4}\,,\ytableaushort{4}\right)\,, \\
        \left(\ytableaushort{124,34}\,,\ytableaushort{123,44}\right) &\supset \left(\ytableaushort{2,4}\,,\ytableaushort{2,4}\right) \times \left(\ytableaushort{14,3}\,,\ytableaushort{13,4}\right)\,, \\
        \left(\ytableaushort{123,34}\,,\ytableaushort{133,24}\right) &\supset \left(\ytableaushort{12,34}\,,\ytableaushort{13,24}\right) \times \left(\ytableaushort{3}\,,\ytableaushort{4}\right)\,, 
    \end{align}
    where the interpretations of all the SSYTs on the right-hand are known. 
    \item[Irrep $(1\,,0)\oplus (0\,,1)$:] The most simple SSYT of this irrep is the one without momentum, whose interpretation needs $\sigma^{\mu\nu}\,,\overline{\sigma}^{\mu\nu}$ to project the two symmetric left-hand and right-handed spinor indices to antisymmetric vector indices,
    \begin{align}
        \left(\ytableaushort{12}\,,\ytableaushort{1,2}\right) &= (\varepsilon_1^\rho \varepsilon^\dagger_2  {}^\kappa - \varepsilon_1^\kappa \varepsilon^\dagger_2  {}^\rho) \text{tr}(\sigma_\rho\overline{\sigma}_\kappa\sigma^{\mu\nu}) = -2i\varepsilon_1^{[\mu}\varepsilon^\dagger_2  {}^{\nu]} -2\epsilon^{\mu\nu\rho\kappa}\varepsilon_1{}_\rho\varepsilon^\dagger_2  {}_\kappa\,, \\
        \left(\ytableaushort{1,2}\,,\ytableaushort{12}\right) &= (\varepsilon_1^\rho \varepsilon^\dagger_2  {}^\kappa - \varepsilon_1^\kappa \varepsilon^\dagger_2  {}^\rho) \text{tr}(\overline{\sigma}_\rho\sigma_\kappa\overline{\sigma}^{\mu\nu}) = -2i\varepsilon_1^{[\mu}\varepsilon^\dagger_2  {}^{\nu]} +2\epsilon^{\mu\nu\rho\kappa}\varepsilon_1{}_\rho\varepsilon^\dagger_2  {}_\kappa\,,
    \end{align}
    thus their corresponding tensors with specific $P$ and $T$ eigenvalues are 
    \begin{equation}
        \left\{\begin{array}{l}
        \left(\ytableaushort{12}\,,\ytableaushort{1,2}\right) + \text{dual} \\
        \left(\ytableaushort{12}\,,\ytableaushort{1,2}\right) - \text{dual}
        \end{array}\right. \longrightarrow \left\{\begin{array}{l}
        \varepsilon_1^{[\mu}\varepsilon^\dagger_2  {}^{\nu]} \\
        \epsilon^{\mu\nu\rho\kappa}\varepsilon_1{}_\rho\varepsilon^\dagger_2  {}_\kappa \,.
        \end{array}\right.
    \end{equation}
    Other SSYTs with more momenta can also be factorized just as the $(\frac{1}{2}\,,\frac{1}{2})$ case, except the SSYTs
    \begin{equation}
        \left(\ytableaushort{123,4}\,,\ytableaushort{12,34}\right) \pm \text{dual}\,,
    \end{equation}
    which of cause can be interpreted via direct evaluation of traces, but here we would like to work it out by a technique of Fock condition,
    \begin{equation}
        \ytableaushort{12,43} = \ytableaushort{12,34} - \ytableaushort{13,24}\,,
    \end{equation}
    So the original SSYT can be expressed as 
    \begin{equation}
        \left(\ytableaushort{123,4}\,,\ytableaushort{12,34}\right) = \left(\ytableaushort{123,4}\,,\ytableaushort{12,43}\right) + \left(\ytableaushort{123,4}\,,\ytableaushort{13,24}\right)\,,
    \end{equation}
    where the last one is not independent because of the relation in Eq.~\eqref{eq:relation_boson_new_2}, thus we can equivalently evaluate the non-SSYTs
    \begin{equation}
        \left(\ytableaushort{123,4}\,,\ytableaushort{12,43}\right) \pm \text{dual}\,,
    \end{equation}
    and they can be factorized.
    \item[Irrep $(1\,,1)$:] The most simple SSYT can be interpreted as 
    \begin{align}
        \left(\ytableaushort{12}\,,\ytableaushort{12}\right) &= (\varepsilon_1^\rho \varepsilon^\dagger_2  {}^\kappa + \varepsilon_1^\kappa \varepsilon^\dagger_2  {}^\rho) \text{tr}(\sigma_\rho\overline{\sigma}^\mu) \text{tr}(\sigma_\kappa\overline{\sigma}^\nu) \rightarrow \varepsilon_1{}^{(\mu} \varepsilon^\dagger_2  {}^{\nu)}\,.
    \end{align}
    Equipped with the factorization technique, all the other SSYTs can be interpreted by the previous results, except one
    \begin{equation}
        \left(\ytableaushort{1244,33}\,,\ytableaushort{1233,44}\right) - \text{dual}\,.
    \end{equation}
    However, its trace can also be factorized, though it is not transparent by its SSYT,
    \begin{align}
        (\varepsilon_1^\rho\varepsilon_2^\dagger {}^\kappa P^\lambda P^\alpha q^\beta q^\delta + \varepsilon_1^\kappa\varepsilon_2^\dagger {}^\rho P^\lambda P^\alpha q^\beta q^\delta)\text{tr}(\sigma^\mu\overline{\sigma}_\beta \sigma_\rho\overline{\sigma}_\lambda)\text{tr}(\sigma^\nu\overline{\sigma}_\delta \sigma_\kappa\overline{\sigma}_\alpha) \,, 
    \end{align}
    so that the SSYT's interpretation is
    \begin{align}
        & \left(\ytableaushort{1244,33}\,,\ytableaushort{1233,44}\right) - \text{dual} \notag \\
        \rightarrow& (\varepsilon^{\mu\rho\lambda\beta}\varepsilon_1{}_\rho P_\lambda q_\beta)(\varepsilon_2^\dagger \cdot q) P^\nu + (\varepsilon^{\mu\rho\lambda\beta}\varepsilon_2^\dagger {}_\rho P_\lambda q_\beta)(\varepsilon_1\cdot q) P^\nu \notag \\
        +& (\varepsilon^{\mu\rho\lambda\beta}\varepsilon_1{}_\rho P_\lambda q_\beta)(\varepsilon_2^\dagger \cdot q) q^\nu - (\varepsilon^{\mu\rho\lambda\beta}\varepsilon_2^\dagger {}_\rho P_\lambda q_\beta)(\varepsilon_1\cdot q) q^\nu\,,
    \end{align}
    with expected $P$ and $T$ eigenvalues.
\end{description}
In summary, we list all the independent tensors of the SSYTs in Tab.~\ref{tab:tensor_boson_1} and Tab.~\ref{tab:tensor_boson_2}, where the tensors correspond to the SSYTs at the same position in Tab.~\ref{tab:SSYTs_boson_1} and Tab.~\ref{tab:SSYTs_boson_2}, respectively. We find that there are only 7 $P$-even and $T$-even tensors of $(1,1)$ representation, not the 8 ones presented in the previous paper~\cite{Cotogno:2019vjb}. This is the first non-trivial result generated by the systematic method here.

\begin{table}[]
\renewcommand\arraystretch{1.5}
\ytableausetup{smalltableaux}
    \centering
\begin{tabularx}{\linewidth}{c|>{\centering\arraybackslash}X|>{\centering\arraybackslash}X|}
    \cline{2-3}
     & \multicolumn{2}{c|}{\textbf{spin-$1$, irrep $(0\,,0)$}} \\
     \cline{2-3}
     \textbf{$T$-even} & $\varepsilon_2^\dagger \cdot\varepsilon_1$, $(\varepsilon_2^\dagger \cdot q)(\varepsilon_1\cdot q)$ & $\epsilon^{\mu\nu\rho\kappa}\varepsilon_1{}_\mu\varepsilon_2^\dagger {}_\nu P_\rho q_\kappa$ \\
     \cline{2-3}
     \textbf{$T$-odd} & & \\
     \cline{2-3}
     \multicolumn{2}{c}{\textbf{$P$-even}} & \multicolumn{1}{c}{\textbf{$P$-odd}} 
\end{tabularx}
\begin{tabularx}{\linewidth}{c| >{\centering\arraybackslash}X|>{\centering\arraybackslash}X|}
    \cline{2-3}
     & \multicolumn{2}{c|}{\textbf{spin-$1$, irrep $(\frac{1}{2}\,,\frac{1}{2})$}} \\
     \cline{2-3}
      \multirow{3}{*}{\textbf{$T$-even}} & $(\varepsilon_1\cdot q)\varepsilon_2^\dagger {}^\mu - (\varepsilon_2^\dagger \cdot q)\varepsilon_1^\mu$ & $\epsilon^{\mu\nu\rho\kappa}\varepsilon_1{}_\nu \varepsilon_2^\dagger {}_\rho q_\kappa$ \\
      & $(\varepsilon_2^\dagger \cdot\varepsilon_1)P^\mu$ & $(\varepsilon_2^\dagger \cdot q)\epsilon^{\mu\nu\rho\kappa}\varepsilon_1{}_\nu P_\rho q_\kappa + (\varepsilon_1\cdot q)\epsilon^{\mu\nu\rho\kappa}\varepsilon_2^\dagger {}_\nu P_\rho q_\kappa$ \\
      & $(\varepsilon_1\cdot q)(\varepsilon_2^\dagger \cdot q) P^\mu$ & $(\varepsilon^{\nu\rho\kappa\lambda}\varepsilon_1{}_\nu \varepsilon^\dagger_2  {}_\rho P_\kappa q_\lambda) P^\mu$ \\
     \cline{2-3}
      \multirow{3}{*}{\textbf{$T$-odd}}& $(\varepsilon_1\cdot q)\varepsilon_2^\dagger {}^\mu + (\varepsilon_2^\dagger \cdot q)\varepsilon_1^\mu$ & 
$\epsilon^{\mu\nu\rho\kappa}\varepsilon_1{}_\nu \varepsilon_2^\dagger {}_\rho P_\kappa$ \\
     & $(\varepsilon_2^\dagger \cdot\varepsilon_1)q^\mu$ & $(\varepsilon_2^\dagger \cdot q)\epsilon^{\mu\nu\rho\kappa}\varepsilon_1{}_\nu P_\rho q_\kappa - (\varepsilon_1\cdot q)\epsilon^{\mu\nu\rho\kappa}\varepsilon_2^\dagger {}_\nu P_\rho q_\kappa$ \\
     & $(\varepsilon_1\cdot q)(\varepsilon_2^\dagger \cdot q) q^\mu$ & $(\varepsilon^{\nu\rho\kappa\lambda}\varepsilon_1{}_\nu \varepsilon^\dagger_2  {}_\rho P_\kappa q_\lambda) q^\mu$ \\
     \cline{2-3}
     \multicolumn{2}{c}{\textbf{$P$-even}} & \multicolumn{1}{c}{\textbf{$P$-odd}} 
\end{tabularx}
\begin{tabularx}{\linewidth}{c| >{\centering\arraybackslash}X|>{\centering\arraybackslash}X|}
    \cline{2-3}
     & \multicolumn{2}{c|}{\textbf{spin-$1$, irrep $(1\,,0)\oplus(0\,,1)$}} \\
     \cline{2-3}
      \multirow{2}{*}{\textbf{$T$-even}} & $(\varepsilon_1\cdot q)\varepsilon^\dagger_2  {}^{[\mu}P^{\nu]} - (\varepsilon_2^\dagger \cdot q)\varepsilon_1{}^{[\mu}P^{\nu]}$ & $\epsilon^{\mu\nu\rho\kappa}[(\varepsilon_1\cdot q)\varepsilon^\dagger_2  {}_\rho P_\kappa - (\varepsilon_2^\dagger \cdot q)\varepsilon_1{}_\rho P_\kappa]$ \\
      & $(\varepsilon_1\cdot q)\varepsilon^\dagger_2  {}^{[\mu}q^{\nu]} + (\varepsilon_2^\dagger \cdot q)\varepsilon_1{}^{[\mu}q^{\nu]}$ & $\epsilon^{\mu\nu\rho\kappa}[(\varepsilon_1\cdot q)\varepsilon^\dagger_2  {}_\rho q_\kappa + (\varepsilon_2^\dagger \cdot q)\varepsilon_1{}_\rho q_\kappa]$ \\
     \cline{2-3}
     \multirow{5}{*}{\textbf{$T$-odd}}& $\varepsilon_1^{[\mu}\varepsilon_2^\dagger {}^{\nu]}$ & $\epsilon^{\mu\nu\rho\kappa}\varepsilon_1{}_\rho \varepsilon_2^\dagger {}_\kappa$ \\
     & $(\varepsilon_1\cdot q)\varepsilon^\dagger_2  {}^{[\mu}q^{\nu]} - (\varepsilon_2^\dagger \cdot q)\varepsilon_1{}^{[\mu}q^{\nu]}$ & $\epsilon^{\mu\nu\rho\kappa}[(\varepsilon_1\cdot q)\varepsilon^\dagger_2  {}_\rho q_\kappa - (\varepsilon_2^\dagger \cdot q)\varepsilon_1{}_\rho q_\kappa]$ \\
     & $(\varepsilon_1\cdot q)\varepsilon^\dagger_2  {}^{[\mu}P^{\nu]} + (\varepsilon_2^\dagger \cdot q)\varepsilon_1{}^{[\mu}P^{\nu]}$ & $\epsilon^{\mu\nu\rho\kappa}[(\varepsilon_1\cdot q)\varepsilon^\dagger_2  {}_\rho P_\kappa + (\varepsilon_2^\dagger \cdot q)\varepsilon_1{}_\rho P_\kappa]$ \\
     & $(\varepsilon_1\cdot\varepsilon_2^\dagger )P^{[\mu}q^{\nu]}$ & $\epsilon^{\mu\nu\rho\kappa}(\varepsilon_1\cdot\varepsilon_2^\dagger )P_\rho q_\kappa$ \\
     & $(\varepsilon_1\cdot q)(\varepsilon_2^\dagger \cdot q)P^{[\mu}q^{\nu]}$ & $\epsilon^{\mu\nu\rho\kappa}(\varepsilon_1\cdot q)(\varepsilon_2^\dagger \cdot q)P_\rho q_\kappa$ \\
     & 
     & $(\epsilon^{\rho\kappa\lambda\alpha}\varepsilon_1{}_\rho \varepsilon^\dagger_2  {}_\kappa P_\lambda q_\alpha)P^{[\mu}q^{\nu]}$ \\
     \cline{2-3}
     \multicolumn{2}{c}{\textbf{$P$-even}} & \multicolumn{1}{c}{\textbf{$P$-odd}} 
\end{tabularx}
    \caption{All the independent tensors of spin-1 boson organized by their $P$ and $T$ eigenvalues. This tabular contains the irreps $(0\,,0)\,,(\frac{1}{2}\,,\frac{1}{2})$, and $(1\,,0)\oplus(0\,,1)$, while the irrep $(1\,,1)$ is presented in Tab.~\ref{tab:tensor_boson_2}. These tensors correspond to the SSYTs at the same positions in Tab.~\ref{tab:SSYTs_boson_1}.}
    \label{tab:tensor_boson_1}
\end{table}

\begin{table}
\begin{center}
\renewcommand\arraystretch{1.5}
\ytableausetup{smalltableaux}
    \begin{tabularx}{\linewidth}{c| >{\centering\arraybackslash}X|>{\centering\arraybackslash}X|}
    \cline{2-3}
     & \multicolumn{2}{c|}{\textbf{spin-$1$, irrep $(1\,,1)$}} \\
     \cline{2-3}
     \multirow{7}{*}{\textbf{$T$-even}} & $\varepsilon_1{}^{(\mu}\varepsilon_2^\dagger {}^{\nu)}$ & $\epsilon^{(\mu\rho\lambda\kappa}\varepsilon_1{}_\rho \varepsilon^\dagger_2  {}_\lambda q_\kappa P^{\nu)}$ \\
     & $(\varepsilon_1\cdot q)\varepsilon_2^\dagger {}^{(\mu} q^{\nu)} + (\varepsilon_2^\dagger \cdot q)\varepsilon_1{}^{(\mu} q^{\nu)}$ & $\epsilon^{(\mu\rho\lambda\kappa}\varepsilon_1{}_\rho \varepsilon^\dagger_2  {}_\lambda P_\kappa q^{\nu)}$ \\
     & $(\varepsilon_1\cdot q)\varepsilon_2^\dagger {}^{(\mu} P^{\nu)} - (\varepsilon_2^\dagger \cdot q)\varepsilon_1{}^{(\mu} P^{\nu)}$ & $(\varepsilon_1\cdot q)\epsilon^{(\mu\rho\kappa\lambda}\varepsilon_2^\dagger {}_\rho P_\kappa q_\lambda P^{\nu)} + (\varepsilon_2^\dagger \cdot q)\epsilon^{(\mu\rho\kappa\lambda}\varepsilon_1{}_\rho P_\kappa q_\lambda P^{\nu)}$ \\
     & $(\varepsilon_1\cdot \varepsilon_2^\dagger )P^{(\mu}P^{\nu)}$ & $(\varepsilon_1\cdot q)\epsilon^{(\mu\rho\kappa\lambda}\varepsilon_2^\dagger {}_\rho P_\kappa q_\lambda q^{\nu)} - (\varepsilon_2^\dagger \cdot q)\epsilon^{(\mu\rho\kappa\lambda}\varepsilon_1{}_\rho P_\kappa q_\lambda q^{\nu)}$ \\
     & $(\varepsilon_1\cdot \varepsilon_2^\dagger )q^{(\mu}q^{\nu)}$ & $(\epsilon^{\rho\kappa\lambda\alpha}\varepsilon_1{}_\rho \varepsilon^\dagger_2  {}_\kappa P_\lambda q_\alpha)P^{(\mu}P^{\nu)}$\\
     & $(\varepsilon_1\cdot q)(\varepsilon_2^\dagger \cdot q)P^{(\mu}P^{\nu)}$ & $(\epsilon^{\rho\kappa\lambda\alpha}\varepsilon_1{}_\rho \varepsilon^\dagger_2  {}_\kappa P_\lambda q_\alpha)q^{(\mu}q^{\nu)}$\\
     & $(\varepsilon_1\cdot q)(\varepsilon_2^\dagger \cdot q)q^{(\mu}q^{\nu)}$ & \\
     \cline{2-3}
     \multirow{6}{*}{\textbf{$T$-odd}}& $(\varepsilon_1\cdot q)\varepsilon_2^\dagger {}^{(\mu} P^{\nu)} + (\varepsilon_2^\dagger \cdot q)\varepsilon_1{}^{(\mu} P^{\nu)}$ & $\epsilon^{(\mu\rho\kappa\lambda}\varepsilon_1{}_\rho P_\kappa q_\lambda \varepsilon_2^\dagger {}^{\nu)} + \epsilon^{(\mu\rho\kappa\lambda}\varepsilon_2^\dagger {}_\rho P_\kappa q_\lambda \varepsilon_1{}^{\nu)}$\\
     & $(\varepsilon_1\cdot q)\varepsilon_2^\dagger {}^{(\mu} q^{\nu)} - (\varepsilon_2^\dagger \cdot q)\varepsilon_1{}^{(\mu} q^{\nu)}$ & $\epsilon^{(\mu\rho\lambda\kappa}\varepsilon_1{}_\rho \varepsilon^\dagger_2  {}_\lambda P_\kappa P^{\nu)}$ \\
     & $(\varepsilon_1\cdot \varepsilon_2^\dagger )P^{(\mu}q^{\nu)}$ & $\epsilon^{(\mu\rho\lambda\kappa}\varepsilon_1{}_\rho \varepsilon^\dagger_2  {}_\lambda q_\kappa q^{\nu)}$ \\
     & $(\varepsilon_1\cdot q)(\varepsilon_2^\dagger \cdot q)P^{(\mu}q^{\nu)}$ & $(\varepsilon_1\cdot q)\epsilon^{(\mu\rho\kappa\lambda}\varepsilon_2^\dagger {}_\rho P_\kappa q_\lambda P^{\nu)} - (\varepsilon_2^\dagger \cdot q)\epsilon^{(\mu\rho\kappa\lambda}\varepsilon_1{}_\rho P_\kappa q_\lambda P^{\nu)}$\\
     & & $(\varepsilon_1\cdot q)\epsilon^{(\mu\rho\kappa\lambda}\varepsilon_2^\dagger {}_\rho P_\kappa q_\lambda q^{\nu)} + (\varepsilon_2^\dagger \cdot q)\epsilon^{(\mu\rho\kappa\lambda}\varepsilon_1{}_\rho P_\kappa q_\lambda q^{\nu)}$ \\
     & & $(\epsilon^{\rho\kappa\lambda\alpha}\varepsilon_1{}_\rho \varepsilon^\dagger_2  {}_\kappa P_\lambda q_\alpha)P^{(\mu}q^{\nu)}$ \\
     \cline{2-3}
     \multicolumn{2}{c}{\textbf{$P$-even}} & \multicolumn{1}{c}{\textbf{$P$-odd}} 
\end{tabularx}
\end{center}
\caption{All the independent tensors of spin-1 boson organized by their $P$ and $T$ eigenvalues. This tabular contains the irrep $(1\,,1)$, while the irreps $(0\,,0)\,,(\frac{1}{2}\,,\frac{1}{2})$, and $(1\,,0)\oplus(0\,,1)$ are presented in Tab.~\ref{tab:tensor_boson_1}. These tensors correspond to the SSYTs at the same positions in Tab.~\ref{tab:SSYTs_boson_2}.}
    \label{tab:tensor_boson_2}
\end{table}

\subsubsection*{Fermion Case}

The interpretations of the fermion case need no more techniques compared to the boson case, but it is different because the fermionic wave functions are direct products of two irreps. For the spin-1/2 particles, the wave functions can be expressed as direct products of Weyl spinors that
\begin{align}
    u_1 &= \xi \oplus \eta \rightarrow \left(\ytableaushort{1}\,,\ytableaushort{1}\right)\,, \\
    \overline{u}_2 &= \eta^\dagger \oplus \xi^\dagger \rightarrow \left(\ytableaushort{2}\,,\ytableaushort{2}\right)\,,
\end{align}
where the Weyl Spinors are projected via the projectors $P_{L/R} = (1+\gamma^5)/2$ that
\begin{equation}
    \xi = P_L u_1 \,,\quad \eta = P_L u_1\,,\quad \eta^\dagger = \overline{u}_2 P_L\,,\quad \xi^\dagger =\overline{u}_2 P_R\,.
\end{equation}
Next, we present some typical examples in terms of their irreps.

\begin{description}
    \item[Irrep $(0\,,0)$:] The two SSYTs can be interpreted directly,
    \begin{equation}
        \left(\ytableaushort{1,2}\,,1\right) = \eta^\dagger \xi \,,\quad \left(1\,,\ytableaushort{1,2}\right) = \xi^\dagger \eta \,,
    \end{equation}
    and their combinations are 
    \begin{align}
        \left(\ytableaushort{1,2}\,,1\right) + \text{dual} &= \overline{u}_2 u_1 \,, \\
        \left(\ytableaushort{1,2}\,,1\right) - \text{dual} &= \overline{u}_2\gamma^5 u_1 \,,
    \end{align}
    which are $P$-even and $P$-odd respectively.
    \item[Irrep $(\frac{1}{2}\,,\frac{1}{2})$:] In addition to the SSYTs that can be factorized, such as
    \begin{equation}
        \left(\ytableaushort{14,2}\,,\ytableaushort{4}\right) \supset \left(\ytableaushort{1,2}\,,1\right) \times \left(\ytableaushort{4}\,,\ytableaushort{4}\right) = q^\mu (\eta^\dagger \xi)\,,
    \end{equation}
    the remaining ones can be interpreted as
    \begin{align}
        \left(\ytableaushort{12,4}\,,\ytableaushort{4}\right) &= q^\nu(\eta^\dagger\sigma_\nu\overline{\sigma}^\mu \xi) = 2q^\mu(\eta^\dagger \xi) + \eta^\dagger\sigma^{\nu\mu}q_\nu\xi \rightarrow \eta^\dagger \sigma^{\mu\nu}q_\nu \xi\,, \\
        \left(\ytableaushort{4}\,,\ytableaushort{12,4}\right) &= q^\nu(\xi^\dagger\overline{\sigma}_\nu\sigma^\mu \eta) = 2q^\mu(\xi^\dagger \eta) + \xi^\dagger\overline{\sigma}^{\nu\mu}q_\nu\eta \rightarrow \xi^\dagger \overline{\sigma}^{\mu\nu}q_\nu \eta\,,
    \end{align}
    and their combinations are
    \begin{align}
        \left(\ytableaushort{12,4}\,,\ytableaushort{4}\right) + \text{dual} &\rightarrow \overline{u}_2 \sigma^{\mu\nu}q_\nu u_1\,, \\
        \left(\ytableaushort{12,4}\,,\ytableaushort{4}\right) - \text{dual} &\rightarrow \epsilon^{\mu\nu\rho\kappa}\overline{u}_2 \sigma_{\nu\rho}q_\kappa u_1\,.
    \end{align}
    \item[Irrep $(1\,,0)\oplus (0\,,1)$:] The interpretation of the SSYT with no momentum is simple
    \begin{equation}
        \left(\ytableaushort{12}\,,1\right) = \eta^\dagger\sigma^{\mu\nu}\xi\,,\quad \left(1\,,\ytableaushort{12}\right) = \xi^\dagger\sigma^{\mu\nu}\eta\,,
    \end{equation}
    with combinations 
    \begin{align}
        \left(\ytableaushort{12}\,,1\right) + \text{dual} &= \overline{u}_2\sigma^{\mu\nu} u_1\,, \\
        \left(\ytableaushort{12}\,,1\right) - \text{dual} &= \epsilon^{\mu\nu\rho\lambda}\overline{u}_2\sigma_{\rho\lambda} u_1\,.
    \end{align}
    Among other SSYTs, there are other SSYTs that can be factorized, 
    \begin{align}
        \left(\ytableaushort{134,2}\,,\ytableaushort{3,4}\right) &\supset \left(\ytableaushort{1,2}\,,1\right)\times \left(\ytableaushort{34}\,,\ytableaushort{3,4}\right) \notag \\
        & = \eta^\dagger \xi (2P^{[\mu}q^{\nu]} + 2i\epsilon^{\mu\nu\rho\kappa}P_\rho q_\kappa) \,,\\
        \left(\ytableaushort{34}\,,\ytableaushort{13,24}\right) &\supset \left(1\,,\ytableaushort{1,2}\right)\times \left(\ytableaushort{34}\,,\ytableaushort{3,4}\right) \notag \\
        & = \xi^\dagger \eta (2P^{[\mu}q^{\nu]} + 2i\epsilon^{\mu\nu\rho\kappa}P_\rho q_\kappa) \,,
    \end{align}
    whose combinations with their dual ones are 
    \begin{align}
        \left(\ytableaushort{134,2}\,,\ytableaushort{3,4}\right) + \text{dual} &\rightarrow P^{[\mu}q^{\nu]}\overline{u}_2u_1 + i\epsilon^{\mu\nu\rho\lambda}P_\rho q_\lambda\overline{u}_2 \gamma^5 u_1 \,, \\
        \left(\ytableaushort{134,2}\,,\ytableaushort{3,4}\right) - \text{dual} &\rightarrow P^{[\mu}q^{\nu]}\overline{u}_2\gamma^5u_1 + i\epsilon^{\mu\nu\rho\lambda}P_\rho q_\lambda\overline{u}_2 u_1 \,, \\
        \left(\ytableaushort{34}\,,\ytableaushort{13,24}\right) + \text{dual} &\rightarrow P^{[\mu}q^{\nu]}\overline{u}_2u_1 - i\epsilon^{\mu\nu\rho\lambda}P_\rho q_\lambda\overline{u}_2 \gamma^5 u_1 \,, \\
        \left(\ytableaushort{34}\,,\ytableaushort{13,24}\right) + \text{dual} &\rightarrow P^{[\mu}q^{\nu]}\overline{u}_2\gamma^5u_1 - i\epsilon^{\mu\nu\rho\lambda}P_\rho q_\lambda\overline{u}_2 u_1 \,.
    \end{align}
    There are also SSYTs that can not be factorized, their evaluations are 
    \begin{align}
        \left(\ytableaushort{123,4}\,,\ytableaushort{3,4}\right) &= P^\rho q^\lambda \eta^\dagger \sigma^{\mu\nu}\overline{\sigma}_\rho\sigma_\lambda \xi \rightarrow P^{[\mu}q^{\nu]}\eta^\dagger \xi + i \epsilon^{\mu\nu\rho\lambda}P_\rho q_\lambda\eta^\dagger \xi \notag \\
        &+ i P^{[\mu}\eta^\dagger\sigma^{\nu]\rho}q_\rho\xi - \epsilon^{\mu\nu\rho\kappa}P_\rho\eta^\dagger\sigma_{\kappa\lambda}q^\lambda \xi \notag \\
        &\rightarrow i P^{[\mu}\eta^\dagger\sigma^{\nu]\rho}q_\rho\xi - \epsilon^{\mu\nu\rho\kappa}P_\rho\eta^\dagger\sigma_{\kappa\lambda}q^\lambda \xi\,,
    \end{align}
    \begin{align}
        \left(\ytableaushort{44}\,,\ytableaushort{12,44}\right) = q^\rho q^\lambda \xi^\dagger \overline{\sigma}_\rho\sigma^{\mu\nu}\sigma_\lambda \eta \rightarrow iq^{[\mu}\xi^\dagger \sigma^{\nu]\rho}q_\rho\eta - \epsilon^{\mu\nu\rho\kappa}q_\rho \xi^\dagger \overline{\sigma}_{\kappa\lambda}q^\lambda \eta\,,
    \end{align}
    whose combinations are
    \begin{align}
        \left(\ytableaushort{123,4}\,,\ytableaushort{3,4}\right) + \text{dual} &\rightarrow P^{[\mu}(\overline{u}_2\sigma^{\nu\rho]}q_\rho u_1) \,,\\
        \left(\ytableaushort{123,4}\,,\ytableaushort{3,4}\right) - \text{dual} &\rightarrow \epsilon^{\mu\nu\rho\kappa}P_\rho(\overline{u}_2\sigma_{\kappa\lambda}q^\lambda u_1) \,,\\
        \left(\ytableaushort{44}\,,\ytableaushort{12,44}\right) + \text{dual} &\rightarrow q^{[\mu}(\overline{u}_2\sigma^{\nu\rho]}q_\rho u_1) \,,\\
        \left(\ytableaushort{44}\,,\ytableaushort{12,44}\right) - \text{dual} &\rightarrow \epsilon^{\mu\nu\rho\kappa}q_\rho(\overline{u}_2\sigma_{\kappa\lambda}q^\lambda u_1)\,.
    \end{align}
    \item[Irrep $(1\,,1)$:] All the SSYTs can be factorized, for example,
    \begin{equation}
        \left(\ytableaushort{133,2}\,,\ytableaushort{33}\right) \supset \left(\ytableaushort{1,2}\,,1\right) \times \left(\ytableaushort{33}\,,\ytableaushort{33}\right) \rightarrow P^{(\mu}P^{\nu)}\eta^\dagger\xi\,,
    \end{equation}
    \begin{align}
        \left(\ytableaushort{123,4}\,,\ytableaushort{34}\right) \supset \left(\ytableaushort{12,4}\,,\ytableaushort{4}\right) \times \left(\ytableaushort{3}\,,\ytableaushort{3}\right) \rightarrow P^{(\mu}\eta^\dagger \sigma^{\nu)\rho}q_\rho \xi\,,
    \end{align}
    whose combinations with specific $P$ and $T$ eigenvalues can be obtained similarly.
\end{description}
In summary, we list all independent tensors in Tab.~\ref{tab:tensor_fermion}, corresponding to the SSYTs at the same positions as in Tab.~\ref{tab:SSYTs_fermion}. 
The non-trivial result implied by the construction above is that the number of the C- and $P$-even tensors of irrep $(1,1)$ for the spin-1 fields is 7, not 8 as previously implied. In particular, the explicit tensors are presented in Tab.~\ref{tab:tensor_boson_2}. We note that the contracted indices within the results are implicitly excluded from the symmetrization and antisymmetrization procedures indicated by parentheses $()$ and brackets $[\,]$.

\begin{table}
\begin{center}
\renewcommand\arraystretch{1.5}
\ytableausetup{smalltableaux}
    \begin{tabularx}{\linewidth}{c|>{\centering\arraybackslash}X|>{\centering\arraybackslash}X|}
     \multicolumn{3}{c}{\textbf{spin-$1/2$, irrep $(0\,,0)$}} \\
     \cline{2-3}
     \textbf{$T$-even} & $(\overline{u}_2u_1)$ & $(\overline{u}_2\gamma^5 u_1)$\\
     \cline{2-3}
     \textbf{$T$-odd} & & \\
     \cline{2-3}
     \multicolumn{2}{c}{\textbf{$P$-even}} & \multicolumn{1}{c}{\textbf{$P$-odd}} 
\end{tabularx}
\begin{tabularx}{\linewidth}{c|>{\centering\arraybackslash}X|>{\centering\arraybackslash}X|}
    \cline{2-3}
     & \multicolumn{2}{c|}{\textbf{spin-$1/2$, irrep $(\frac{1}{2}\,,\frac{1}{2})$}} \\
     \cline{2-3}
     \multirow{2}{*}{\textbf{$T$-even}} & $(\overline{u}_2\sigma^{\mu\nu}q_\nu u_1)$ & $\epsilon^{\mu\nu\rho\lambda}(\overline{u}_2\sigma_{\nu\rho}q_\lambda u_1)$\\
     & $P^\mu(\overline{u}_2u_1)$ & $P^\mu(\overline{u}_2\gamma^5 u_1)$\\
     \cline{2-3}
     \textbf{$T$-odd} & $q^\mu(\overline{u}_2u_1)$ & $q^\mu(\overline{u}_2\gamma^5 u_1)$\\
     \cline{2-3}
     \multicolumn{2}{c}{\textbf{$P$-even}} & \multicolumn{1}{c}{\textbf{$P$-odd}} 
\end{tabularx}
\begin{tabularx}{\linewidth}{c|>{\centering\arraybackslash}X|>{\centering\arraybackslash}X|}
    \cline{2-3}
     & \multicolumn{2}{c|}{\textbf{spin-$1/2$, irrep $(1\,,0)\oplus (0\,,1)$}} \\
     \cline{2-3}
     \textbf{$T$-even} & $P^{[\mu}(\overline{u}_2\sigma^{\nu]\rho}q_\rho u_1)$ & $\epsilon^{\mu\nu\rho\lambda}P_\rho(\overline{u}_2\sigma_{\lambda\kappa}q^\kappa u_1)$\\
     \cline{2-3}
     \multirow{4}{*}{\textbf{$T$-odd}} & $(\overline{u}_2\sigma^{\mu\nu}u_1)$ & $\epsilon^{\mu\nu\rho\lambda}(\overline{u}_2\sigma_{\rho\lambda}u_1)$\\
     & $P^{[\mu}q^{\nu]}(\overline{u}_2 u_1) + \epsilon^{\mu\nu\rho\lambda}P_\rho q_\lambda(\overline{u}_2\gamma^5 u_1)$ & $P^{[\mu}q^{\nu]}(\overline{u}_2\gamma^5 u_1) + \epsilon^{\mu\nu\rho\lambda}P_\rho q_\lambda(\overline{u}_2 u_1)$\\
     & $P^{[\mu}q^{\nu]}(\overline{u}_2 u_1) - \epsilon^{\mu\nu\rho\lambda}P_\rho q_\lambda(\overline{u}_2\gamma^5 u_1)$ & $P^{[\mu}q^{\nu]}(\overline{u}_2\gamma^5 u_1) - \epsilon^{\mu\nu\rho\lambda}P_\rho q_\lambda(\overline{u}_2 u_1)$\\
     & $q^{[\mu}(\overline{u}_2\sigma^{\nu]\rho}q_\rho u_1)$ & $\epsilon^{\mu\nu\rho\lambda}q_\rho(\overline{u}_2\sigma_{\lambda\kappa}q^\kappa u_1)$\\
     \cline{2-3}
     \multicolumn{2}{c}{\textbf{$P$-even}} & \multicolumn{1}{c}{\textbf{$P$-odd}} 
\end{tabularx}
\begin{tabularx}{\linewidth}{c|>{\centering\arraybackslash}X|>{\centering\arraybackslash}X|}
    \cline{2-3}
     & \multicolumn{2}{c|}{\textbf{spin-$1/2$, irrep $(1\,,1)$}} \\
     \cline{2-3}
     \multirow{3}{*}{\textbf{$T$-even}} & $P^{(\mu}(\overline{u}_2\sigma^{\nu)\rho} q_\rho u_1)$ & $P^{(\mu}\epsilon^{\nu)\rho\lambda\kappa}(\overline{u}_2\sigma_{\lambda\kappa}q_\rho u_1)$ \\
     & $P^{(\mu}P^{\nu)}(\overline{u}_2 u_1)$ & $P^{(\mu}P^{\nu)}(\overline{u}_2 \gamma^5 u_1)$\\
     & $q^{(\mu}q^{\nu)}(\overline{u}_2 u_1)$ & $q^{(\mu}q^{\nu)}(\overline{u}_2 \gamma^5 u_1)$\\
     \cline{2-3}
     \multirow{2}{*}{\textbf{$T$-odd}} & $P^{(\mu}q^{\nu)}(\overline{u}_2 u_1)$ & $P^{(\mu}q^{\nu)}(\overline{u}_2 \gamma^5 u_1)$\\
     & $q^{(\mu}(\overline{u}_2\sigma^{\nu)\rho} q_\rho u_1)$ & $q^{(\mu}\epsilon^{\nu)\rho\lambda\kappa}(\overline{u}_2\sigma_{\lambda\kappa}q_\rho u_1)$ \\
     \cline{2-3}
     \multicolumn{2}{c}{\textbf{$P$-even}} & \multicolumn{1}{c}{\textbf{$P$-odd}} 
\end{tabularx}
\end{center}
\caption{All the nonequivalent SSYTs of spin-1/2 fermion organized by their $P$ and $T$ eigenvalues. }
    \label{tab:tensor_fermion}
\end{table}

\begingroup
\raggedbottom
\allowdisplaybreaks
\setlength{\abovedisplayskip}{3pt}
\setlength{\belowdisplayskip}{6pt}
\setlength{\abovedisplayshortskip}{0pt}
\setlength{\belowdisplayshortskip}{4pt}
\setlength{\parskip}{0pt}

\section{Form Factors of Higher-Spin Particles}
\label{sec:highspin}
As implied before, the Young tensor method can be generalized to any spin.
In this section, we apply the technique to the FFs for spin-$3/2$ and spin-$2$ particles. 
Performing the rigorous procedure, we present the complete tensors classified by the P- and T-eigenvalues for the first time.
Given these irreducible tensors, any tensor of rank $n$, $T^{\mu\nu\dots}$, can be expressed as 
\begin{equation}
    T^{\mu\nu\dots} = \sum_{\mathbf{r},i} F_{\mathbf{r},i} \times T_{\mathbf{r},i}^{\mu\nu\dots}\,,
\end{equation}
where the sum runs over the irreps $\mathbf{r}$ and all the independent tensors $i$. The coefficients $F_{\mathbf{r},i}$ are the associated FFs.
For simplicity, we adopt the convention that contracted indices within this paragraph are implicitly excluded from the symmetrization and antisymmetrization procedures indicated by parentheses $(\,)$ and brackets $[\,]$ below.
\subsection{Generalized FFs of Spin-2 Boson}

For spin-2 bosons, the matrix elements are parameterized by incorporating the completely symmetric and traceless polarization tensors $\varepsilon_{\mu\nu}$ as the specific wave functions. The complete sets of the tensors, classified by their irreps and P- and T-eigenvalues, are given as follows:

\vspace{1em}
\noindent{\large\textbf{Irrep $(0\,,0)$}}\par\nopagebreak\vspace{4pt}
\noindent\textbf{$P$-even, $T$-even:}\nopagebreak
\begin{align}
    \left(\ytableaushort{11,22}\,,\ytableaushort{11,22}\right) &\rightarrow \varepsilon_2^*{}^{\mu\nu}\varepsilon_1{}_{\mu\nu}\, ,\label{eq:2-00-i}\\
    \left(\ytableaushort{112,244}\,,\ytableaushort{112,244}\right) &\rightarrow (\varepsilon_2^*{}^{\mu\nu}q_\nu) (\varepsilon_1{}_{\mu\rho}q^\rho)\,, \\
    \left(\ytableaushort{1122,4444}\,,\ytableaushort{1122,4444}\right) &\rightarrow (\varepsilon^*_2{}^{\mu\nu}q_\mu q_\nu)(\varepsilon_1^{\rho\kappa}q_\rho q_\kappa)\,,
\end{align} 
\noindent\textbf{$P$-odd, $T$-even:}\nopagebreak
\begin{align}
    \left(\ytableaushort{112,234}\,,\ytableaushort{113,224}\right) - \text{dual} &\rightarrow \epsilon_{\nu\rho\kappa\lambda}(\varepsilon_2^*{}^{\mu\nu}\varepsilon_1{}_{\mu}^{\rho}P^\kappa q^\lambda)\,, \\
    \left(\ytableaushort{1122,3344}\,,\ytableaushort{1133,2244}\right) - \text{dual} &\rightarrow \epsilon_{\mu\nu\alpha\beta}(\varepsilon_2^*{}^{\mu\rho}q_\rho)(\varepsilon_1{}^{\nu\lambda}q_\lambda) P^\alpha q^\beta \,\label{eq:2-00-f}.
\end{align} 

\vspace{1em}
\noindent{\large\textbf{Irrep $(\frac{1}{2}\,,\frac{1}{2})$}}\par\nopagebreak\vspace{4pt}
\noindent\textbf{$P$-even, $T$-even:}\nopagebreak
\begin{align}
    \left(\ytableaushort{113,22}\,,\ytableaushort{113,22}\right) &\rightarrow (\varepsilon_2^*{}^{\nu\rho}\varepsilon_1{}_{\nu\rho})P^\mu\,,\label{eq:2-1/21/2-i} \\
    \left(\ytableaushort{112,23}\,,\ytableaushort{112,23}\right) &\rightarrow \varepsilon_2^*{}^{\mu\nu}(\varepsilon_1{}_{\nu\rho}q^\rho) - \varepsilon_1{}^{\mu\nu}(\varepsilon_2^*{}_{\nu\rho}q^\rho) \\
    \left(\ytableaushort{1123,244}\,,\ytableaushort{1123,244}\right) &\rightarrow (\varepsilon_2^*{}^{\nu\rho}q_\rho)(\varepsilon_1{}_{\nu\lambda}q^\lambda) P^\mu \\
    \left(\ytableaushort{1122,344}\,,\ytableaushort{1122,344}\right) &\rightarrow (\varepsilon_2^*{}^{\nu\rho}q_\nu q_\rho)(\varepsilon_1{}^{\mu\kappa}q_\kappa) -(\varepsilon_1{}^{\nu\rho}q_\nu q_\rho)(\varepsilon_2^*{}^{\mu\kappa}q_\kappa) \\
    \left(\ytableaushort{11223,4444}\,,\ytableaushort{11223,4444}\right) &\rightarrow (\varepsilon_2^*{}^{\nu\rho}q_\nu q_\rho)(\varepsilon_1{}^{\lambda\kappa}q_\lambda q_\kappa) P^\mu
\end{align}
\noindent\textbf{$P$-odd, $T$-even:}\nopagebreak
\begin{align}
    \left(\ytableaushort{112,24}\,,\ytableaushort{114,22}\right) -\text{dual} &\rightarrow \epsilon^{\mu\rho\lambda\kappa}\varepsilon_2^*{}^\nu_\rho\varepsilon_1{}_{\nu\lambda}q_\kappa \,,\\
    \left(\ytableaushort{1123,234}\,,\ytableaushort{1133,224}\right) - \text{dual} &\rightarrow \epsilon_{\rho\lambda\kappa\alpha}(\varepsilon_2^*{}^{\nu\rho}\varepsilon_1{}_\nu^\lambda P^\kappa q^\alpha) P^\mu \,,\\
    \left(\ytableaushort{1123,234}\,,\ytableaushort{1124,233}\right) -\text{dual} &\rightarrow \epsilon^{\mu\lambda\alpha\beta}(\varepsilon_2^*{}^{\nu\rho}q_\rho)\varepsilon_1{}_{\nu\lambda} P_\alpha q_\beta + \epsilon^{\mu\lambda\alpha\beta}(\varepsilon_1{}^{\nu\rho}q_\rho)\varepsilon_2^*{}_{\nu\lambda} P_\alpha q_\beta \,, \\
    \left(\ytableaushort{1122,444}\,,\ytableaushort{1124,244}\right) -\text{dual} &\rightarrow \epsilon^{\mu\nu\rho\kappa}(\varepsilon_2^*{}_{\nu\alpha}q^\alpha)(\varepsilon_1{}_{\rho\beta}q^\beta)q_\kappa \,,\\
    \left(\ytableaushort{11223,4444}\,,\ytableaushort{11224,3444}\right) - \text{dual} &\rightarrow \epsilon^{\mu\nu\rho\lambda}(\varepsilon_2^*{}_{\nu\kappa}q^\kappa)(\varepsilon_1{}^{\alpha\beta}q_\alpha q_\beta)P_\rho q_\lambda + \epsilon^{\mu\nu\rho\lambda}(\varepsilon_1{}_{\nu\kappa}q^\kappa)(\varepsilon_2^*{}^{\alpha\beta}q_\alpha q_\beta)P_\rho q_\lambda \,,\\
    \left(\ytableaushort{11223,3334}\,,\ytableaushort{11233,2334}\right) -\text{dual} &\rightarrow \epsilon^{\nu\rho\lambda\kappa}(\varepsilon_2^*{}_{\nu\alpha}q^\alpha)(\varepsilon_1{}_{\rho\beta}q^\beta) P_\lambda q_\kappa P^\mu \,. 
\end{align}
\noindent\textbf{$P$-even, $T$-odd:}\nopagebreak
\begin{align}
    \left(\ytableaushort{114,22}\,,\ytableaushort{114,22}\right) &\rightarrow (\varepsilon_2^*{}^{\nu\rho}\varepsilon_1{}_{\nu\rho})q^\mu\,, \\
    \left(\ytableaushort{112,24}\,,\ytableaushort{112,24}\right) &\rightarrow \varepsilon_2^*{}^{\mu\nu}(\varepsilon_1{}_{\nu\rho}q^\rho) + \varepsilon_1{}^{\mu\nu}(\varepsilon_2^*{}_{\nu\rho}q^\rho) \,,\\
    \left(\ytableaushort{1124,244}\,,\ytableaushort{1124,244}\right) &\rightarrow (\varepsilon_2^*{}^{\nu\rho}q_\rho)(\varepsilon_1{}_{\nu\lambda}q^\lambda) q^\mu \,,\\
    \left(\ytableaushort{1122,444}\,,\ytableaushort{1122,444}\right) &\rightarrow (\varepsilon_2^*{}^{\nu\rho}q_\nu q_\rho)(\varepsilon_1{}^{\mu\kappa}q_\kappa) + (\varepsilon_1{}^{\nu\rho}q_\nu q_\rho)(\varepsilon_2^*{}^{\mu\kappa}q_\kappa) \,,\\
    \left(\ytableaushort{11224,4444}\,,\ytableaushort{11224,4444}\right) &\rightarrow (\varepsilon_2^*{}^{\nu\rho}q_\nu q_\rho)(\varepsilon_1{}^{\lambda\kappa}q_\lambda q_\kappa) q^\mu\,.
\end{align}
\noindent\textbf{$P$-odd, $T$-odd:}\nopagebreak
\begin{align}
    \left(\ytableaushort{112,23}\,,\ytableaushort{113,22}\right) -\text{dual} &\rightarrow \epsilon^{\mu\rho\lambda\kappa}\varepsilon_2^*{}^\nu_\rho\varepsilon_1{}_{\nu\lambda}P_\kappa \,,\\
    \left(\ytableaushort{1124,234}\,,\ytableaushort{1134,224}\right) - \text{dual} &\rightarrow \epsilon_{\rho\lambda\kappa\alpha}(\varepsilon_2^*{}^{\nu\rho}\varepsilon_1{}_\nu^\lambda P^\kappa q^\alpha) q^\mu \,,\\
    \left(\ytableaushort{1123,234}\,,\ytableaushort{1124,233}\right) -\text{dual} &\rightarrow \epsilon^{\mu\lambda\alpha\beta}(\varepsilon_2^*{}^{\nu\rho}q_\rho)\varepsilon_1{}_{\nu\lambda} P_\alpha q_\beta - \epsilon^{\mu\lambda\alpha\beta}(\varepsilon_1{}^{\nu\rho}q_\rho)\varepsilon_2^*{}_{\nu\lambda} P_\alpha q_\beta \,,\\
    \left(\ytableaushort{1122,344}\,,\ytableaushort{1123,244}\right) -\text{dual} &\rightarrow \epsilon^{\mu\nu\rho\kappa}(\varepsilon_2^*{}_{\nu\alpha}q^\alpha)(\varepsilon_1{}_{\rho\beta}q^\beta)P_\kappa \,,\\
    \left(\ytableaushort{11223,3334}\,,\ytableaushort{11224,3333}\right) - \text{dual} &\rightarrow \epsilon^{\mu\nu\rho\lambda}(\varepsilon_2^*{}_{\nu\kappa}q^\kappa)(\varepsilon_1{}^{\alpha\beta}q_\alpha q_\beta)P_\rho q_\lambda - \epsilon^{\mu\nu\rho\lambda}(\varepsilon_1{}_{\nu\kappa}q^\kappa)(\varepsilon_2^*{}^{\alpha\beta}q_\alpha q_\beta)P_\rho q_\lambda \,,\\
    \left(\ytableaushort{11224,3444}\,,\ytableaushort{11234,2444}\right) -\text{dual} &\rightarrow \epsilon^{\nu\rho\lambda\kappa}(\varepsilon_2^*{}_{\nu\alpha}q^\alpha)(\varepsilon_1{}_{\rho\beta}q^\beta) P_\lambda q_\kappa q^\mu \,.\label{eq:2-1/21/2-f}
\end{align}

\vspace{1em}
\noindent{\large\textbf{Irrep $(1\,,0)\oplus (0\,,1)$}}\par\nopagebreak\vspace{4pt}
\noindent\textbf{$P$-even, $T$-even:}\nopagebreak
\begin{align}
    \left(\ytableaushort{1123,23}\,,\ytableaushort{112,233}\right) + \text{dual} &\rightarrow P^{[\mu}\varepsilon_2^*{}^{\nu]\rho}(\varepsilon_1{}_{\rho\lambda}q^\lambda) - P^{[\mu}\varepsilon_1{}^{\nu]\rho}(\varepsilon_2^*{}_{\rho\lambda}q^\lambda) \label{eq:2-1001-i}\,,\\
    \left(\ytableaushort{1124,24}\,,\ytableaushort{112,244}\right) + \text{dual} &\rightarrow q^{[\mu}\varepsilon_2^*{}^{\nu]\rho}(\varepsilon_1{}_{\rho\lambda}q^\lambda) + q^{[\mu}\varepsilon_1{}^{\nu]\rho}(\varepsilon_2^*{}_{\rho\lambda}q^\lambda)\,, \\
    \left(\ytableaushort{11223,344}\,,\ytableaushort{1122,3443}\right) + \text{dual}  &\rightarrow (\varepsilon_2^*{}^{\rho\lambda}q_\rho q_\lambda) (\varepsilon_1{}^{[\mu\kappa}q_\kappa) P^{\nu]} - (\varepsilon_1{}^{\rho\lambda}q_\rho q_\lambda) (\varepsilon_2^*{}^{[\mu\kappa}q_\kappa) P^{\nu]} \,,\\
    \left(\ytableaushort{11224,444}\,,\ytableaushort{1122,4444}\right) + \text{dual} &\rightarrow (\varepsilon_2^*{}^{\rho\lambda}q_\rho q_\lambda) (\varepsilon_1{}^{[\mu\kappa}q_\kappa) q^{\nu]} + (\varepsilon_1{}^{\rho\lambda}q_\rho q_\lambda) (\varepsilon_2^*{}^{[\mu\kappa}q_\kappa) q^{\nu]}\,.
\end{align}
\noindent\textbf{$P$-odd, $T$-even:}\nopagebreak
\begin{align}
    \left(\ytableaushort{1123,23}\,,\ytableaushort{112,233}\right) - \text{dual} &\rightarrow \epsilon^{\mu\nu\alpha\beta}\left[P_\alpha\varepsilon_2^*{}^{\rho}_\beta(\varepsilon_1{}_{\rho\lambda}q^\lambda) - P_\alpha\varepsilon_1{}^{\rho}_\beta(\varepsilon_2^*{}_{\rho\lambda}q^\lambda)\right] \,,\\
    \left(\ytableaushort{1124,24}\,,\ytableaushort{112,244}\right) - \text{dual} &\rightarrow \epsilon^{\mu\nu\alpha\beta}\left[q_\alpha\varepsilon_2^*{}^{\rho}_\beta(\varepsilon_1{}_{\rho\lambda}q^\lambda) + q_\alpha\varepsilon_1{}^{\rho}_\beta(\varepsilon_2^*{}_{\rho\lambda}q^\lambda)\right] \,,\\
    \left(\ytableaushort{11223,344}\,,\ytableaushort{1122,3443}\right) - \text{dual}  &\rightarrow \epsilon^{\mu\nu\alpha\beta}\left[(\varepsilon_2^*{}^{\rho\lambda}q_\rho q_\lambda) (\varepsilon_1{}^{\kappa}_\alpha q_\kappa) P_\beta - (\varepsilon_1{}^{\rho\lambda}q_\rho q_\lambda) (\varepsilon_2^*{}^{\kappa}_\alpha q_\kappa) P_\beta\right] \,,\\
    \left(\ytableaushort{11224,444}\,,\ytableaushort{1122,4444}\right) - \text{dual} &\rightarrow \epsilon^{\mu\nu\alpha\beta}\left[(\varepsilon_2^*{}^{\rho\lambda}q_\rho q_\lambda) (\varepsilon_1{}^{\kappa}_\alpha q_\kappa) q_\beta + (\varepsilon_1{}^{\rho\lambda}q_\rho q_\lambda) (\varepsilon_2^*{}^{\kappa}_\alpha q_\kappa) q_\beta\right]\,.
\end{align}
\noindent\textbf{$P$-even, $T$-odd:}\nopagebreak
\begin{align}
    \left(\ytableaushort{112,2}\,,\ytableaushort{11,22}\right) + \text{dual} &\rightarrow \varepsilon_2^*{}^{[\mu\rho}\varepsilon_1{}_\rho^{\nu]}\,, \\
    \left(\ytableaushort{1122,44}\,,\ytableaushort{112,244}\right) + \text{dual} &\rightarrow (\varepsilon_2^*{}^{[\mu\rho}q_\rho)(\varepsilon_1{}^{\nu]\lambda}q_\lambda) \,,\\
    \left(\ytableaushort{1123,24}\,,\ytableaushort{112,234}\right) + \text{dual} &\rightarrow P^{[\mu}\varepsilon_2^*{}^{\nu]\rho}(\varepsilon_1{}_{\rho\lambda}q^\lambda) + P^{[\mu}\varepsilon_1{}^{\nu]\rho}(\varepsilon_2^*{}_{\rho\lambda}q^\lambda) \,,\\
    \left(\ytableaushort{1124,23}\,,\ytableaushort{112,234}\right) + \text{dual} &\rightarrow q^{[\mu}\varepsilon_2^*{}^{\nu]\rho}(\varepsilon_1{}_{\rho\lambda}q^\lambda) - q^{[\mu}\varepsilon_1{}^{\nu]\rho}(\varepsilon_2^*{}_{\rho\lambda}q^\lambda) \,,\\
    \left(\ytableaushort{1134,22}\,,\ytableaushort{113,224}\right) + \text{dual} &\rightarrow P^{[\mu}q^{\nu]}(\varepsilon_2^*{}^{\rho\lambda}\varepsilon_1{}_{\rho\lambda}) \,,\\
    \left(\ytableaushort{11223,444}\,,\ytableaushort{1122,3444}\right) + \text{dual} &\rightarrow (\varepsilon_2^*{}^{\rho\lambda}q_\rho q_\lambda) (\varepsilon_1{}^{[\mu\kappa}q_\kappa) P^{\nu]} + (\varepsilon_1{}^{\rho\lambda}q_\rho q_\lambda) (\varepsilon_2^*{}^{[\mu\kappa}q_\kappa) P^{\nu]} \,,\\
    \left(\ytableaushort{11224,333}\,,\ytableaushort{1122,3334}\right) + \text{dual} &\rightarrow (\varepsilon_2^*{}^{\rho\lambda}q_\rho q_\lambda) (\varepsilon_1{}^{[\mu\kappa}q_\kappa) q^{\nu]} - (\varepsilon_1{}^{\rho\lambda}q_\rho q_\lambda) (\varepsilon_2^*{}^{[\mu\kappa}q_\kappa) q^{\nu]} \,,\\
    \left(\ytableaushort{11234,244}\,,\ytableaushort{1123,2444}\right) + \text{dual} &\rightarrow (\varepsilon_2^*{}^{\rho\lambda}q_\lambda)(\varepsilon_1{}_{\rho\kappa}q^\kappa)P^{[\mu}q^{\nu]} \,,\\
    \left(\ytableaushort{112234,3333}\,,\ytableaushort{11223,33334}\right) + \text{dual} &\rightarrow (\varepsilon_2^*{}^{\rho\lambda}q_\rho q_\lambda)(\varepsilon_1{}^{\kappa\delta}q_\kappa q_\delta)P^{[\mu}q^{\nu]}\,.
\end{align}
\noindent\textbf{$P$-odd, $T$-odd:}\nopagebreak
\begin{align}
    \left(\ytableaushort{112,2}\,,\ytableaushort{11,22}\right) - \text{dual} &\rightarrow \epsilon^{\mu\nu\alpha\beta}\varepsilon_2^*{}^{\rho}_\alpha\varepsilon_1{}_{\rho\beta} \,,\\
    \left(\ytableaushort{1122,44}\,,\ytableaushort{112,244}\right) - \text{dual} &\rightarrow \epsilon^{\mu\nu\alpha\beta}(\varepsilon_2^*{}^{\rho}_\alpha q_\rho)(\varepsilon_1{}^{\lambda}_\beta q_\lambda) \,,\\
    \left(\ytableaushort{1123,24}\,,\ytableaushort{112,234}\right) - \text{dual} &\rightarrow \epsilon^{\mu\nu\alpha\beta}\left[P_\alpha\varepsilon_2^*{}^{\rho}_\beta(\varepsilon_1{}_{\rho\lambda}q^\lambda) + P_\alpha\varepsilon_1{}^{\rho}_\beta(\varepsilon_2^*{}_{\rho\lambda}q^\lambda)\right] \,,\\
    \left(\ytableaushort{1124,23}\,,\ytableaushort{112,234}\right) - \text{dual} &\rightarrow \epsilon^{\mu\nu\alpha\beta}\left[q_\alpha\varepsilon_2^*{}^{\rho}_\beta(\varepsilon_1{}_{\rho\lambda}q^\lambda) - q_\alpha\varepsilon_1{}^{\rho}_\beta(\varepsilon_2^*{}_{\rho\lambda}q^\lambda)\right] \,,\\
    \left(\ytableaushort{1134,22}\,,\ytableaushort{113,224}\right) - \text{dual} &\rightarrow \epsilon^{\mu\nu\alpha\beta}P_\alpha q_\beta (\varepsilon_2^*{}^{\rho\lambda}\varepsilon_1{}_{\rho\lambda}) \,,\\
    \left(\ytableaushort{11223,444}\,,\ytableaushort{1122,3444}\right) - \text{dual} &\rightarrow \epsilon^{\mu\nu\alpha\beta}\left[(\varepsilon_2^*{}^{\rho\lambda}q_\rho q_\lambda) (\varepsilon_1{}^{\kappa}_\alpha q_\kappa) P_\beta + (\varepsilon_1{}^{\rho\lambda}q_\rho q_\lambda) (\varepsilon_2^*{}^{\kappa}_\alpha q_\kappa) P_\beta\right] \,,\\
    \left(\ytableaushort{11224,333}\,,\ytableaushort{1122,3334}\right) - \text{dual} &\rightarrow \epsilon^{\mu\nu\alpha\beta}\left[(\varepsilon_2^*{}^{\rho\lambda}q_\rho q_\lambda) (\varepsilon_1{}^{\kappa}_\alpha q_\kappa) q_\beta - (\varepsilon_1{}^{\rho\lambda}q_\rho q_\lambda) (\varepsilon_2^*{}^{\kappa}_\alpha q_\kappa) q_\beta\right] \,,\\
    \left(\ytableaushort{11234,244}\,,\ytableaushort{1123,2444}\right) - \text{dual} &\rightarrow \epsilon^{\mu\nu\alpha\beta}(\varepsilon_2^*{}^{\rho\lambda}q_\lambda)(\varepsilon_1{}_{\rho\kappa}q^\kappa)P_\alpha q_\beta \,,\\
    \left(\ytableaushort{11234,234}\,,\ytableaushort{1133,2244}\right) -\text{dual} &\rightarrow (\epsilon^{\lambda\kappa\alpha\beta}\varepsilon_2^*{}^{\rho}_\lambda\varepsilon_1{}_{\rho\kappa}P_\alpha q_\beta) P^{[\mu}q^{\nu]} \,,\\
    \left(\ytableaushort{112234,3333}\,,\ytableaushort{11223,33334}\right) - \text{dual} &\rightarrow \epsilon^{\mu\nu\alpha\beta}(\varepsilon_2^*{}^{\rho\lambda}q_\rho q_\lambda)(\varepsilon_1{}^{\kappa\delta}q_\kappa q_\delta)P_\alpha q_\beta \,,\\
    \left(\ytableaushort{112234,3334}\,,\ytableaushort{11233,23344}\right) - \text{dual} &\rightarrow \epsilon^{\delta\rho\lambda\kappa}(\varepsilon^*_2{}_{\delta\alpha}q^\alpha)(\varepsilon_1{}_{\rho\beta}q^\beta)P_\lambda q_\kappa P^{[\mu}q^{\nu]}\label{eq:2-1001-f}\,.
\end{align}

\vspace{1em}
\noindent{\large\textbf{Irrep $(1\,,1)$}}\par\nopagebreak\vspace{4pt}
\noindent\textbf{$P$-even, $T$-even:}\nopagebreak
\begin{align}
    \left(\ytableaushort{112,2}\,,\ytableaushort{112,2}\right) &\rightarrow \varepsilon^*_2{}^{(\mu\rho}\varepsilon_2{}_\rho^{\nu)} \label{eq:2-11-i}\,,\\
    \left(\ytableaushort{1133,22}\,,\ytableaushort{1133,22}\right) &\rightarrow P^{(\mu}P^{\nu)}(\varepsilon_2^*{}^{\rho\lambda}\varepsilon_1{}_{\rho\lambda}) \,,\\
    \left(\ytableaushort{1144,22}\,,\ytableaushort{1144,22}\right) &\rightarrow q^{(\mu}q^{\nu)}(\varepsilon_2^*{}^{\rho\lambda}\varepsilon_1{}_{\rho\lambda}) \,,\\
    \left(\ytableaushort{1123,23}\,,\ytableaushort{1123,23}\right) &\rightarrow (\varepsilon_2^*{}^{\rho\lambda}q_\lambda)\varepsilon_1{}^{(\mu}_\rho P^{\nu)} - (\varepsilon_1{}^{\rho\lambda}q_\lambda)\varepsilon_2^*{}^{(\mu}_\rho P^{\nu)} \,,\\
    \left(\ytableaushort{1124,24}\,,\ytableaushort{1124,24}\right) &\rightarrow (\varepsilon_2^*{}^{\rho\lambda}q_\lambda)\varepsilon_1{}^{(\mu}_\rho q^{\nu)} + (\varepsilon_1{}^{\rho\lambda}q_\lambda)\varepsilon_2^*{}^{(\mu}_\rho q^{\nu)} \,,\\
    \left(\ytableaushort{1122,33}\,,\ytableaushort{1122,33}\right) &\rightarrow (\varepsilon_2^*{}^{\rho\lambda}q_\rho q_\lambda) \varepsilon_1{}^{\mu\nu} + (\varepsilon_1{}^{\rho\lambda}q_\rho q_\lambda) \varepsilon_2^*{}^{\mu\nu} \,,\\
    \left(\ytableaushort{11223,344}\,,\ytableaushort{11223,344}\right) &\rightarrow (\varepsilon_2^*{}_{\rho\lambda}q^\rho q^\lambda)\varepsilon_1{}^{(\mu\kappa}q_\kappa P^{\nu)} - (\varepsilon_1{}_{\rho\lambda}q^\rho q^\lambda)\varepsilon_2^*{}^{(\mu\kappa}q_\kappa P^{\nu)} \,,\\
    \left(\ytableaushort{11224,444}\,,\ytableaushort{11224,444}\right) &\rightarrow (\varepsilon_2^*{}_{\rho\lambda}q^\rho q^\lambda)\varepsilon_1{}^{(\mu\kappa}q_\kappa q^{\nu)} + (\varepsilon_1{}_{\rho\lambda}q^\rho q^\lambda)\varepsilon_2^*{}^{(\mu\kappa}q_\kappa q^{\nu)} \,,\\
    \left(\ytableaushort{11233,244}\,,\ytableaushort{11233,244}\right) &\rightarrow (\varepsilon_2^*{}^{\rho\lambda}q_\lambda)(\varepsilon_1{}_{\rho\kappa}q^\kappa)P^{(\mu}P^{\nu)} \,,\\
    \left(\ytableaushort{11244,244}\,,\ytableaushort{11244,244}\right) &\rightarrow (\varepsilon_2^*{}^{\rho\lambda}q_\lambda)(\varepsilon_1{}_{\rho\kappa}q^\kappa)q^{(\mu}q^{\nu)} \,,\\
    \left(\ytableaushort{112233,4444}\,,\ytableaushort{112233,4444}\right) &\rightarrow (\varepsilon_2^*{}^{\rho\lambda}q_\rho q_\lambda)(\varepsilon_1{}^{\alpha\beta}q_\alpha q_\beta)P^{(\mu}P^{\nu)} \,,\\
    \left(\ytableaushort{112244,4444}\,,\ytableaushort{112244,4444}\right) &\rightarrow (\varepsilon_2^*{}^{\rho\lambda}q_\rho q_\lambda)(\varepsilon_1{}^{\alpha\beta}q_\alpha q_\beta)q^{(\mu}q^{\nu)}\,.
\end{align}
\noindent\textbf{$P$-odd, $T$-even:}\nopagebreak
\begin{align}
    \left(\ytableaushort{1134,22}\,,\ytableaushort{1123,24}\right) - \text{dual} &\rightarrow \epsilon^{(\mu\lambda\kappa\alpha}\varepsilon_2^*{}^{\rho}_\lambda \varepsilon_1{}_{\rho\kappa}P_\alpha q^{\nu)} - \epsilon^{(\mu\lambda\kappa\alpha}\varepsilon_1{}^{\rho}_\lambda \varepsilon_2^*{}_{\rho\kappa}P_\alpha q^{\nu)} \,,\\
    \left(\ytableaushort{1134,22}\,,\ytableaushort{1124,23}\right) - \text{dual} &\rightarrow \epsilon^{(\mu\lambda\kappa\alpha}\varepsilon_2^*{}^{\rho}_\lambda \varepsilon_1{}_{\rho\kappa}q_\alpha P^{\nu)} - \epsilon^{(\mu\lambda\kappa\alpha}\varepsilon_1{}^{\rho}_\lambda \varepsilon_2^*{}_{\rho\kappa}q_\alpha P^{\nu)} \,,\\
    \left(\ytableaushort{1133,22}\,,\ytableaushort{1122,33}\right) - \text{dual} &\rightarrow \epsilon^{(\mu\lambda\kappa\alpha}(\varepsilon_2^*{}_{\rho\lambda}q^\rho) \varepsilon_1{}_{\kappa}^{\nu)}P_\alpha - \epsilon^{(\mu\lambda\kappa\alpha}(\varepsilon_1{}_{\rho\lambda}q^\rho) \varepsilon_2^*{}_{\kappa}^{\nu)}P_\alpha \,,\\
    \left(\ytableaushort{1144,22}\,,\ytableaushort{1122,44}\right) - \text{dual} &\rightarrow \epsilon^{(\mu\lambda\kappa\alpha}(\varepsilon_2^*{}_{\rho\lambda}q^\rho) \varepsilon_1{}_{\kappa}^{\nu)}q_\alpha + \epsilon^{(\mu\lambda\kappa\alpha}(\varepsilon_1{}_{\rho\lambda}q^\rho) \varepsilon_2^*{}_{\kappa}^{\nu)}q_\alpha \,,\\
    \left(\ytableaushort{11224,334}\,,\ytableaushort{11223,344}\right) - \text{dual} &\rightarrow \epsilon^{(\mu\lambda\alpha\beta}(\varepsilon_2^*{}_{\rho\lambda}q^\rho)(\varepsilon_1{}^{\nu)\kappa}q_\kappa)P_\alpha q_\beta - \epsilon^{(\mu\lambda\alpha\beta}(\varepsilon_1{}_{\rho\lambda}q^\rho)(\varepsilon_2^*{}^{\nu)\kappa}q_\kappa)P_\alpha q_\beta \,,\\
    \left(\ytableaushort{11233,244}\,,\ytableaushort{11244,233}\right) - \text{dual} &\rightarrow \epsilon^{(\mu\kappa\alpha\beta}(\varepsilon_2^*{}^{\rho\lambda} q_\lambda)\varepsilon_1{}_{\rho\kappa}P_\alpha q_\beta q^{\nu)} - \epsilon^{(\mu\kappa\alpha\beta}(\varepsilon_1{}^{\rho\lambda} q_\lambda)\varepsilon_2^*{}_{\rho\kappa}P_\alpha q_\beta q^{\nu)} \,,\\
    \left(\ytableaushort{11233,244}\,,\ytableaushort{11234,234}\right) - \text{dual} &\rightarrow \epsilon^{(\mu\kappa\alpha\beta}(\varepsilon_2^*{}^{\rho\lambda} q_\lambda)\varepsilon_1{}_{\rho\kappa}P_\alpha q_\beta P^{\nu)} + \epsilon^{(\mu\kappa\alpha\beta}(\varepsilon_1{}^{\rho\lambda} q_\lambda)\varepsilon_2^*{}_{\rho\kappa}P_\alpha q_\beta P^{\nu)} \,,\\
    \left(\ytableaushort{11233,234}\,,\ytableaushort{11333,224}\right) - \text{dual} &\rightarrow \epsilon^{\rho\lambda\kappa\delta}\varepsilon_2^*{}_{\alpha\rho}\varepsilon_1{}^\alpha_\lambda P_\kappa q_\delta P^{(\mu}P^{\nu)} \,,\\
    \left(\ytableaushort{11244,234}\,,\ytableaushort{11344,224}\right) - \text{dual} &\rightarrow \epsilon^{\rho\lambda\kappa\delta}\varepsilon_2^*{}_{\alpha\rho}\varepsilon_1{}^\alpha_\lambda P_\kappa q_\delta q^{(\mu}q^{\nu)} \,,\\
    \left(\ytableaushort{11224,344}\,,\ytableaushort{11234,244}\right) - \text{dual} &\rightarrow \epsilon^{(\mu\lambda\kappa\alpha}(\varepsilon_2^*{}_{\rho\lambda}q^\lambda)(\varepsilon_1{}_{\kappa\delta}q^\delta)P_\alpha q^{\nu)} \,,\\
    \left(\ytableaushort{11223,444}\,,\ytableaushort{11234,244}\right) - \text{dual} &\rightarrow \epsilon^{(\mu\lambda\kappa\alpha}(\varepsilon_2^*{}_{\rho\lambda}q^\lambda)(\varepsilon_1{}_{\kappa\delta}q^\delta)q_\alpha P^{\nu)} \,,\\
    \left(\ytableaushort{11224,334}\,,\ytableaushort{11334,224}\right) - \text{dual} &\rightarrow \epsilon^{\rho\lambda\kappa\delta}(\varepsilon_2^*{}_{\rho\alpha}q^\alpha)\varepsilon_1{}^{(\mu}_\lambda P_\kappa q_\delta q^{\nu)} - \epsilon^{\rho\lambda\kappa\delta}(\varepsilon_1{}_{\rho\alpha}q^\alpha)\varepsilon_2^*{}^{(\mu}_\lambda P_\kappa q_\delta q^{\nu)}\,,\\
    \left(\ytableaushort{11223,344}\,,\ytableaushort{11334,224}\right) - \text{dual} &\rightarrow \epsilon^{\rho\lambda\kappa\delta}(\varepsilon_2^*{}_{\rho\alpha}q^\alpha)\varepsilon_1{}^{(\mu}_\lambda P_\kappa q_\delta q^{\nu)} + \epsilon^{\rho\lambda\kappa\delta}(\varepsilon_1{}_{\rho\alpha}q^\alpha)\varepsilon_2^*{}^{(\mu}_\lambda P_\kappa q_\delta P^{\nu)} \,,\\
    \left(\ytableaushort{112233,3444}\,,\ytableaushort{112333,2444}\right) -\text{dual} &\rightarrow \epsilon^{\rho\lambda\kappa\delta}(\varepsilon_2^*{}_{\rho\alpha}q^\alpha)(\varepsilon_1{}_{\lambda\beta}q^\beta)P_\kappa q_\delta P^{(\mu}P^{\nu)} \,,\\
    \left(\ytableaushort{112244,3444}\,,\ytableaushort{112344,2444}\right) -\text{dual} &\rightarrow \epsilon^{\rho\lambda\kappa\delta}(\varepsilon_2^*{}_{\rho\alpha}q^\alpha)(\varepsilon_1{}_{\lambda\beta}q^\beta)P_\kappa q_\delta q^{(\mu}q^{\nu)} \,.
\end{align}
\noindent\textbf{$P$-even, $T$-odd:}\nopagebreak
\begin{align}
    \left(\ytableaushort{1134,22}\,,\ytableaushort{1134,22}\right) &\rightarrow P^{(\mu}q^{\nu)}(\varepsilon_2^*{}^{\rho\lambda}\varepsilon_1{}_{\rho\lambda}) \,,\\
    \left(\ytableaushort{1123,24}\,,\ytableaushort{1123,24}\right) &\rightarrow (\varepsilon_2^*{}^{\rho\lambda}q_\lambda)\varepsilon_1{}^{(\mu}_\rho P^{\nu)} + (\varepsilon_1{}^{\rho\lambda}q_\lambda)\varepsilon_2^*{}^{(\mu}_\rho P^{\nu)} \,,\\
    \left(\ytableaushort{1124,23}\,,\ytableaushort{1124,23}\right) &\rightarrow (\varepsilon_2^*{}^{\rho\lambda}q_\lambda)\varepsilon_1{}^{(\mu}_\rho q^{\nu)} - (\varepsilon_1{}^{\rho\lambda}q_\lambda)\varepsilon_2^*{}^{(\mu}_\rho q^{\nu)} \,,\\
    \left(\ytableaushort{11224,344}\,,\ytableaushort{11224,344}\right) &\rightarrow (\varepsilon_2^*{}_{\rho\lambda}q^\rho q^\lambda)\varepsilon_1{}^{(\mu\kappa}q_\kappa q^{\nu)} - (\varepsilon_1{}_{\rho\lambda}q^\rho q^\lambda)\varepsilon_2^*{}^{(\mu\kappa}q_\kappa q^{\nu)} \,,\\
    \left(\ytableaushort{11223,444}\,,\ytableaushort{11223,444}\right) &\rightarrow (\varepsilon_2^*{}_{\rho\lambda}q^\rho q^\lambda)\varepsilon_1{}^{(\mu\kappa}q_\kappa P^{\nu)} + (\varepsilon_1{}_{\rho\lambda}q^\rho q^\lambda)\varepsilon_2^*{}^{(\mu\kappa}q_\kappa P^{\nu)} \,,\\
    \left(\ytableaushort{11234,244}\,,\ytableaushort{11234,244}\right) &\rightarrow (\varepsilon_2^*{}^{\rho\lambda}q_\lambda)(\varepsilon_1{}_{\rho\kappa}q^\kappa)P^{(\mu}q^{\nu)} \,,\\
    \left(\ytableaushort{112234,4444}\,,\ytableaushort{112234,4444}\right) &\rightarrow (\varepsilon_2^*{}^{\rho\lambda}q_\rho q_\lambda)(\varepsilon_1{}^{\alpha\beta}q_\alpha q_\beta)P^{(\mu}q^{\nu)}\,.
\end{align}
\noindent\textbf{$P$-odd, $T$-odd:}\nopagebreak
\begin{align}
    \left(\ytableaushort{1124,23}\,,\ytableaushort{1123,24}\right) - \text{dual} &\rightarrow \epsilon^{(\mu\lambda\alpha\beta}\varepsilon_2^*{}^{\rho}_\lambda \varepsilon_1{}^{\nu)}_{\rho}P_\alpha q_\beta + \epsilon^{(\mu\lambda\alpha\beta}\varepsilon_1{}^{\rho}_\lambda \varepsilon_2^*{}^{\nu)}_{\rho}P_\alpha q_\beta \,,\\
    \left(\ytableaushort{1133,22}\,,\ytableaushort{1123,23}\right) - \text{dual} &\rightarrow \epsilon^{(\mu\lambda\kappa\alpha}\varepsilon_2^*{}^{\rho}_\lambda \varepsilon_1{}_{\rho\kappa}P_\alpha P^{\nu)} - \epsilon^{(\mu\lambda\kappa\alpha}\varepsilon_1{}^{\rho}_\lambda \varepsilon_2^*{}_{\rho\kappa}P_\alpha P^{\nu)} \,,\\
    \left(\ytableaushort{1144,22}\,,\ytableaushort{1124,24}\right) - \text{dual} &\rightarrow \epsilon^{(\mu\lambda\kappa\alpha}\varepsilon_2^*{}^{\rho}_\lambda \varepsilon_1{}_{\rho\kappa}q_\alpha q^{\nu)} - \epsilon^{(\mu\lambda\kappa\alpha}\varepsilon_1{}^{\rho}_\lambda \varepsilon_2^*{}_{\rho\kappa}q_\alpha q^{\nu)} \,,\\
    \left(\ytableaushort{1134,22}\,,\ytableaushort{1122,34}\right) - \text{dual} &\rightarrow \epsilon^{(\mu\lambda\kappa\alpha}(\varepsilon_2^*{}_{\rho\lambda}q^\rho) \varepsilon_1{}_{\kappa}^{\nu)}P_\alpha + \epsilon^{(\mu\lambda\kappa\alpha}(\varepsilon_1{}_{\rho\lambda}q^\rho) \varepsilon_2^*{}_{\kappa}^{\nu)}P_\alpha \,,\\
    \left(\ytableaushort{1124,24}\,,\ytableaushort{1122,44}\right) - \text{dual}&\rightarrow \epsilon^{(\mu\lambda\kappa\alpha}(\varepsilon_2^*{}_{\rho\lambda}q^\rho) \varepsilon_1{}_{\kappa}^{\nu)}q_\alpha - \epsilon^{(\mu\lambda\kappa\alpha}(\varepsilon_1{}_{\rho\lambda}q^\rho) \varepsilon_2^*{}_{\kappa}^{\nu)}q_\alpha \,,\\
    \left(\ytableaushort{11223,334}\,,\ytableaushort{11224,333}\right) - \text{dual} &\rightarrow (\varepsilon_2^*{}^{\rho\lambda}q_\rho q_\lambda)\epsilon^{(\mu\kappa\alpha\beta}\varepsilon_1{}_\kappa^{\nu)}P_\alpha q_\beta + (\varepsilon_1{}^{\rho\lambda}q_\rho q_\lambda)\epsilon^{(\mu\kappa\alpha\beta}\varepsilon_2^*{}_\kappa^{\nu)}P_\alpha q_\beta \,,\\
    \left(\ytableaushort{11223,444}\,,\ytableaushort{11224,344}\right) - \text{dual} &\rightarrow \epsilon^{(\mu\lambda\alpha\beta}(\varepsilon_2^*{}_{\rho\lambda}q^\rho)(\varepsilon_1{}^{\nu)\kappa}q_\kappa)P_\alpha q_\beta + \epsilon^{(\mu\lambda\alpha\beta}(\varepsilon_1{}_{\rho\lambda}q^\rho)(\varepsilon_2^*{}^{\nu)\kappa}q_\kappa)P_\alpha q_\beta \,,\\
    \left(\ytableaushort{11234,233}\,,\ytableaushort{11233,234}\right) - \text{dual} &\rightarrow \epsilon^{(\mu\kappa\alpha\beta}(\varepsilon_2^*{}^{\rho\lambda} q_\lambda)\varepsilon_1{}_{\rho\kappa}P_\alpha q_\beta P^{\nu)} - \epsilon^{(\mu\kappa\alpha\beta}(\varepsilon_1{}^{\rho\lambda} q_\lambda)\varepsilon_2^*{}_{\rho\kappa}P_\alpha q_\beta P^{\nu)} \,,\\
    \left(\ytableaushort{11234,244}\,,\ytableaushort{11244,234}\right) - \text{dual} &\rightarrow \epsilon^{(\mu\kappa\alpha\beta}(\varepsilon_2^*{}^{\rho\lambda} q_\lambda)\varepsilon_1{}_{\rho\kappa}P_\alpha q_\beta q^{\nu)} + \epsilon^{(\mu\kappa\alpha\beta}(\varepsilon_1{}^{\rho\lambda} q_\lambda)\varepsilon_2^*{}_{\rho\kappa}P_\alpha q_\beta q^{\nu)} \,,\\
    \left(\ytableaushort{11234,234}\,,\ytableaushort{11334,224}\right) - \text{dual} &\rightarrow \epsilon^{\rho\lambda\kappa\delta}\varepsilon_2^*{}_{\alpha\rho}\varepsilon_1{}^\alpha_\lambda P_\kappa q_\delta P^{(\mu}q^{\nu)}\,, \\
    \left(\ytableaushort{11223,344}\,,\ytableaushort{11233,244}\right) - \text{dual} &\rightarrow \epsilon^{(\mu\lambda\kappa\alpha}(\varepsilon_2^*{}_{\rho\lambda}q^\lambda)(\varepsilon_1{}_{\kappa\delta}q^\delta)P_\alpha P^{\nu)} \,,\\
    \left(\ytableaushort{11224,444}\,,\ytableaushort{11244,244}\right) - \text{dual} &\rightarrow \epsilon^{(\mu\lambda\kappa\alpha}(\varepsilon_2^*{}_{\rho\lambda}q^\lambda)(\varepsilon_1{}_{\kappa\delta}q^\delta)q_\alpha q^{\nu)}\,, \\
    \left(\ytableaushort{11224,344}\,,\ytableaushort{11344,224}\right) - \text{dual} &\rightarrow \epsilon^{\rho\lambda\kappa\delta}(\varepsilon_2^*{}_{\rho\alpha}q^\alpha)\varepsilon_1{}^{(\mu}_\lambda P_\kappa q_\delta q^{\nu)} + \epsilon^{\rho\lambda\kappa\delta}(\varepsilon_1{}_{\rho\alpha}q^\alpha)\varepsilon_2^*{}^{(\mu}_\lambda P_\kappa q_\delta q^{\nu)}\,,\\
    \left(\ytableaushort{11223,334}\,,\ytableaushort{11333,224}\right) - \text{dual} &\rightarrow \epsilon^{\rho\lambda\kappa\delta}(\varepsilon_2^*{}_{\rho\alpha}q^\alpha)\varepsilon_1{}^{(\mu}_\lambda P_\kappa q_\delta q^{\nu)} - \epsilon^{\rho\lambda\kappa\delta}(\varepsilon_1{}_{\rho\alpha}q^\alpha)\varepsilon_2^*{}^{(\mu}_\lambda P_\kappa q_\delta P^{\nu)} \,,\\
    \left(\ytableaushort{112234,3444}\,,\ytableaushort{112334,2444}\right) -\text{dual} &\rightarrow \epsilon^{\rho\lambda\kappa\delta}(\varepsilon_2^*{}_{\rho\alpha}q^\alpha)(\varepsilon_1{}_{\lambda\beta}q^\beta)P_\kappa q_\delta P^{(\mu}q^{\nu)} \label{eq:2-11-f}\,.
\end{align}

\subsection{Generalized FFs of Spin-3/2 Fermion}

Moving to spin-3/2 fermions, the corresponding invariant bases are constructed by incorporating the Rarita-Schwinger spinors $u_\mu$ as the constituent wave functions for the covariant representations, yielding the following complete sets of structures:

\vspace{1em}
\noindent{\large\textbf{Irrep $(0\,,0)$}}\par\nopagebreak\vspace{4pt}
\noindent\textbf{$P$-even, $T$-even:}\nopagebreak
\begin{align}
    \left(\ytableaushort{11,22}\,,\ytableaushort{1,2}\right) + \text{dual} &\rightarrow \overline{u}_2^\mu u_1{}_\mu\,, \label{eq:3/2-00-i}\,,\\
    \left(\ytableaushort{112,244}\,,\ytableaushort{12,44}\right) + \text{dual} &\rightarrow (\overline{u}_2^\mu q_\mu)(u_1{}_\nu q^\nu) .
\end{align}
\noindent\textbf{$P$-odd, $T$-even:}\nopagebreak
\begin{align}
    \left(\ytableaushort{11,22}\,,\ytableaushort{1,2}\right) - \text{dual} &\rightarrow \overline{u}_2^\mu \gamma^5 u_1{}_\mu\,, \\
    \left(\ytableaushort{112,244}\,,\ytableaushort{12,44}\right) - \text{dual} &\rightarrow (\overline{u}_2^\mu q_\mu)\gamma^5 (u_1{}_\nu q^\nu) \,, \\
    \left(\ytableaushort{113,234}\,,\ytableaushort{12,34}\right) - \text{dual} &\rightarrow \epsilon_{\mu\nu\rho\lambda}\overline{u}_2^\mu u_1^\nu P^\rho q^\lambda\,.\label{eq:3/2-00-f}
\end{align}

\vspace{1em}
\noindent{\large\textbf{Irrep $(\frac{1}{2}\,,\frac{1}{2})$}}\par\nopagebreak\vspace{4pt}
\noindent\textbf{$P$-even, $T$-even:}\nopagebreak
\begin{align}
    \left(\ytableaushort{113,22}\,,\ytableaushort{13,2}\right) + \text{dual} &\rightarrow P^\mu(\overline{u}_2^\nu u_1{}_\nu)\,, \label{eq:3/2-1/21/2-i}\\
    \left(\ytableaushort{14,2}\,,\ytableaushort{112,24}\right) + \text{dual} &\rightarrow \overline{u}_2^\nu \sigma^{\mu\rho}q_\rho u_1{}_\nu\,, \\
    \left(\ytableaushort{1123,244}\,,\ytableaushort{123,44}\right) + \text{dual} &\rightarrow P^\mu (\overline{u}_2\cdot q)(q\cdot u_1) \,, \\
    \left(\ytableaushort{124,44}\,,\ytableaushort{1212,444}\right) + \text{dual} &\rightarrow (\overline{u}_2\cdot q) \sigma^{\mu\rho}q_\rho (q\cdot u_1)\,,
\end{align}
\noindent\textbf{$P$-even, $T$-odd:}\nopagebreak
\begin{align}
    \left(\ytableaushort{114,22}\,,\ytableaushort{14,2}\right) + \text{dual} &\rightarrow q^\mu(\overline{u}_2^\nu u_2{}_\nu)\,, \\
    \left(\ytableaushort{1124,244}\,,\ytableaushort{124,44}\right) + \text{dual} &\rightarrow q^\mu (\overline{u}_2\cdot q)(q\cdot u_1) \,,
\end{align}
\noindent\textbf{$P$-odd, $T$-even:}\nopagebreak
\begin{align}
    \left(\ytableaushort{113,22}\,,\ytableaushort{13,2}\right) - \text{dual} &\rightarrow P^\mu(\overline{u}_2^\nu\gamma^5 u_1{}_\nu)\,, \\
    \left(\ytableaushort{114,22}\,,\ytableaushort{12,4}\right) - \text{dual} &\rightarrow \epsilon^{\mu\nu\rho\lambda}q_\nu\overline{u}_2{}_{\rho}u_1{}_\lambda \,, \\
    \left(\ytableaushort{14,2}\,,\ytableaushort{112,24}\right) - \text{dual} &\rightarrow \epsilon^{\mu\rho\lambda\kappa}\overline{u}_2^\nu \sigma_{\rho\lambda}q_\kappa u_1{}_\nu\,, \\
    \left(\ytableaushort{124,44}\,,\ytableaushort{1212,444}\right) - \text{dual} &\rightarrow \epsilon^{\mu\nu\rho\lambda}(\overline{u}_2\cdot q) \sigma_{\nu\rho}q_\lambda (q\cdot u_1)\,, \\
    \left(\ytableaushort{112,23}\,,\ytableaushort{12,3}\right) - \text{dual} &\rightarrow \overline{u}_2^\mu \gamma^5 (q\cdot u_1) - (\overline{u}_2\cdot q) \gamma^5 u_1^\mu\,, \\
    \left(\ytableaushort{1123,244}\,,\ytableaushort{123,44}\right) - \text{dual} &\rightarrow P^\mu (\overline{u}_2\cdot q)\gamma^5 (q\cdot u_1) \,, \\
    \left(\ytableaushort{1123,234}\,,\ytableaushort{124,33}\right) - \text{dual} &\rightarrow \epsilon^{\mu\nu\rho\lambda} (\overline{u}_2\cdot q) u_1{}_\nu P_\rho q_\lambda + \epsilon^{\mu\nu\rho\lambda} \overline{u}_2{}_\nu (q\cdot u_1) P_\rho q_\lambda\,, \\ 
    \left(\ytableaushort{1133,224}\,,\ytableaushort{123,34}\right) - \text{dual} &\rightarrow P^\mu(\epsilon^{\alpha\beta\rho\lambda} \overline{u}_2^\alpha u_1^\beta P^\rho q^\lambda) \,, \\
    \left(\ytableaushort{1122,344}\,,\ytableaushort{134,24}\right) - \text{dual} &\rightarrow \epsilon^{\alpha\beta\rho\lambda}\overline{u}_2{}_{\alpha} \sigma^{\mu\nu}q_\nu u_1{}_\beta P_\rho q_\lambda\,, \\
    \left(\ytableaushort{1122,444}\,,\ytableaushort{124,44}\right) - \text{dual} &\rightarrow \epsilon^{\mu\nu\rho\lambda}(\overline{u}_2\cdot q) \sigma_{\nu\alpha}q^\alpha (q\cdot u_1)\,,
\end{align}
\noindent\textbf{$P$-odd, $T$-odd:}\nopagebreak
\begin{align}
    \left(\ytableaushort{114,22}\,,\ytableaushort{14,2}\right) - \text{dual} &\rightarrow q^\mu(\overline{u}_2^\nu\gamma^5 u_2{}_\nu)\,, \\
    \left(\ytableaushort{113,22}\,,\ytableaushort{12,3}\right) - \text{dual} &\rightarrow \epsilon^{\mu\nu\rho\lambda}P_\nu\overline{u}_2{}_{\rho}u_1{}_\lambda \,, \\
    \left(\ytableaushort{112,24}\,,\ytableaushort{12,4}\right) - \text{dual} &\rightarrow \overline{u}_2^\mu \gamma^5 (q\cdot u_1) + (\overline{u}_2\cdot q) \gamma^5 u_1^\mu\,, \\
    \left(\ytableaushort{1124,244}\,,\ytableaushort{124,44}\right) - \text{dual} &\rightarrow q^\mu (\overline{u}_2\cdot q)\gamma^5 (q\cdot u_1) \,, \\
    \left(\ytableaushort{1123,234}\,,\ytableaushort{124,33}\right) - \text{dual} &\rightarrow \epsilon^{\mu\nu\rho\lambda} (\overline{u}_2\cdot q) u_1{}_\nu P_\rho q_\lambda - \epsilon^{\mu\nu\rho\lambda} \overline{u}_2{}_\nu (q\cdot u_1) P_\rho q_\lambda\,, \\ 
    \left(\ytableaushort{1134,224}\,,\ytableaushort{124,34}\right) - \text{dual} &\rightarrow q^\mu(\epsilon_{\alpha\beta\rho\lambda} \overline{u}_2^\alpha u_1^\beta P^\rho q^\lambda) \,.\label{eq:3/2-1/21/2-f}
\end{align}

\vspace{1em}
\noindent{\large\textbf{Irrep $(1\,,0)\oplus (0\,,1)$}}\par\nopagebreak\vspace{4pt}
\noindent\textbf{$P$-even, $T$-even:}\nopagebreak
\begin{align}
    \left(\ytableaushort{1123,24}\,,\ytableaushort{13,24}\right) + \text{dual} &\rightarrow P^{[\mu}(\overline{u}_2^\rho \sigma^{\nu]\lambda}q_\lambda u_1{}_\rho) \,, \label{eq:3/2-1001-i}\,,\\
    \left(\ytableaushort{1123,23}\,,\ytableaushort{12,33}\right) + \text{dual} &\rightarrow P^{[\mu}(\overline{u}_2\cdot q) u_1^{\nu]} - P^{[\mu}\overline{u}_2^{\nu]}(q\cdot u_1)\,, \\
    \left(\ytableaushort{1124,24}\,,\ytableaushort{12,44}\right) + \text{dual} &\rightarrow q^{[\mu}(\overline{u}_2\cdot q) u_1^{\nu]} + q^{[\mu}\overline{u}_2^{\nu]}(q\cdot u_1)\,, \\
    \left(\ytableaushort{12124,343}\,,\ytableaushort{123,344}\right) + \text{dual} &\rightarrow P^{[\mu}(\overline{u}_2\cdot q) \sigma^{\nu]\rho}q_\rho (q\cdot u_1)\,.
\end{align}
\noindent\textbf{$P$-odd, $T$-even:}\nopagebreak
\begin{align}
    \left(\ytableaushort{1123,24}\,,\ytableaushort{13,24}\right) - \text{dual} &\rightarrow \epsilon^{\mu\nu\alpha\beta}P_\alpha(\overline{u}_2^\rho \sigma_{\beta\lambda}q^\lambda u_1{}_\rho) \,, \\
    \left(\ytableaushort{1123,23}\,,\ytableaushort{12,33}\right) - \text{dual} &\rightarrow \epsilon^{\mu\nu\rho\lambda}(P_\rho(\overline{u}_2\cdot q) u_1{}_\lambda - P_\rho\overline{u}_2{}_\lambda(q\cdot u_1))\,, \\
    \left(\ytableaushort{1124,24}\,,\ytableaushort{12,44}\right) - \text{dual} &\rightarrow \epsilon^{\mu\nu\rho\lambda}(q_\rho(\overline{u}_2\cdot q) u_1{}_\lambda - q_\rho\overline{u}_2{}_\lambda(q\cdot u_1))\,, \\
    \left(\ytableaushort{12123,344}\,,\ytableaushort{133,244}\right) - \text{dual} &\rightarrow P^{[\mu}u_2^\alpha \sigma^{\nu]\rho}q_\rho u_1^\beta P^\kappa q^\lambda \epsilon_{\alpha\beta\kappa\lambda}\,, \\
    \left(\ytableaushort{12124,343}\,,\ytableaushort{123,344}\right) - \text{dual} &\rightarrow \epsilon^{\mu\nu\kappa\lambda}P_\rho(\overline{u}_2\cdot q) \sigma_{\kappa\rho}q^\rho (q\cdot u_1)\,.
\end{align}
\noindent\textbf{$P$-even, $T$-odd:}\nopagebreak
\begin{align}
    \left(\ytableaushort{12}\,,\ytableaushort{11,22}\right) + \text{dual} &\rightarrow \overline{u}_1^{[\mu}u_1^{\nu]} \,, \\
    \left(\ytableaushort{112,2}\,,\ytableaushort{1,2}\right) + \text{dual} &\rightarrow \overline{u}_2^\rho \sigma^{\mu\nu} u_1{}_\rho \,, \\
    \left(\ytableaushort{1134,22}\,,\ytableaushort{13,24}\right) + \text{dual} &\rightarrow P^{[\mu}q^{\nu]}(\overline{u}_2^\rho u_1{}_\rho) + \epsilon^{\mu\nu\alpha\beta}P_\alpha q_\beta(\overline{u}_2^\rho \gamma^5 u_1{}_\rho)\,, \\
    \left(\ytableaushort{134,2}\,,\ytableaushort{112,224}\right) + \text{dual} &\rightarrow P^{[\mu}q^{\nu]}(\overline{u}_2^\rho u_1{}_\rho) -  \epsilon^{\mu\nu\alpha\beta}P_\alpha q_\beta(\overline{u}_2^\rho \gamma^5 u_1{}_\rho)\,, \\
    \left(\ytableaushort{144,2}\,,\ytableaushort{112,244}\right) + \text{dual} &\rightarrow q^{[\mu}(\overline{u}_2^\rho \sigma^{\nu]\lambda}q_\lambda u_1{}_\rho)\,, \\ 
    \left(\ytableaushort{1123,24}\,,\ytableaushort{12,34}\right) + \text{dual} &\rightarrow q^{[\mu}(\overline{u}_2\cdot q) u_1^{\nu]} - q^{[\mu}\overline{u}_2^{\nu]}(q\cdot u_1)\,, \\
    \left(\ytableaushort{1124,23}\,,\ytableaushort{12,34}\right) + \text{dual} &\rightarrow P^{[\mu}(\overline{u}_2\cdot q) u_1^{\nu]} + P^{[\mu}\overline{u}_2^{\nu]}(q\cdot u_1)\,, \\
    \left(\ytableaushort{11234,244}\,,\ytableaushort{123,444}\right) + \text{dual} &\rightarrow P^{[\mu}q^{\nu]}(\overline{u}_2\cdot q)(q\cdot u_1) + \epsilon^{\mu\nu\rho\lambda}P_\rho q_\lambda (\overline{u}_2\cdot q)\gamma^5 (q\cdot u_1)\,, \\
    \left(\ytableaushort{1234,44}\,,\ytableaushort{1123,2444}\right) + \text{dual} &\rightarrow P^{[\mu}q^{\nu]}(\overline{u}_2\cdot q)(q\cdot u_1) - \epsilon^{\mu\nu\rho\lambda}P_\rho q_\lambda (\overline{u}_2\cdot q)\gamma^5 (q\cdot u_1)\,, \\
    \left(\ytableaushort{1122,34}\,,\ytableaushort{12,34}\right) + \text{dual} &\rightarrow (\overline{u}_2\cdot q)\sigma^{\mu\nu}(q\cdot u_1)\,.
\end{align}
\noindent\textbf{$P$-odd, $T$-odd:}\nopagebreak
\begin{align}
    \left(\ytableaushort{12}\,,\ytableaushort{11,22}\right) - \text{dual} &\rightarrow \epsilon^{\mu\nu\rho\lambda}\overline{u}_1{}_\rho u_1{}_\lambda \,, \\
    \left(\ytableaushort{112,2}\,,\ytableaushort{1,2}\right) - \text{dual} &\rightarrow \epsilon^{\mu\nu\alpha\beta}\overline{u}_2^\rho \sigma_{\alpha\beta} u_1{}_\rho \,, \\
    \left(\ytableaushort{1134,22}\,,\ytableaushort{13,24}\right) - \text{dual} &\rightarrow \epsilon^{\mu\nu\alpha\beta}P_{\alpha}q_{\beta}(\overline{u}_2^\rho u_1{}_\rho) + P^{[\mu} q^{\nu]}(\overline{u}_2^\rho \gamma^5 u_1{}_\rho)\,, \\
    \left(\ytableaushort{134,2}\,,\ytableaushort{112,224}\right) - \text{dual} &\rightarrow \epsilon^{\mu\nu\alpha\beta}P_{\alpha}q_{\beta}(\overline{u}_2^\rho u_1{}_\rho) - P^{[\mu} q^{\nu]}(\overline{u}_2^\rho \gamma^5 u_1{}_\rho)\,, \\ 
    \left(\ytableaushort{144,2}\,,\ytableaushort{112,244}\right) - \text{dual} &\rightarrow \epsilon^{\mu\nu\alpha\beta}q_\alpha(\overline{u}_2^\rho \sigma_{\beta\lambda}q^\lambda u_1{}_\rho)\,, \\ 
    \left(\ytableaushort{1123,24}\,,\ytableaushort{12,34}\right) - \text{dual} &\rightarrow \epsilon^{\mu\nu\rho\lambda}(q_\rho(\overline{u}_2\cdot q) u_1{}_\lambda - q_\rho\overline{u}_2{}_\lambda(q\cdot u_1))\,, \\
    \left(\ytableaushort{1124,23}\,,\ytableaushort{12,34}\right) - \text{dual} &\rightarrow \epsilon^{\mu\nu\rho\lambda}(P_\rho(\overline{u}_2\cdot q) u_1{}_\lambda + P_\rho\overline{u}_2{}_\lambda(q\cdot u_1))\,, \\
    \left(\ytableaushort{1122,34}\,,\ytableaushort{13,24}\right) - \text{dual} &\rightarrow \epsilon_{\rho\lambda\alpha\beta}P^\rho q^\lambda (\overline{u}_2^\alpha \sigma^{\mu\nu} u_1^\beta)\,, \\
    \left(\ytableaushort{11234,234}\,,\ytableaushort{133,244}\right) - \text{dual} &\rightarrow P^{[\mu}q^{\nu]}(\epsilon_{\alpha\beta\rho\lambda}\overline{u}_2^\alpha u_1^\beta P^\rho q^\lambda) + P^{[\mu}q^{\nu]}(\overline{u}_2\cdot q)\gamma^5 (q\cdot u_1) \,,\\
    \left(\ytableaushort{1234,34}\,,\ytableaushort{1133,2244}\right) - \text{dual} &\rightarrow P^{[\mu}q^{\nu]}(\epsilon_{\alpha\beta\rho\lambda}\overline{u}_2^\alpha u_1^\beta P^\rho q^\lambda) - P^{[\mu}q^{\nu]}(\overline{u}_2\cdot q)\gamma^5 (q\cdot u_1) \,,\\
    \left(\ytableaushort{11234,244}\,,\ytableaushort{123,444}\right) - \text{dual} &\rightarrow \epsilon^{\mu\nu\rho\lambda}P_\rho q_\lambda (\overline{u}_2\cdot q)(q\cdot u_1) + P^{[\mu} q^{\nu]} (\overline{u}_2\cdot q)\gamma^5 (q\cdot u_1)\,, \\
    \left(\ytableaushort{1234,44}\,,\ytableaushort{1123,2444}\right) - \text{dual} &\rightarrow \epsilon^{\mu\nu\rho\lambda}P_\rho q_\lambda (\overline{u}_2\cdot q)(q\cdot u_1) - P^{[\mu} q^{\nu]} (\overline{u}_2\cdot q)\gamma^5 (q\cdot u_1)\,, \\
    \left(\ytableaushort{1122,34}\,,\ytableaushort{12,34}\right) - \text{dual} &\rightarrow \epsilon^{\mu\nu\rho\lambda}(\overline{u}_2\cdot q)\sigma_{\rho\lambda}(q\cdot u_1)\,, \\
    \left(\ytableaushort{1344,24}\,,\ytableaushort{1212,3444}\right) - \text{dual} &\rightarrow q^{[\mu}\overline{u}_2^\alpha \sigma^{\nu]\rho}q_\rho u_1^\beta P^\kappa p^\lambda \epsilon_{\alpha\beta\kappa\lambda}\,.\label{eq:3/2-1001-f}
\end{align}

\vspace{1em}
\noindent{\large\textbf{Irrep $(1\,,1)$}}\par\nopagebreak\vspace{4pt}
\noindent\textbf{$P$-even, $T$-even:}\nopagebreak
\begin{align}
    \left(\ytableaushort{112,2}\,,\ytableaushort{12}\right) + \text{dual} &\rightarrow \overline{u}_2^{(\mu}u_1^{\nu)} \,, \label{eq:3/2-11-i}\\
    \left(\ytableaushort{1133,22}\,,\ytableaushort{133,2}\right) + \text{dual} &\rightarrow P^{(\mu}P^{\nu)}(\overline{u}_2^\rho u_1{}_\rho) \,, \\
    \left(\ytableaushort{1144,22}\,,\ytableaushort{144,2}\right) + \text{dual} &\rightarrow q^{(\mu}q^{\nu)}(\overline{u}_2^\rho u_1{}_\rho) \,, \\
    \left(\ytableaushort{134,2}\,,\ytableaushort{1123,24}\right) + \text{dual} &\rightarrow P^{(\mu}\overline{u}_2^\rho \sigma^{\nu)\lambda}q_\lambda u_1{}_\rho\,, \\
    \left(\ytableaushort{1124,24}\,,\ytableaushort{124,4}\right) + \text{dual} &\rightarrow (\overline{u}_2\cdot q) u_1^{(\mu}q^{\nu)} + \overline{u}_2^{(\mu}(q\cdot u_1) q^{\nu)}\,, \\
    \left(\ytableaushort{1123,23}\,,\ytableaushort{123,3}\right) + \text{dual} &\rightarrow (\overline{u}_2\cdot q) u_1^{(\mu}P^{\nu)} - \overline{u}_2^{(\mu}(q\cdot u_1) P^{\nu)}\,, \\
    \left(\ytableaushort{11233,244}\,,\ytableaushort{1233,44}\right) + \text{dual} &\rightarrow (\overline{u}_2\cdot q)(q\cdot u_1) P^{(\mu}P^{\nu)}\,, \\
    \left(\ytableaushort{11244,244}\,,\ytableaushort{1244,44}\right) + \text{dual} &\rightarrow (\overline{u}_2\cdot q)(q\cdot u_1) q^{(\mu}q^{\nu)}\,.
\end{align}
\noindent\textbf{$P$-odd, $T$-even:}\nopagebreak
\begin{align}
    \left(\ytableaushort{112,2}\,,\ytableaushort{12}\right) - \text{dual} &\rightarrow \overline{u}_2^{(\mu}\gamma^5 u_1^{\nu)} \,, \\
    \left(\ytableaushort{1133,22}\,,\ytableaushort{133,2}\right) - \text{dual} &\rightarrow P^{(\mu}P^{\nu)}(\overline{u}_2^\rho \gamma^5 u_1{}_\rho) \,, \\
    \left(\ytableaushort{1144,22}\,,\ytableaushort{144,2}\right) - \text{dual} &\rightarrow q^{(\mu}q^{\nu)}(\overline{u}_2^\rho \gamma^5 u_1{}_\rho) \,, \\
    \left(\ytableaushort{134,2}\,,\ytableaushort{1123,24}\right) - \text{dual} &\rightarrow P^{(\mu}\epsilon^{\nu)\lambda\alpha\beta}\overline{u}_2^\rho \sigma_{\lambda\alpha}q_\beta u_1{}_\rho\,, \\
    \left(\ytableaushort{1124,24}\,,\ytableaushort{124,4}\right) - \text{dual} &\rightarrow (\overline{u}_2\cdot q) \gamma^5 u_1^{(\mu}q^{\nu)} + \overline{u}_2^{(\mu}\gamma^5(q\cdot u_1) q^{\nu)}\,, \\
    \left(\ytableaushort{1123,23}\,,\ytableaushort{123,3}\right) - \text{dual} &\rightarrow (\overline{u}_2\cdot q)\gamma^5 u_1^{(\mu}P^{\nu)} - \overline{u}_2^{(\mu}\gamma^5 (q\cdot u_1) P^{\nu)}\,, \\
    \left(\ytableaushort{11333,224}\,,\ytableaushort{1233,34}\right) - \text{dual} &\rightarrow P^{(\mu}P^{\nu)}(\epsilon_{\rho\lambda\alpha\beta}\overline{u}_2^\rho u_1^\lambda P^\alpha q^\beta ) + P^{(\mu}P^{\nu)}(\overline{u}_2\cdot q)\gamma^5 (q\cdot u_1) \,,\\
    \left(\ytableaushort{11344,224}\,,\ytableaushort{1244,34}\right) - \text{dual} &\rightarrow q^{(\mu}q^{\nu)}(\epsilon_{\rho\lambda\alpha\beta}\overline{u}_2^\rho u_1^\lambda P^\alpha q^\beta ) + q^{(\mu}q^{\nu)}(\overline{u}_2\cdot q)\gamma^5 (q\cdot u_1) \,,\\
    \left(\ytableaushort{1333,24}\,,\ytableaushort{11233,234}\right) - \text{dual} &\rightarrow P^{(\mu}P^{\nu)}(\epsilon_{\rho\lambda\alpha\beta}\overline{u}_2^\rho u_1^\lambda P^\alpha q^\beta ) - P^{(\mu}P^{\nu)}(\overline{u}_2\cdot q)\gamma^5 (q\cdot u_1) \,,\\
    \left(\ytableaushort{1344,24}\,,\ytableaushort{11244,234}\right) - \text{dual} &\rightarrow q^{(\mu}q^{\nu)}(\epsilon_{\rho\lambda\alpha\beta}\overline{u}_2^\rho u_1^\lambda P^\alpha q^\beta ) - q^{(\mu}q^{\nu)}(\overline{u}_2\cdot q)\gamma^5 (q\cdot u_1)\,,\\
    \left(\ytableaushort{1134,22}\,,\ytableaushort{123,4}\right) - \text{dual} &\rightarrow \epsilon^{(\mu\rho\lambda\kappa}\overline{u}_2{}_\rho u_1{}_\lambda q_\kappa P^{\nu)}\,, \\
    \left(\ytableaushort{1134,22}\,,\ytableaushort{124,3}\right) - \text{dual} &\rightarrow \epsilon^{(\mu\rho\lambda\kappa}\overline{u}_2{}_\rho u_1{}_\lambda P_\kappa q^{\nu)}\,, \\
    \left(\ytableaushort{11233,244}\,,\ytableaushort{1233,44}\right) - \text{dual} &\rightarrow (\overline{u}_2\cdot q)\gamma^5(q\cdot u_1) P^{(\mu}P^{\nu)}\,, \\
    \left(\ytableaushort{11244,244}\,,\ytableaushort{1244,44}\right) - \text{dual} &\rightarrow (\overline{u}_2\cdot q)\gamma^5(q\cdot u_1) q^{(\mu}q^{\nu)}\,, \\
    \left(\ytableaushort{11234,234}\,,\ytableaushort{1233,44}\right) - \text{dual} &\rightarrow \epsilon^{(\mu\rho\lambda\kappa}(\overline{u}_2\cdot q)u_1{}_\rho P_\lambda q_\kappa P^{\nu)} + \epsilon^{(\mu\rho\lambda\kappa}\overline{u}_2{}_\rho (q\cdot u_1)P_\lambda q_\kappa P^{\nu)} \,, \\
    \left(\ytableaushort{11244,233}\,,\ytableaushort{1233,44}\right) - \text{dual} &\rightarrow \epsilon^{(\mu\rho\lambda\kappa}(\overline{u}_2\cdot q)u_1{}_\rho P_\lambda q_\kappa q^{\nu)} - \epsilon^{(\mu\rho\lambda\kappa}\overline{u}_2{}_\rho (q\cdot u_1)P_\lambda q_\kappa q^{\nu)} \,, \\
    \left(\ytableaushort{11223,344}\,,\ytableaushort{1344,24}\right) - \text{dual} &\rightarrow \epsilon_{\alpha\beta\rho\lambda}\overline{u}_2^\alpha \sigma^{(\mu\kappa}q_\kappa u_1^\beta P^\rho q^\lambda P^{\nu)}\,.
\end{align}
\noindent\textbf{$P$-even, $T$-odd:}\nopagebreak
\begin{align}
    \left(\ytableaushort{1134,22}\,,\ytableaushort{134,2}\right) + \text{dual} &\rightarrow P^{(\mu}q^{\nu)}(\overline{u}_2^\rho u_1{}_\rho) \,, \\
    \left(\ytableaushort{144,2}\,,\ytableaushort{1124,24}\right) + \text{dual} &\rightarrow q^{(\mu}\overline{u}_2^\rho \sigma^{\nu)\lambda}q_\lambda u_1{}_\rho\,, \\
    \left(\ytableaushort{1123,24}\,,\ytableaushort{123,4}\right) + \text{dual} &\rightarrow (\overline{u}_2\cdot q) u_1^{(\mu}P^{\nu)} + \overline{u}_2^{(\mu}(q\cdot u_1) P^{\nu)}\,, \\
    \left(\ytableaushort{1124,23}\,,\ytableaushort{124,3}\right) + \text{dual} &\rightarrow (\overline{u}_2\cdot q) u_1^{(\mu}q^{\nu)} - \overline{u}_2^{(\mu}(q\cdot u_1) q^{\nu)}\,, \\
    \left(\ytableaushort{11234,244}\,,\ytableaushort{1234,44}\right) + \text{dual} &\rightarrow (\overline{u}_2\cdot q)(q\cdot u_1) P^{(\mu}q^{\nu)}\,.
\end{align}
\noindent\textbf{$P$-odd, $T$-odd:}\nopagebreak
\begin{align}
    \left(\ytableaushort{1134,22}\,,\ytableaushort{134,2}\right) - \text{dual} &\rightarrow P^{(\mu}q^{\nu)}(\overline{u}_2^\rho \gamma^5 u_1{}_\rho) \,, \\
    \left(\ytableaushort{144,2}\,,\ytableaushort{1124,24}\right) - \text{dual} &\rightarrow q^{(\mu}\epsilon^{\nu\lambda\alpha\beta}\overline{u}_2^\rho \sigma_{\lambda\alpha}q_\beta u_1{}_\rho\,, \\
    \left(\ytableaushort{1123,24}\,,\ytableaushort{123,4}\right) - \text{dual} &\rightarrow (\overline{u}_2\cdot q) \gamma^5 u_1^{(\mu}P^{\nu)} + \overline{u}_2^{(\mu}\gamma^5 (q\cdot u_1) P^{\nu)}\,, \\
    \left(\ytableaushort{1124,23}\,,\ytableaushort{124,3}\right) - \text{dual} &\rightarrow (\overline{u}_2\cdot q) \gamma^5 u_1^{(\mu}q^{\nu)} - \overline{u}_2^{(\mu}\gamma^5 (q\cdot u_1) q^{\nu)}\,, \\
    \left(\ytableaushort{1123,24}\,,\ytableaushort{124,3}\right) - \text{dual} &\rightarrow \epsilon^{(\mu\rho\kappa\lambda} \overline{u}_2{}_\rho u_2^{\nu)}P_\kappa q_\lambda + \epsilon^{(\mu\rho\kappa\lambda} \overline{u}_2^{\nu)}u_1{}_\rho P_\kappa q_\lambda u_\rho\,,\\ 
    \left(\ytableaushort{11334,224}\,,\ytableaushort{1234,34}\right) - \text{dual} &\rightarrow P^{(\mu}q^{\nu)}(\epsilon_{\rho\lambda\alpha\beta}\overline{u}_2^\rho u_1^\lambda P^\alpha q^\beta ) + P^{(\mu}q^{\nu)}(\overline{u}_2\cdot q)\gamma^5 (q\cdot u_1) \,,\\
    \left(\ytableaushort{1334,24}\,,\ytableaushort{11234,234}\right) - \text{dual} &\rightarrow P^{(\mu}q^{\nu)}(\epsilon_{\rho\lambda\alpha\beta}\overline{u}_2^\rho u_1^\lambda P^\alpha q^\beta ) - P^{(\mu}q^{\nu)}(\overline{u}_2\cdot q)\gamma^5 (q\cdot u_1)\,,\\
    \left(\ytableaushort{11234,244}\,,\ytableaushort{1234,44}\right) - \text{dual} &\rightarrow (\overline{u}_2\cdot q)\gamma^5(q\cdot u_1) P^{(\mu}q^{\nu)}\,, \\
    \left(\ytableaushort{11234,233}\,,\ytableaushort{1233,34}\right) - \text{dual} &\rightarrow \epsilon^{(\mu\rho\lambda\kappa}(\overline{u}_2\cdot q)u_1{}_\rho P_\lambda q_\kappa q^{\nu)} + \epsilon^{(\mu\rho\lambda\kappa}\overline{u}_2{}_\rho (q\cdot u_1)P_\lambda q_\kappa q^{\nu)} \,, \\
    \left(\ytableaushort{11244,233}\,,\ytableaushort{1234,44}\right) - \text{dual} &\rightarrow \epsilon^{(\mu\rho\lambda\kappa}(\overline{u}_2\cdot q)u_1{}_\rho P_\lambda q_\kappa P^{\nu)} - \epsilon^{(\mu\rho\lambda\kappa}\overline{u}_2{}_\rho (q\cdot u_1)P_\lambda q_\kappa P^{\nu)} \,, \\
    \left(\ytableaushort{11224,344}\,,\ytableaushort{1344,24}\right) - \text{dual} &\rightarrow \epsilon_{\alpha\beta\rho\lambda}\overline{u}_2^\alpha \sigma^{(\mu\kappa}q_\kappa u_1^\beta P^\rho q^\lambda q^{\nu)}\,.
    \label{eq:3/2-11-f}
\end{align}

\section{Nonlocal Form Factors}
\label{sec:nonlocal}
Nonlocal FFs serve as a fundamental framework for parameterizing the matrix elements of nonlocal operators. These operators capture the emission and absorption of fields at arbitrary separated spacetime points, say $z_1$ and $z_2$. By exploiting translational invariance, any two such spacetime points can be generically parameterized relative to an origin as $z_1 = 0$ and $z_2 = x$, where $x^\mu$ defines the four-dimensional spacetime separation vector (direction). 
To make this parameterization explicit, the matrix element of a generic nonlocal operator $\mathcal{O}^\Gamma(0, x)$ evaluated between initial and final physical states is systematically expanded as a sum of invariant scalar functions multiplied by covariant tensor structures:
\begin{align}
    \langle p_2, s_2 | \mathcal{O}^\Gamma(0, x) | p_1, s_1 \rangle = \sum_i \mathcal{F}_i(x^2, x \cdot P, x \cdot \Delta, \dots) \, \mathcal{T}_i^\Gamma(x, P, \Delta, s_1, s_2)\,,
\end{align}
where $\mathcal{F}_i$ represent the scalar nonlocal FFs that encapsulate the underlying interaction dynamics of the specific quantum field theory. Conversely, $\mathcal{T}_i^\Gamma$ are the universally applicable and completely model-independent Lorentz covariant bases constructed from external momenta ($p = p_1 + p_2$, $\Delta = p_2 - p_1$) and wave functions of initial and final states. 

\subsection{Decomposition of nonlocal operators}
To rigorously classify the physical degrees of freedom within such a nonlocal operator, one must decompose it according to its twist $\tau$, fundamentally defined as the difference between the operator's canonical dimension $d$ and its Lorentz spin $j$~\cite{Gross:1971wn}:
\begin{align}
    \text{twist } (\tau) = \text{canonical dimension } (d) - \text{Lorentz spin } (j)\,.
\end{align}
Where the Lorentz spin in this equation for an operator transforming under the irreducible Lorentz representation $(j_L, j_R)$ is given by:
\begin{align}
    j = j_L + j_R \quad \,.
\end{align}
Physically, higher-twist contributions ($\tau \ge 3$) are dynamically suppressed by inverse powers of the hard momentum transfer scale $Q$ in the deep inelastic limit, but remain essential at moderate $Q^2$. 

Equipped with the definition and the basic properties of twist above, the twist decomposition and tensor reduction for nonlocal operators can be explicitly illustrated. Firstly, the Taylor expansion of a general bilocal operator  $\mathcal{O}^\Gamma(0, x) = \Phi^\dagger(0) \Gamma U(0, x) \Phi( x)$ at the origin is given by:
\begin{align}
    \mathcal{O}^\Gamma(0, x)
    &= \sum_{n=0}^\infty \frac{1}{n!}\,
    x^{\mu_1}\cdots x^{\mu_n}
    \left[
    \Phi^\dagger(0)\,\Gamma\,
    \overset{\leftrightarrow}{D}_{\mu_1}\cdots
    \overset{\leftrightarrow}{D}_{\mu_n}\,
    \Phi(0)
    \right]
    \nonumber\\
    &\equiv
    \sum_{n=0}^\infty \frac{1}{n!}\,
    \mathcal{O}_{\Gamma n}(x)\,.
    \label{Tayor}
\end{align}
When $\Phi$ is taken to be a Dirac field, the matrix $\Gamma$ runs over the set $\Gamma=\{1,\gamma^\mu,\sigma^{\mu\nu},\gamma_5\gamma^\mu,\gamma_5\}$, where $\sigma^{\mu\nu}=\frac{i}{2}[\gamma^\mu,\gamma^\nu]$. Here, contracting the local tensor operator with the completely symmetric coordinate product $x^{\mu_1} \dots x^{\mu_n}$ inherently acts as a natural total-symmetrization projector. Since the tensor structure of the local operator $\Phi^\dagger(0)\,\Gamma\,
    \overset{\leftrightarrow}{D}_{(\mu_1}\cdots
    \overset{\leftrightarrow}{D}_{\mu_n)}\,
    \Phi(0)$ are uniquely determined by $\Gamma$ after the indexes $\mu_1\dots \mu_n$ are totally symmetrized, we adopt the lower index $\Gamma$ of $\mathcal{O}_{\Gamma n}(x)$ to label the relevant tensor structure. The local tensor operators generated by the Taylor expansion form finite-dimensional, but generally reducible, representations of the orthochronous Lorentz group $SO(1, 3)$. To guarantee that the extracted local operators are irreducible, it must simultaneously satisfy two differential constraints~\cite{Geyer:1999uq}:
\begin{align}
    \Box \mathcal{O}_{\alpha n}^{\circ \Gamma}(x) &= 0\,, \\
    \partial^\alpha \mathcal{O}_{\alpha n}^{\circ \Gamma}(x) &= 0\,,
\end{align}
where 
\begin{align}
\mathcal{O}_{\alpha n}^{\circ \Gamma}(x)=\frac{1}{n!}\,
    x^{\mu_1}\cdots x^{\mu_n}
{\left[\Phi^\dagger(0)\,\Gamma_{\alpha}\,
    \overset{\leftrightarrow}{D}_{\mu_1}\cdots
    \overset{\leftrightarrow}{D}_{\mu_n}\,
    \Phi(0)-\text{trace terms}\right]}\,.
\end{align}
In practice, the trace terms of $\Phi^\dagger(0)\,\Gamma_{\alpha}\,
    \overset{\leftrightarrow}{D}_{\mu_1}\cdots
    \overset{\leftrightarrow}{D}_{\mu_n}\,
    \Phi(0)$ containing the metric tensor $g_{\mu\nu}$, contracting with $x^{\mu_1}\cdots x^{\mu_n}$ , have no contribution to $\mathcal{O}^\Gamma(0, x)$ upon light-cone projection ($x^2 \to 0$). At the same time, on projection onto the light-cone ($x \to \tilde{x}$), where $\tilde{x}^2=0$, only a finite number of twist contributions are left. Subsequently, we can perform the decomposition of the local operator by tensor reduction and twist decomposition:
\begin{align}
    \mathcal{O}^{\circ}_{\Gamma n}(x) = \bigoplus_{\tau} c_{n\Gamma }^{(\tau)\Gamma' }(x) \mathcal{O}_{\Gamma' n}^{\circ(\tau)}(x)\,,
\end{align}
where the lower index $\Gamma' $ represents the specific irreducible tensor structure of the local operator contained in $\mathcal{O}_{\Gamma' n}^{\circ(\tau)}(x)$. Finally, $\mathcal{O}_{\Gamma' n'}^{\circ(\tau)}(x) $ can be expanded in the complete basis constructed by the methods introduced by us in previous sections.
 Substituting the fully decomposed operator back into the Taylor expansion, we obtain:
\begin{align}
    \langle p_2, s_2 | \mathcal{O}^\Gamma(0, \tilde{x}) | p_1, s_1 \rangle = \sum_{n=0}^\infty \frac{1}{n!} \bigoplus_{\tau} c_{n\Gamma }^{(\tau)\Gamma' }(x) \, \tilde{x}^{\mu_1} \dots \tilde{x}^{\mu_{n}} \left[ \sum_i \mathcal{F}_{i, n,\Gamma'}^{(\tau)} \cdot \mathcal{T}_{\Gamma' \mu_1 \dots \mu_{n}}^{(i)} \right]\,,
\end{align}
where $\sum_i \mathcal{F}_{i, n,\Gamma'}^{(\tau)} \cdot \mathcal{T}_{\Gamma' \mu_1 \dots \mu_{n}}^{(i)} $ represents the matrix element of the local operator that is a part of 
\begin{align}
\left[\Phi^\dagger(0)\,\Gamma_{\alpha}\,
    \overset{\leftrightarrow}{D}_{\mu_1}\cdots
    \overset{\leftrightarrow}{D}_{\mu_n}\,
    \Phi(0)-\text{trace terms}\right]
 \end{align}
    with definite twist $\tau$ and $\mathcal{T}_{\Gamma' \mu_1 \dots \mu_{n'}}^{(i)} $ is used to represent the bases constructed in our framework, multiplied by FFs $f_i$ formed from external kinematic variables. 
The geometric twist $\tau$ of each term is determined straightforwardly by:
\begin{align}
    \tau \left[ \mathcal{F}_{i, n,\Gamma'} \cdot \mathcal{T}_{\Gamma' \mu_1 \dots \mu_{n}}^{(i)} \right] &= \text{dim}\left[\mathcal{F}_{i, n,\Gamma'}\right] + \text{dim}\left[\mathcal{T}_{\Gamma' \mu_1 \dots \mu_{n}}^{(i)}\right] - j\left[\mathcal{T}_{\Gamma' \mu_1 \dots \mu_{n}}^{(i)}\right]\,.
\end{align}
 It is worth mentioning that the matrix elements of the trace terms can also be expanded in our basis if needed, which act as the redundancies in the construction of the basis of its matrix elements because their Lorentz representations are equivalent to some traceless local operators with lower-rank  tensor structures from Taylor expansion in Eq (\ref{Tayor}).
\subsection{Tensor Reduction of Generated Local Operators}
To systematically apply our construction and results to nonlocal matrix elements, we will next introduce the tensor reduction of operators containing multiple gauge covariant derivatives.

Building upon our established achievement that, given the initial and final state particle information and the irrep $(j_L, j_R)$ of any matrix element or operator, we can systematically construct all Lorentz representation structures, we now detail how this group-theoretical foundation rigorously defines the aforementioned covariant bases. Specifically, when a local tensor $T$ and $n$ covariant derivatives are combined via a tensor product, the representation decomposes into a direct sum of irreps:
\begin{align}
    T \otimes D^{\mu_1}D^{\mu_2}\dots D^{\mu_n} \sim (j_L,j_R)\otimes \left(\frac{n}{2},\frac{n}{2}\right) = \bigoplus (j_1,j_2)\,,
\end{align}
where $|j_L-n/2|\leq j_1 \leq j_L+n/2$ and $|j_R-n/2|\leq j_2 \leq j_R+n/2$. All the allowed representations compose a matrix:
\begin{align}
    \mathcal{M} = \begin{pmatrix}
    (\frac{n}{2}+j_L,\frac{n}{2}+j_R) & \dots & (\frac{n}{2}+j_L, |j_R-\frac{n}{2}|) \\
    \vdots & \ddots & \vdots \\
    (|\frac{n}{2}-j_L|,\frac{n}{2}+j_R) & \dots & (|j_L-\frac{n}{2}|, |j_R-\frac{n}{2}|)
    \end{pmatrix}\,.
\end{align}
Considering $j_L=j_R$, the matrix $\mathcal{M}$ is a square matrix. The elements on its main diagonal are of the form:
\begin{align}
    (J,J)\,,\quad (J-1,J-1)\,,\quad (J-2,J-2)\,,\dots
\end{align}
all of which but the first one are redundant. We know that the number of Lorentz indices of a representation $(j_L,j_R)$ is $2\times \text{Max}(j_L,j_R)$, so the main diagonal elements but the first correspond to tensors with fewer Lorentz indices. This means there are internal contractions among the derivatives and the tensor $T$, explicitly shown as:
\begin{align}
    T^{\mu\dots} \sigma_\mu{}_{a\dot{b}} D^\nu\sigma_{\nu}{}_{c\dot{d}} \epsilon^{ac}\epsilon^{\dot{b}\dot{d}} \propto T^{\mu\dots} D_\mu\,.
\end{align}
For the off-diagonal elements, only the ones in the first row (or column) should be reserved because of a similar argument, and we can only consider half of them since the matrix $\mathcal{M}$ is symmetrical. Applying this to the $(0,0)$ representation, there are two irreducible tensors:
\begin{align}
    \overline{\psi}\psi\,, \quad \overline{\psi}\gamma^5 \psi \in (0,0)\,.
\end{align}
We evaluate the tensor product:
\begin{align}
    (0,0)\otimes \left(\frac{n}{2}, \frac{n}{2}\right) = \left(\frac{n}{2}, \frac{n}{2}\right)\,,
\end{align}
meaning the matrix $\mathcal{M}$ is a $1\times 1$ matrix with a single element. Thus, there is only one irrep of the terms in the expansion of scalar FFs, corresponding to which the $SO(4)$ tensor SYD is:
\begin{align}
    \underbrace{\ydiagram{2}\dots\ydiagram{1}}_{n}\,.
\end{align}
For the $(\frac{1}{2},\frac{1}{2})$ representation, the matrix $\mathcal{M}$ is a $2\times 2$ matrix:
\begin{align}
    \begin{pmatrix}
        (\frac{n+1}{2},\frac{n+1}{2}) & (\frac{n+1}{2},\frac{n-1}{2}) \\
        (\frac{n-1}{2},\frac{n+1}{2}) & (\frac{n-1}{2},\frac{n-1}{2})
    \end{pmatrix}\,,
\end{align}
where $(\frac{n-1}{2},\frac{n-1}{2})$ is redundant, $(\frac{n+1}{2},\frac{n+1}{2})$ corresponds to tensors totally symmetrical, and $(\frac{n+1}{2},\frac{n-1}{2})$ and $(\frac{n-1}{2},\frac{n+1}{2})$ correspond to tensors with a pair of asymmetrical indices. The corresponding $SO(4)$ SYDs are:
\begin{align}
    (\frac{n+1}{2},\frac{n+1}{2}): & \quad \underbrace{\ydiagram{2}\dots \ydiagram{1}}_{n+1} \\
    (\frac{n-1}{2},\frac{n+1}{2}) + (\frac{n+1}{2},\frac{n-1}{2}): & \quad \ydiagram{1,1}\underbrace{\ydiagram{1}\dots \ydiagram{1}}_{n-1} \\
    (\frac{n-1}{2},\frac{n+1}{2}) - (\frac{n+1}{2},\frac{n-1}{2}): & \quad \overline{\ydiagram{1,1}}\underbrace{\ydiagram{1}\dots \ydiagram{1}}_{n-1}\,.
\end{align}
For the $(1,1)$ representation, if $n=1$, $\mathcal{M}$ is a $2\times 2$ matrix:
\begin{align}
    \begin{pmatrix}
        (\frac{3}{2},\frac{3}{2}) & (\frac{3}{2},\frac{1}{2}) \\
        (\frac{1}{2},\frac{3}{2}) & (\frac{1}{2},\frac{1}{2})
    \end{pmatrix}\,,
\end{align}
and the corresponding $SO(4)$ SYDs are:
\begin{align}
    (\frac{3}{2},\frac{3}{2}): & \quad \ydiagram{3} \\
    (\frac{3}{2},\frac{1}{2}) + (\frac{1}{2},\frac{3}{2}): & \quad \ydiagram{2,1} \\
    (\frac{3}{2},\frac{1}{2}) - (\frac{1}{2},\frac{3}{2}): & \quad \overline{\ydiagram{1,1}}\ydiagram{1}\,.
\end{align}
If $n\geq 2$, $\mathcal{M}$ is a $3\times 3$ matrix:
\begin{align}
    \begin{pmatrix}
        (\frac{n+2}{2},\frac{n+2}{2}) & (\frac{n+2}{2},\frac{n}{2}) & (\frac{n+2}{2},\frac{n-2}{2}) \\
        (\frac{n}{2},\frac{n+2}{2}) & (\frac{n}{2},\frac{n}{2}) & (\frac{n}{2},\frac{n-2}{2}) \\ 
        (\frac{n-2}{2},\frac{n+2}{2}) & (\frac{n-2}{2},\frac{n}{2}) & (\frac{n-2}{2},\frac{n-2}{2})
    \end{pmatrix}\,,
\end{align}
yielding the following SYDs:
\begin{align}
    (\frac{n+2}{2},\frac{n+2}{2}):&\quad \underbrace{\ydiagram{2}\dots\ydiagram{1}}_{n+2} \\
    (\frac{n+2}{2},\frac{n}{2}) + (\frac{n}{2},\frac{n+2}{2}) :&\quad \ydiagram{1,1}\underbrace{\ydiagram{1}\dots\ydiagram{1}}_{n} \\
    (\frac{n+2}{2},\frac{n}{2}) - (\frac{n}{2},\frac{n+2}{2}) :&\quad \overline{\ydiagram{1,1}}\underbrace{\ydiagram{1}\dots\ydiagram{1}}_{n} \\
    (\frac{n+2}{2},\frac{n-2}{2}) + (\frac{n-2}{2},\frac{n+2}{2}) :&\quad \ydiagram{2,2}\underbrace{\ydiagram{1}\dots\ydiagram{1}}_{n-2} + \overline{\ydiagram{2,2}}\underbrace{\ydiagram{1}\dots\ydiagram{1}}_{n-2} \\
    (\frac{n+2}{2},\frac{n-2}{2}) - (\frac{n-2}{2},\frac{n+2}{2}) :&\quad \overline{\ydiagram{1,1}}\ydiagram{1,1}\underbrace{\ydiagram{1}\dots\ydiagram{1}}_{n-2} + \ydiagram{1,1}\overline{\ydiagram{1,1}}\underbrace{\ydiagram{1}\dots\ydiagram{1}}_{n-2}\,.
\end{align}
Considering a general representation $(j_L, j_L-m)$ satisfying $2j_L-m \in \mathbb{Z}$, the tensor product $(j_L, j_L-m) \otimes (\frac{n}{2},\frac{n}{2})$ can be discussed for two different cases:
\begin{align}
    (1) \quad \frac{n}{2} \leq j_L-m\,,\quad (2) \quad \frac{n}{2} > j_L-m\,.
\end{align}
For the (1) case, the matrix $\mathcal{M}$ is a square matrix with dimension $(n+1)\times(n+1)$, which features elements residing on a specific subdiagonal and can be explicitly structured as:
\begin{align}
    \mathcal{M} = \begin{pmatrix}
    (\frac{n}{2}+j_L,\frac{n}{2}+j_L-m) & \dots & (\frac{n}{2}+j_L, j_L-m-\frac{n}{2}) \\
    \vdots & \ddots & \vdots \\
    (j_L-\frac{n}{2},\frac{n}{2}+j_L-m) & \dots & (j_L-\frac{n}{2}, j_L-m-\frac{n}{2})
    \end{pmatrix}\,.
\end{align}
Applying this framework to the $(1,0)$ representation, if $n=1$, the matrix $\mathcal{M}$ is:
\begin{align}
    \begin{pmatrix}
        (\frac{3}{2},\frac{1}{2}) \\
        (\frac{1}{2},\frac{1}{2}) 
    \end{pmatrix}\,,
\end{align}
both of which are allowed, and the corresponding $SO(4)$ SYDs are:
\begin{align}
    (\frac{3}{2},\frac{1}{2}):\quad & \ydiagram{1,1}\ydiagram{1} + \overline{\ydiagram{1,1}}\ydiagram{1} \\
    (\frac{1}{2},\frac{1}{2}):\quad & \overline{\ydiagram{1}}=\ydiagram{1,1,1}\,.
\end{align}
If $n\geq 2$, the matrix is:
\begin{align}
    \begin{pmatrix}
        (\frac{n}{2}+1,\frac{n}{2}) \\
        (\frac{n}{2},\frac{n}{2}) \\
        (\frac{n}{2}-1,\frac{n}{2}) 
    \end{pmatrix}\,,
\end{align}
only the first two of which are allowed. The corresponding $SO(4)$ SYDs are:
\begin{align}
    (\frac{n}{2}+1,\frac{n}{2}): \quad & \ydiagram{1,1}\underbrace{\ydiagram{1}\dots\ydiagram{1}}_{n} + \overline{\ydiagram{1,1}}\underbrace{\ydiagram{1}\dots\ydiagram{1}}_{n} \\
    (\frac{n}{2},\frac{n}{2}): \quad & \overline{\ydiagram{1}}\underbrace{\ydiagram{1}\dots\ydiagram{1}}_{n-1} = \ydiagram{1,1,1}\underbrace{\ydiagram{1}\dots\ydiagram{1}}_{n-1}\,.
\end{align}
\subsection{Expansion on the tensor Basis}

To demonstrate how the theoretical results constructed in this work are practically applied to physical matrix elements and FFs via tensor reduction, we present two explicit examples.

Let us first consider a tensor current defined as $T^{\mu\nu} = \overline{\psi}\sigma^{\mu\nu}\psi + i\varepsilon^{\mu\nu\rho\lambda}\overline{\psi}\sigma_{\rho\lambda}\psi$. When coupled with $n$ covariant derivatives, the corresponding matrix element is defined by inserting the derivatives between the two $\psi$ fields of the explicit expression:
\begin{equation}
    \mathcal{M}^{\mu\nu\mu_1\dots\mu_n} \equiv \langle P_2, S_2 | \overline{\psi}\sigma^{\mu\nu} iD^{\mu_1} \dots iD^{\mu_n} \psi + i\varepsilon^{\mu\nu\rho\lambda}\overline{\psi}\sigma_{\rho\lambda} iD^{\mu_1} \dots iD^{\mu_n} \psi | P_1, S_1 \rangle\,.
\end{equation}
This physical matrix element is parameterized as a linear superposition of our constructed irreducible tensor bases. The total symmetrization of indices leads to the following explicit examples for different highest-weight representations $(j_L, j_R)$:
\begin{align}
    (\frac{n}{2}+1,\frac{n}{2}): \quad & T^{(\mu|\nu|}D^{\mu_1}D^{\mu_2}\dots D^{\mu_n)} \notag \\
    &= \overline{\psi}\sigma^{(\mu|\nu|}D^{\mu_1}D^{\mu_2}\dots D^{\mu_n)}\psi + i\varepsilon^{(\mu|\nu\rho\lambda|}\overline{\psi}\sigma_{\rho\lambda}D^{\mu_1}D^{\mu_2}\dots D^{\mu_n)}\psi \\
    (\frac{n}{2},\frac{n}{2}): \quad & T^{(\mu|\nu}D_\nu D^{\mu_1}\dots D^{\mu_{n-1})} \notag \\
    &= \overline{\psi}\sigma^{(\mu|\nu|}D_\nu D^{\mu_1}\dots D^{\mu_{n-1})}\psi + i\varepsilon^{(\mu|\nu\rho\lambda|}\overline{\psi}\sigma_{\rho\lambda}D_\nu D^{\mu_1}\dots D^{\mu_{n-1})}\psi \notag \\
    &\rightarrow \overline{\psi}\sigma^{(\mu|\nu|}D_\nu D^{\mu_1}\dots D^{\mu_{n-1})}\psi + i\varepsilon^{(\mu|\nu\rho\lambda|}\overline{\psi}\sigma_{\lambda\delta}D^\delta D^{\mu_1}\dots D^{\mu_{n-1})}\psi\,.
\end{align}
Here, the resulting tensor structure for $(\frac{n}{2}+1,\frac{n}{2})$ corresponds to a mixed-symmetry Lorentz tensor. In the spinor formalism, this irreducible structure is constructed by the pair of spinor SYDs: $(\underbrace{\ydiagram{1}\dots\ydiagram{1}}_{n+2}, \underbrace{\ydiagram{1}\dots\ydiagram{1}}_{n})$, signifying an asymmetric dotted-undotted spinor configuration. The representation $(\frac{n}{2},\frac{n}{2})$ shown above is simply an example of a redundant trace term that must be eliminated. Although this first example is illustrated in terms of $s=1/2$ particles, the procedure applies universally to any spin, effectively mapping foundational group-theoretical constructions directly to the fully-parameterized, irreducible Lorentz structures required for modeling arbitrary nonlocal FFs.

Now, let us consider the second example. Taking the quark-antiquark vector operator ($\Gamma = \gamma^\mu$) as a prototypical case:
\begin{equation}
    \bar{\psi}(0) \gamma^\mu \psi(z) = \sum_{k=0}^{\infty} \frac{1}{k!} z_{\nu_1} \dots z_{\nu_k} \bar{\psi}(0) \gamma^\mu \overset{\leftrightarrow}{D}^{\nu_1} \dots \overset{\leftrightarrow}{D}^{\nu_k} \psi(0)\,.
\end{equation}
Now we can perform the tensor reduction for the matrix elements of $\bar{\psi}(0) \gamma^\mu \overset{\leftrightarrow}{D}^{\nu_1} \dots \overset{\leftrightarrow}{D}^{\nu_k} \psi(0)$. Assuming that all trace components have been systematically subtracted (i.e., working entirely with traceless tensors), we now examine the tensor reduction of the matrix element of the local operator at order $k$. The matrix element takes the form 
\begin{align}
\mathcal{M}^{\mu \nu_1 \dots \nu_k} \equiv\langle P_2, S_2 | \bar{\psi}(0) \gamma^\mu i\overset{\leftrightarrow}{D}^{\nu_1} \dots i\overset{\leftrightarrow}{D}^{\nu_k} \psi(0) | P_1, S_1 \rangle_{\text{traceless}}\,, 
\end{align}
where the $k$ indices of the covariant derivatives are totally symmetric, transforming as a single-row spinor SYD with $k$ boxes. The single Lorentz index from the Dirac matrix $\gamma^\mu$ corresponds to a single-box spinor SYD.

According to the L-R rule, the tensor product of these two representations is decomposed by adding the single box to the row of $k$ boxes in all permissible ways. Explicitly, this decomposition is given by:
\begin{equation}
    \ydiagram{1} \quad \otimes \quad \underbrace{\ydiagram{2}\dots\ydiagram{1}}_{k} \quad = \quad \underbrace{\ydiagram{2}\dots\ydiagram{1}}_{k+1} \quad \oplus \quad \ydiagram{1,1}\underbrace{\ydiagram{1}\dots\ydiagram{1}}_{k-1}\,.
\end{equation}
The spinor SYD decomposition then dictates the following exact splitting:
\begin{equation}
    \mathcal{M}^{\mu \nu_1 \dots \nu_k} z_{\nu_1} \dots z_{\nu_k} = \mathcal{M}^{\{\mu \nu_1 \dots \nu_k\}} z_{\nu_1} \dots z_{\nu_k} + \mathcal{M}^{[\mu \nu_1] \nu_2 \dots \nu_k} z_{\nu_1} z_{\nu_2} \dots z_{\nu_k}\,.
\end{equation}
Here, the resulting tensor structures of the matrix elements are mapped to their corresponding highest-weight representations $(j_L, j_R)$ and their spinor SYDs in the $SL(2, \mathbb{C})$ formalism:
\begin{itemize}
    \item The totally symmetric tensor SYD $\underbrace{\ydiagram{2}\dots\ydiagram{1}}_{k+1}$ corresponds to the $(\frac{k+1}{2}, \frac{k+1}{2})$ representation. In the spinor formalism, this irreducible matrix element structure is represented by the spinor SYD $(\underbrace{\ydiagram{1}\dots\ydiagram{1}}_{k+1}, \underbrace{\ydiagram{1}\dots\ydiagram{1}}_{k+1})$, signifying a symmetric dotted-undotted spinor.
    \item The hook-shaped tensor SYD $\ydiagram{1,1}\underbrace{\ydiagram{1}\dots\ydiagram{1}}_{k-1}$ corresponds to the mixed-symmetry representation $(\frac{k+1}{2}, \frac{k-1}{2}) \oplus (\frac{k-1}{2}, \frac{k+1}{2})$. Its spinor equivalent is given by the direct sum of spinor SYDs: $(\underbrace{\ydiagram{1}\dots\ydiagram{1}}_{k+1}, \underbrace{\ydiagram{1}\dots\ydiagram{1}}_{k-1}) \oplus (\underbrace{\ydiagram{1}\dots\ydiagram{1}}_{k-1}, \underbrace{\ydiagram{1}\dots\ydiagram{1}}_{k+1})$.
\end{itemize}

Crucially, all these irrep structures can be systematically and efficiently constructed within the spinor representation using the general methodology developed in this work, ensuring their applicability to target particles of any arbitrary spin.

For convenience, we list several other commonly used quark and gluon twist two local operators obtained from the expansion of nonlocal quark-antiquark and gluon operators and the tensor reduction, together with their corresponding highest weight Lorentz representations $(j_L,j_R)$.\\
Firstly, the twist-2 local operator generated by bilocal gluon operator,
\begin{equation}
    F^{\{\mu\alpha}
    i\overleftrightarrow{\mathcal{D}}^{\mu_1}
    \dots
    i\overleftrightarrow{\mathcal{D}}^{\mu_n}
    F^{\nu\}\beta}\,,
\end{equation}
 transforms in the irreducible Lorentz representation
$\left(\frac{n+4}{2},\frac{n}{2}\right)
\oplus
\left(\frac{n}{2},\frac{n+4}{2}\right)$.\\
Secondly, the twist-2 local operator generated by bilocal quark-antiquark antisymmetric tensor operator,
\begin{equation}
    \overline{\psi}
    i\overleftrightarrow{\mathcal{D}}^{\{\mu_1}
    \dots
    i\overleftrightarrow{\mathcal{D}}^{\mu_{n-1}}
    \sigma^{\mu_n\}\alpha}
    \psi\,,
\end{equation}
transforms in the mixed symmetry irreducible Lorentz representation
$\left(\frac{n+1}{2},\frac{n-1}{2}\right)
\oplus
\left(\frac{n-1}{2},\frac{n+1}{2}\right)$. \\
Furthermore, the twist-2 local operator generated by bilocal quark-antiquark  axial vector operator,
\begin{equation}
    \overline{\psi}_q
    i\overleftrightarrow{\mathcal{D}}^{\{\mu_1}
    \dots
    i\overleftrightarrow{\mathcal{D}}^{\mu_{n-1}}
    \gamma^{\mu_n\}}
    \gamma_5
    \psi_q\,,
\end{equation}
transforms in the totally symmetric Lorentz representation
$\left(\frac{n}{2},\frac{n}{2}\right)$. 

\section{Conclusion}
\label{sec:conclusion}

In this study, we have developed a comprehensive group-theoretical technique to construct fully independent, Lorentz-covariant tensor bases for arbitrary matrix elements. Instead of bypassing traditional tensor reductions, we translate the tensor indexes into the $SL(2,\mathbb{C})$ spinor indexes. By exploiting spinor Young tableaux, we avoid the cumbersome redundancy elimination required in the direct tensor representation. This allows for a streamlined and exact identification of independent structures before reverting to the standard tensor representation.
Furthermore, the integration of the Hilbert series with non-relativistic counting techniques provides the exact number of allowed structures, ensuring the validity of the results.

Leveraging this algebraic machinery, we have successfully derived the exact covariant structures for targets of spin-$1/2$, $1$, $3/2$, and $2$. Building on these explicit constructions, our constraint analysis identifies a redundancy in the existing spin-$2$ parameterization of matrix elements of rank-2 tensor operators: while Ref.~\cite{Cotogno:2019vjb} presents 20 structures, our construction yields only 19 independent structures. This result agrees with the nonrelativistic counting scheme, which provides a more efficient way to determine the number of independent structures under $P$ and $T$ conservation than the Hilbert series approach.
More importantly, we provided the inaugural complete parameterizations for higher-spin systems, specifically spin-$3/2$ and spin-$2$ particles. To maximize their utility, all derived bases have been explicitly engineered to be eigenstates under $P$ and $T$ transformations.

We summarize our main results here. For spin-1/2 particles, the matrix elements of the scalar operator $\mathcal{O}$, vector operator $\mathcal{J}^{\mu}$, the antisymmetric tensor operator $\mathcal{A}^{[\mu\nu]}$,the symmetric tensor operator $\mathcal{O}^{(\mu\nu)}$ with non-zero trace, and the tensor operator $O^{\mu\nu}$ with no index symmetry in arbitrary forms, are given by
\begin{align}
\langle p_1, \sigma | \mathcal{O} | p_2, \sigma' \rangle
&= \{ \text{irrep}(0,0)\, \text{of Tab.}\, \ref{tab:tensor_fermion} \}\,, \\
\langle p_1, \sigma | \mathcal{J}^{\mu} | p_2, \sigma' \rangle
&= \{ \text{irrep}(\frac{1}{2},\frac{1}{2})\, \text{of Tab.}\ref{tab:tensor_fermion} \}\,,\\
\langle p_1, \sigma | \mathcal{A}^{[\mu\nu]} | p_2, \sigma' \rangle
&=   \{ \text{irrep}(1\,,0)\oplus (0\,,1)\, \text{of Tab.}\ref{tab:tensor_fermion} \}\,, \\
\langle p_1, \sigma | \mathcal{O}^{(\mu\nu)} | p_2, \sigma' \rangle
&= g^{\mu\nu} \{  \text{irrep}(0,0)\, \text{of Tab.}\, \ref{tab:tensor_fermion} \}\text{(trace terms)}\notag\\
   &+ \{ \text{irrep}(1,1)\, \text{of Tab.}\, \ref{tab:tensor_fermion} \}\text{(traceless terms)}\,,\\
\langle p_1, \sigma | \mathcal{O}^{\mu\nu} | p_2, \sigma' \rangle
&= g^{\mu\nu} \{  \text{irrep}(0,0)\, \text{of Tab.}\, \ref{tab:tensor_fermion} \}
  + \{ \text{irrep}(1\,,0)\oplus (0\,,1)\, \text{of Tab.}\, \ref{tab:tensor_fermion} \} \notag \\
&+ \{ \text{irrep}(1,1)\, \text{of Tab.}\, \ref{tab:tensor_fermion} \}\,,
\end{align}
For spin-1 particles, we have
\begin{align}
\langle p_1, \sigma | \mathcal{O} | p_2, \sigma' \rangle
&= \{ \text{irrep}(0,0)\, \text{of Tab.}\, \ref{tab:tensor_boson_1} \}\,, \\
\langle p_1, \sigma | \mathcal{J}^{\mu} | p_2, \sigma' \rangle
&= \{ \text{irrep}(\frac{1}{2},\frac{1}{2})\, \text{of Tab.}\ref{tab:tensor_boson_1} \}\,,\\
\langle p_1, \sigma | \mathcal{A}^{[\mu\nu]} | p_2, \sigma' \rangle
&=   \{ \text{irrep}(1\,,0)\oplus (0\,,1)\, \text{of Tab.}\ref{tab:tensor_boson_1} \}\,, \\
\langle p_1, \sigma | \mathcal{O}^{(\mu\nu)} | p_2, \sigma' \rangle
&= g^{\mu\nu} \{  \text{irrep}(0,0)\, \text{of Tab.}\, \ref{tab:tensor_boson_1} \}\text{(trace terms)}\notag\\
   &+ \{ \text{irrep}(1,1)\, \text{of Tab.}\, \ref{tab:tensor_boson_2} \}\text{(traceless terms)}\,,\\
\langle p_1, \sigma | \mathcal{O}^{\mu\nu} | p_2, \sigma' \rangle
&= g^{\mu\nu} \{  \text{irrep}(0,0)\, \text{of Tab.}\, \ref{tab:tensor_boson_1} \}
  + \{ \text{irrep}(1\,,0)\oplus (0\,,1)\, \text{of Tab.}\, \ref{tab:tensor_boson_1} \} \notag \\
&+ \{ \text{irrep}(1,1)\, \text{of Tab.}\, \ref{tab:tensor_boson_2} \}\,,
\end{align}
For spin-3/2 particles, we have
\begin{align}
\langle p_1, \sigma | \mathcal{O} | p_2, \sigma' \rangle
&= \{ (\ref{eq:3/2-00-i}) \,\text{to}\, (\ref{eq:3/2-00-f}) \} \\
\langle p_1, \sigma | \mathcal{O}^{\mu} | p_2, \sigma' \rangle
&= \{ \text{from Eq.}(\ref{eq:3/2-1/21/2-i}) \,\text{to}\, (\ref{eq:3/2-1/21/2-f}) \} \\
\langle p_1, \sigma | \mathcal{A}^{\mu\nu} | p_2, \sigma' \rangle
&=  \{ \text{from Eq.}(\ref{eq:3/2-1001-i}) \,\text{to}\, (\ref{eq:3/2-1001-f}) \} \,\\
\langle p_1, \sigma | \mathcal{O}^{(\mu\nu)} | p_2, \sigma' \rangle
&= g^{\mu\nu} \{ \text{from Eq.}(\ref{eq:3/2-00-i}) \,\text{to}\, (\ref{eq:3/2-00-f}) \}\text{(trace terms)}\notag\\
  &+ \{ \text{from Eq.}(\ref{eq:3/2-11-i}) \,\text{to}\, (\ref{eq:3/2-11-f}) \}\text{(traceless terms)}\,\\
\langle p_1, \sigma | \mathcal{O}^{\mu\nu} | p_2, \sigma' \rangle
&= g^{\mu\nu} \{ \text{from Eq.}(\ref{eq:3/2-00-i}) \,\text{to}\, (\ref{eq:3/2-00-f}) \}
  + \{ \text{from Eq.}(\ref{eq:3/2-1001-i}) \,\text{to}\, (\ref{eq:3/2-1001-f}) \} \notag \\
&\quad + \{ \text{from Eq.}(\ref{eq:3/2-11-i}) \,\text{to}\, (\ref{eq:3/2-11-f}) \}\,.
\end{align}
For spin-2 particles, we have
\begin{align}
\langle p_1, \sigma | \mathcal{O} | p_2, \sigma' \rangle
&= \{ (\ref{eq:2-00-i}) \,\text{to}\, (\ref{eq:2-00-f}) \}\,, \\
\langle p_1, \sigma | \mathcal{J}^{\mu} | p_2, \sigma' \rangle
&= \{ \text{from Eq.}(\ref{eq:2-1/21/2-i}) \,\text{to}\, (\ref{eq:2-1/21/2-f}) \} \,,\\
\langle p_1, \sigma | \mathcal{A}^{[\mu\nu]} | p_2, \sigma' \rangle
&=  \{ \text{from Eq.}(\ref{eq:2-1001-i}) \,\text{to}\, (\ref{eq:2-1001-f}) \} \,, \\
\langle p_1, \sigma | \mathcal{O}^{(\mu\nu)} | p_2, \sigma' \rangle
&= g^{\mu\nu} \{ \text{from Eq.}(\ref{eq:2-00-i}) \,\text{to}\, (\ref{eq:2-00-f}) \}\text{(trace terms)}\notag\\
   &+ \{ \text{from Eq.}(\ref{eq:2-11-i}) \,\text{to}\, (\ref{eq:2-11-f}) \}\text{(traceless terms)}\,,\\
\langle p_1, \sigma | \mathcal{O}^{\mu\nu} | p_2, \sigma' \rangle
&= g^{\mu\nu} \{ \text{from Eq.}(\ref{eq:2-00-i}) \,\text{to}\, (\ref{eq:2-00-f}) \}
  + \{ \text{from Eq.}(\ref{eq:2-1001-i}) \,\text{to}\, (\ref{eq:2-1001-f}) \} \notag \\
&\quad + \{ \text{from Eq.}(\ref{eq:2-11-i}) \,\text{to}\, (\ref{eq:2-11-f}) \}\,,
\end{align}
where $\{ \dots \}$ means the linear superpositions of the basis elements included in the brace bracket with the superposition coefficients as the FFs.

As the application of the bases, we discussed the expansion of the matrix elements of general nonlocal operators on the bases constructed by our technique as well as the tensor reduction of the nonlocal operators, which are applicable to particles of arbitrary spin. At the same time, several common operators  such as gauge invariant bilocal quark-antiquark and gluon operators and the tower of operators generated by them via Taylor expansion are taken as examples for explicit illustration. 

Overall, we generalize the algebraic methods for invariants, such as the Young tensor method and the Hilbert series, to the covariants under Lorentz transformation, broadening their validity. Experimentally, as experimental facilities push towards higher precision and increasingly probe higher-spin targets, these theoretical tools will be vital for accurately extracting fundamental multidimensional distributions, including GPDs and TMDs, from corresponding scattering amplitudes.


\acknowledgments

We would like to thank Yi-Ning Wang and Zhite Yu for careful and valuable comments on the draft. This work is supported by the National Science Foundation of China under Grants No. 12347105, No. 12375099 and No. 12447101, and the National Key Research and Development Program of China Grant No. 2020YFC2201501, No. 2021YFA0718304.


\appendix 

\section{Redundancy Reduction of Hilbert Series}
\label{app:hs}
In this appendix, we will illustrate the calculations of the Hilbert series that we are interested in for the spin-1 FFs, and show how they are reduced by the relations such as Eq.~\eqref{eq:relation1} to the final expressions. 

The 4 building blocks are $P\,,q\,,\varepsilon_2^\dagger$ and $\varepsilon_1$, which are all in the irrep $(\frac{1}{2},\frac{1}{2})$ of the Lorentz group,
\begin{equation}
    g_P = g_q = g_{\varepsilon_2^\dagger} = g_{\varepsilon_1} = \text{diag}\left(xy,\frac{x}{y},\frac{y}{x},\frac{1}{xy}\right)\,.
\end{equation}
The Molien-Weyl formula becomes
\begin{equation}
    \mathcal{HS}^{R}_G = \int_G d\mu_G \frac{\chi_R}{\det[1-P g_P]\det[1-q g_q]\det[1-\varepsilon_2^\dagger g_{\varepsilon_2^\dagger}]\det[1-\varepsilon_1 g_{\varepsilon_1}]}\,,
\end{equation}
where the Haar measure $d\mu_G$ is
\begin{equation}
d\mu_G = \frac{1}{4}\frac{1}{(2\pi i)^2}\frac{dx dy}{xy} (1-x^2)(1-\frac{1}{x^2})(1-y^2)(1-\frac{1}{y^2})\,.
\end{equation}
Consequently, the integral over the maximal torus is reduced to the contour integral along the unit circle in the complex $x$- and $y$-planes, respectively. To evaluate the integral, we assume the variables $P, q, \varepsilon_2^\dagger$ and $\varepsilon_1$ are all small, then the theorem of residues applies. The resultant Hilbert series shares the same denominator,
\begin{equation}
    D' = (1-P^2)(1-Pq)(1-q^2)(1-P\varepsilon_1)(1-q\varepsilon_1)(1-P\varepsilon_2^\dagger)(1-q\varepsilon_2^\dagger)(1-\varepsilon_1^2)(1-\varepsilon_2^\dagger{}^2)(1-\varepsilon_1\varepsilon_2^\dagger)\,,
\end{equation}
which means the covariant tensors form modules over the same ring. Each factor corresponds to an invariant composed of the indicating variables, for example, the factor $(1-P^2)$ corresponds to the invariant $P^2=P^\mu P_\mu$. Apparently, not all the invariants are needed to compose nonequivalent tensors corresponding to the FFs. The ones that are composed solely of momenta or contain more than one $\varepsilon_1/\varepsilon_2^\dagger$ correspond to no FFs, thus should be eliminated. Moreover, the relations in Eq.~\eqref{eq:relation1} eliminate all the contractions between the momentum $P$ and the wave functions.
Thus, the reduced denominator is
\begin{equation}
    D'\rightarrow D =(1-q\varepsilon_1)(1-q\varepsilon_2^\dagger)(1-\varepsilon_1\varepsilon_2^\dagger)\,.
\end{equation}
As for the numerators, different characters $\chi_R$ give different results,
\begin{align}
    N_{(\frac{1}{2},\frac{1}{2})} &= P+q + \varepsilon_1 + \varepsilon_2^\dagger + Pq\varepsilon_1 + Pq\varepsilon_2^\dagger + P\varepsilon_1\varepsilon_2^\dagger + q\varepsilon_1\varepsilon_2^\dagger\,, \\
    N_{(1,0)} &= Pq + P\varepsilon_1 + q\varepsilon_1 + P\varepsilon_2^\dagger + q\varepsilon_2^\dagger + \varepsilon_1 \varepsilon_2^\dagger\,, \\
    N_{(0,1)} &= Pq + P\varepsilon_1 + q\varepsilon_1 + P\varepsilon_2^\dagger + q\varepsilon_2^\dagger + \varepsilon_1 \varepsilon_2^\dagger\,, \\
    N_{(1,1)} &= P^2 + Pq + q^2 + P\varepsilon_1 + q\varepsilon_1 + P\varepsilon_2^\dagger + q\varepsilon_2^\dagger + P^2q\varepsilon_1 + q^2P \varepsilon_1 + P^2q\varepsilon_2^\dagger \notag \\
    &+ q^2P \varepsilon_2^\dagger + \varepsilon_1 \varepsilon_2^\dagger + P^2\varepsilon_1 \varepsilon_2^\dagger + 3Pq \varepsilon_1\varepsilon_2^\dagger + q^2\varepsilon_1\varepsilon_2^\dagger\,,
\end{align}
Similar reductions give the numerators of the irreps $(\frac{1}{2},\frac{1}{2}),(1,1)$. The representation $(1,0)\oplus (0,1)$ is special, since it is not irreducible. The associate numerator is not simply the sum of the ones of the irreps $(1,0)$ and $(0,1)$. Actually, the relation 
\begin{equation}
    -(\epsilon^{\mu\nu\rho\lambda}P_\mu q_\nu \varepsilon_{1\rho}\varepsilon_{2\lambda} ^\dagger) \times (\epsilon^{\alpha\beta\delta\kappa}P_\delta q_\kappa) = P^\alpha q^\beta (P\cdot \varepsilon_1)(q\cdot \varepsilon_2^\dagger) + \text{permutations}\,,
\end{equation}
eliminate a covariant of degree $P^2 q^2 \varepsilon_1 \varepsilon_2^\dagger$. Thus, there is a minus term in the numerator.

\section{Tensor-Representation Construction}
\label{app:tensor}

As discussed before, the tensor representations are characterized by the permutation symmetries of the indices, which are described by SYDs. In particular, the tensor SYDs are related to the spinor ones via the products of the two Young operators corresponding to the left- and right-handed spinors, respectively. For example, the correspondence of the SYDs between the two representations up to rank-2 tensors is presented in Tab.~\ref{tab:corres1}.
\begin{table}[]
\ytableausetup{centertableaux}
\renewcommand{\arraystretch}{1.8}
    \centering
    \begin{tabular}{|c|c|c|}
\hline
irrep & tensor SYD & spinor SYD \\
\hline
$(0,0)$ & 1 & 1 \\
\hline
$(\frac{1}{2},\frac{1}{2})$ & $\ydiagram{1}$ & $(\ydiagram{1},\ydiagram{1})$ \\
\hline
$(1,0)$ & $\ydiagram{1,1}$ & $(\ydiagram{2},\ydiagram{1,1})$ \\
\hline
$(0,1)$ & $\overline{\ydiagram{1,1}}$ & $(\ydiagram{1,1},\ydiagram{2})$ \\
\hline
$(1,1)$ & $\ydiagram{2}$ & $(\ydiagram{2},\ydiagram{2})$ \\
\hline
    \end{tabular}
    \caption{The correspondence between the tensor SYDs and the spinor SYDs of the Lorentz irreps.}
    \label{tab:corres1}
\end{table}

Because of the invariant tensor $\delta_{\mu\nu}$ or $g_{\mu\nu}$, the outer product of the $SO(N)$ SYDs is more complicated than $SU(N)$, since the L-R rules can not apply simply. Alternatively, we specify the SYD for each irrep and find out all the SSYTs as the irreducible tensors. For example, the $(1,1)$ representation corresponds to the tensor SYD $\ydiagram{2}$, implying the two indices are symmetric.
However, this single diagram is not enough, since for the tensor fields with indices exceeding the boxes of the SYD, their contractions may contribute as trivial representations, 
\begin{equation}
    1 \sim \ydiagram{1,1,1,1}\,.
\end{equation} 
Due to similar reasons, the dual SYDs are also needed. Thus, the necessary SYDs of the symmetric irrep $(j,j)$ are 
\begin{equation}
\ytableausetup{aligntableaux=top}
    \overbrace{\ydiagram{1}\dots \ydiagram{1}}^{2j}\,,\quad \overbrace{\ydiagram{2,1,1}\dots\ydiagram{1}}^{2j}\,, \quad \ydiagram{1,1,1,1}\dots \overbrace{\ydiagram{1}\dots\ydiagram{1}}^{2j}\,,\quad \ydiagram{1,1,1,1}\dots \ydiagram{1,1,1,1} \!\overbrace{\ydiagram{2,1,1}\dots\ydiagram{1}}^{2j}\,,
\end{equation}
where the last two are included as outer products, for example,
\begin{equation}
    \ydiagram{1,1,1,1}\dots \ydiagram{1,1,1,1} \!\overbrace{\ydiagram{2,1,1}\dots\ydiagram{1}}^{2j} = \ydiagram{1,1,1,1}\dots \ydiagram{1,1,1,1}\times \!\overbrace{\ydiagram{2,1,1}\dots\ydiagram{1}}^{2j}\,.
\end{equation}
The next step is to fill the SYDs with the indices $1\sim 4$ and find all the SSYTs. However, such a program generates redundant tensors, which can be reduced in principle by the relations among tensors. Consequently, the interpretation of the tesnor SYDs is of no disadvantage. Especially for the tensor fields of spin $>1$, the redundancy is severe. 

For example, the contributing diagrams of the irrep $(\frac{1}{2},\frac{1}{2})$ of the spin-$2$ fields are 
\begin{equation}
    \ydiagram{1}\,,\quad \ydiagram{1,1,1}\,,\quad \ydiagram{2,1,1,1}\,,
\end{equation}
and the resultant tensors are 
\begin{align}
    \ytableaushort{1} &\rightarrow \left\{\begin{array}{l}
        T_{(\frac{1}{2}\,,\frac{1}{2})}^1 = q^\mu ({\varepsilon_2^\dagger}{}^{\rho\nu}{\varepsilon_1}_{\rho\nu}) \\
        T_{(\frac{1}{2}\,,\frac{1}{2})}^2 = q^\mu (q_\lambda{\varepsilon_2^\dagger}{}^{\lambda\nu})(q^\rho{\varepsilon_1}_{\rho\nu}) \\
        T_{(\frac{1}{2}\,,\frac{1}{2})}^3 = q^\mu (q_\delta q_\nu {\varepsilon_2^\dagger}{}^{\delta\nu})(q_\rho q_\lambda \varepsilon_1^{\rho\lambda})
    \end{array}\right. \,,\\
    \ytableaushort{2} &\rightarrow \left\{\begin{array}{l}
        T_{(\frac{1}{2}\,,\frac{1}{2})}^4 = P^\mu ({\varepsilon_2^\dagger}{}^{\rho\nu}{\varepsilon_1}_{\rho\nu}) \\
        T_{(\frac{1}{2}\,,\frac{1}{2})}^5 = P^\mu (q_\lambda{\varepsilon_2^\dagger}{}^{\lambda\nu})(q^\rho{\varepsilon_1}_{\rho\nu}) \\
        T_{(\frac{1}{2}\,,\frac{1}{2})}^6 = P^\mu (q_\delta q_\nu {\varepsilon_2^\dagger}{}^{\delta\nu})(q_\rho q_\lambda \varepsilon_1^{\rho\lambda})
    \end{array}\right. \,,\\
    \ytableaushort{3} &\rightarrow \left\{\begin{array}{l}
        T_{(\frac{1}{2}\,,\frac{1}{2})}^7 = (\varepsilon_1^{\mu\nu}q_\nu)({\varepsilon_2^\dagger}{}^{\rho\lambda}q_\rho q_\lambda) \\
        T_{(\frac{1}{2}\,,\frac{1}{2})}^8 = \varepsilon_1^{\mu\nu}({\varepsilon_2^\dagger}{}_{\nu\rho}q^\rho)
    \end{array}\right. \,,\\
    \ytableaushort{4} &\rightarrow \left\{\begin{array}{l}
        T_{(\frac{1}{2}\,,\frac{1}{2})}^9 = ({\varepsilon_2^\dagger}{}^{\mu\nu}q_\nu)(\varepsilon_1^{\rho\lambda}q_\rho q_\lambda) \\
        T_{(\frac{1}{2}\,,\frac{1}{2})}^{10} = {\varepsilon_2^\dagger}{}^{\mu\nu}({\varepsilon_1}_{\nu\rho}q^\rho)
    \end{array}\right. \,,
\end{align}
\begin{align}
    \ytableaushort{1,2,3} &\rightarrow \left\{\begin{array}{l}
        T_{(\frac{1}{2}\,,\frac{1}{2})}^{11} = \epsilon^{\mu\nu\rho\lambda}q_\nu P_\rho ({\epsilon_1}_{\lambda\alpha}q^\alpha)({\varepsilon_2^\dagger}{}^{\beta\gamma}q_\beta q_\gamma) \\
        T_{(\frac{1}{2}\,,\frac{1}{2})}^{12} = \epsilon^{\mu\nu\rho\lambda}q_\nu P_\rho {\varepsilon_1}_{\lambda\alpha}({\varepsilon_2^\dagger}{}^{\alpha\beta}q_\beta)
    \end{array}\right. \,,\\
    \ytableaushort{1,2,4} &\rightarrow \left\{\begin{array}{l}
        T_{(\frac{1}{2}\,,\frac{1}{2})}^{13} = \epsilon^{\mu\nu\rho\lambda}q_\nu P_\rho ({\varepsilon_2^\dagger}_{\lambda\alpha}q^\alpha)(\varepsilon_1^{\beta\gamma}q_\beta q_\gamma) \\
        T_{(\frac{1}{2}\,,\frac{1}{2})}^{14} = \epsilon^{\mu\nu\rho\lambda}q_\nu P_\rho {\varepsilon_2^\dagger}_{\lambda\alpha}(\varepsilon_1^{\alpha\beta}q_\beta)
    \end{array}\right. \,,\\
    \ytableaushort{1,3,4} &\rightarrow \left\{\begin{array}{l}
        T_{(\frac{1}{2}\,,\frac{1}{2})}^{15} = \epsilon^{\mu\nu\rho\lambda}q_\nu ({\varepsilon_2^\dagger}_{\rho\alpha}q^\alpha)({\varepsilon_1}_{\lambda\beta}q^\beta) \\
        T_{(\frac{1}{2}\,,\frac{1}{2})}^{16} = \epsilon^{\mu\nu\rho\lambda}q_\nu {\varepsilon_2^\dagger}_{\rho\alpha}{\varepsilon_1}_{\lambda}{}^\alpha
    \end{array}\right. \,,\\
    \ytableaushort{2,3,4} &\rightarrow \left\{\begin{array}{l}
        T_{(\frac{1}{2}\,,\frac{1}{2})}^{17} = \epsilon^{\mu\nu\rho\lambda}P_\nu ({\varepsilon_2^\dagger}_{\rho\alpha}q^\alpha)({\varepsilon_1}_{\lambda\beta}q^\beta) \\
        T_{(\frac{1}{2}\,,\frac{1}{2})}^{18} = \epsilon^{\mu\nu\rho\lambda}P_\nu {\varepsilon_2^\dagger}_{\rho\alpha}{\varepsilon_1}_{\lambda}{}^\alpha
    \end{array}\right.\,,
\end{align}
\begin{align}
    \ytableaushort{11,2,3,4} &\rightarrow \left\{\begin{array}{l}
        T_{(\frac{1}{2}\,,\frac{1}{2})}^{19} = (\epsilon^{\alpha\beta\rho\lambda}q_\alpha P_\beta {\varepsilon_1}_{\rho\nu}{\varepsilon_2^\dagger}_\lambda{}^\nu) q^\mu \\
        T_{(\frac{1}{2}\,,\frac{1}{2})}^{20} = (\epsilon^{\alpha\beta\rho\lambda}q_\alpha P_\beta ({\varepsilon_1}_{\rho\nu}q^\nu)({\varepsilon_2^\dagger}_{\lambda\gamma}q^\gamma)) q^\mu \\
    \end{array}\right. \,,\\
    \ytableaushort{12,2,3,4} &\rightarrow \left\{\begin{array}{l}
        T_{(\frac{1}{2}\,,\frac{1}{2})}^{21} = (\epsilon^{\alpha\beta\rho\lambda}q_\alpha P_\beta {\varepsilon_1}_{\rho\nu}{\varepsilon_2^\dagger}_\lambda{}^\nu) P^\mu \\
        T_{(\frac{1}{2}\,,\frac{1}{2})}^{22} = (\epsilon^{\alpha\beta\rho\lambda}q_\alpha P_\beta ({\varepsilon_1}_{\rho\nu}q^\nu)({\varepsilon_2^\dagger}_{\lambda\gamma}q^\gamma)) P^\mu \\
    \end{array}\right. \,,\\
    \ytableaushort{13,2,3,4} &\rightarrow T_{(\frac{1}{2}\,,\frac{1}{2})}^{23} = \epsilon^{\alpha\beta\rho\lambda}q_\alpha P_\beta{\varepsilon_1}_\rho{}^\mu ({\varepsilon_2^\dagger}_{\lambda\nu}q^\nu) \\
    \ytableaushort{14,2,3,4} &\rightarrow T_{(\frac{1}{2}\,,\frac{1}{2})}^{24} = \epsilon^{\alpha\beta\rho\lambda}q_\alpha P_\beta{\varepsilon_2^\dagger}{}_\rho{}^\mu ({\varepsilon_1}_{\lambda\nu}q^\nu)\,.
\end{align}
These 24 tensors exceed the 22 ones obtained from the spinor SYDs, which means redundancies certainly exist. However, all the SSYTs and the associated tensors are on the same footing, and redundancies are difficult to identify and eliminate. On the other hand, the spinor SYDs are easy to do the outer product, thus such redundancies can be constructively eliminated, as shown previously.




\bibliography{biblio}

\end{document}